\documentclass[10pt]{article}
\usepackage{lineno,hyperref}
\usepackage{amssymb}
\usepackage{amsmath}
\usepackage{cite}
\usepackage{newtxtext}       %
\usepackage[varvw]{newtxmath} 
\numberwithin{equation}{section}
\usepackage[bottom]{footmisc}
\usepackage[nottoc]{tocbibind}
\renewcommand{\vec}[1]{\boldsymbol{#1}}
\begin{document}
	\author{Alexander I. Zhmakin\footnote{Ioffe Physical Technical  Institute \& SoftImpact, Ltd., e-mail: a.zhmakin0@gmail.com}}
	\title{The Zoo of Non-Fourier Heat Conduction Models}
	
	\maketitle
	\tableofcontents
	\pagenumbering{arabic}

\begin{abstract}
	The Fourier heat conduction  model 
	is valid for most macroscopic problems. However, it fails when the wave nature of the heat propagation or time lags become dominant and the memory or/and spatial  non-local effects significant  ---  
	in ultrafast heating (pulsed laser heating and melting), rapid solidification of liquid metals, processes in 		glassy polymers near the glass transition temperature, 
	in heat transfer at nanoscale,  in heat transfer in a solid state laser medium at  the high pump density or under the ultra-short pulse duration,  in granular and porous  materials including  polysilicon, at extremely high values of the heat flux, in heat transfer in biological tissues. 
	In common materials the relaxation time ranges from $10^{-8}$ to $10^{-14}$ sec, however, it   could be as high as 1 sec in the degenerate cores of aged stars  and  its reported  values in granular and biological objects varies up to 30 sec.  The paper considers numerous  non-Fourier heat conduction models   that incorporate time non-locality for materials with memory (hereditary materials, including fractional hereditary materials) or/and  spatial non-locality, i.e. materials with non-homogeneous inner structure.
\end{abstract}





\section{Introduction}

The heat conduction is probably the
most important dissipative phenomenon; its distinctive property
is that it does not have a reversible part \cite{sho21}.
The  heat conduction model suggested in the beginning of the  XIX century by Jean Baptiste Joseph Fourier \cite{nar99} is based on the constitutive relation
\begin{equation}
\label{four}
\vec{q} (\vec{r},t) = - \lambda \nabla T(\vec{r},t)
\end{equation} 
\noindent
leading through the energy conservation equation for the solid body in rest (the first law of thermodynamics)
\begin{equation}
\label{cond}
\frac{\partial (\varrho c T)}{\partial t} = \nabla \cdot \vec{q} +Q
\end{equation} 
to the parabolic  equation (called the "Fourier-Kirchhof equation",  "Maxwell-Fourier law" \cite{her19}, or the "Fourier heat conduction equation" (FHCE) \cite{zec15})
\begin{equation}
\label{cond1}
\frac{\partial (\varrho c T)}{\partial t} = \nabla \cdot (\lambda \nabla T) +Q
\end{equation} 
that  for the case of constant thermophysical properties is written as
\begin{equation}
	\frac{\partial T}{\partial t} = \kappa \Delta T
\end{equation}
where $\kappa = \lambda / \varrho c$ is the thermal diffusivity. 
The thermal  conductivity $\lambda$ is an  property that does not depend on the size and geometry of the sample when phonon
transport is diffusive; when the sample size along the transport direction is much smaller than the phonon mean free path (MFP), phonons propagate ballistically across the sample without experiencing appreciable scattering \cite{gu17}. 

The Fourier law can be derived from classical irreversible thermodynamics (CIT) based on local equilibrium hypothesis \cite{gya,leb11}. The entropy production is  
\begin{equation}
	\Sigma = \vec{q} \cdot \nabla \frac{1}{T} \ge 0.
\end{equation}

From the viewpoint of the irreversible thermodynamics, the Fourier law describes a linear relationship between the generalized force (temperature gradient) and the generalized flux (heat flux) \cite{guo07}. For isotropic materials
\begin{equation}
	\vec{q} = \lambda_T \nabla \frac{1}{T} = - \frac{\lambda_T}{T^2} \nabla T. 
\end{equation}
Thus $\lambda = \lambda_T / T^2$ is the Fourier heat conduction coefficient with $\lambda_T$  being a thermodynamic conduction coefficient \cite{sho21}.

The derivation of the Fourier equation is based on the
thermodynamic equilibrium and the continuum
hypothesis, according to which each small element
of the medium  has a local equilibrium
state and  can be described
by the local thermodynamic potentials that are dependent on
the spatial variable and time only through the thermodynamic
parameters. Accepting the local equilibrium principle is
possible only if the rate of change in the system macroparameters
due to the external influences is much less than the
rate of system relaxation to the local equilibrium \cite{sob97}. It is assumed that the
transport laws are valid not only for the entire system, but
also for any arbitrarily small part of it. Thus it is possible to perform a limit transition  in the integral conservation laws
and obtain these laws in the form of differential equations.

If the characteristic microscale of the system and time of its relaxation to equilibrium are significantly less than the characteristic macroscale and total time of the process, then the differential equations derived based on the local equilibrium
principle and continuum hypothesis will be local both in
space and in time. Thus we obtain transport equations that do not contain the relaxation time $\tau$ and the characteristic scale of the microstructure $l$.

The drawback of the FHCE is the instantaneous propagation of disturbances \cite{str11} ("paradox of propagation of thermal signals" \cite{sob18a}, "paradox of heat conduction"  \cite{chr05} or "heat transfer paradox" \cite{gha15}) called  {\em acausality} - it does not forecasts propagation of disturbances along characteristic causal light-cones \cite{her01}.
Onsager  \cite{on} pointed out that "... Fourier law is only approximate description of the process of conduction neglecting the time needed for acceleration of the heat flow".

The Fourier law is valid if \cite{tw,sob94b}
{${L}/{\Lambda} \gg O (1)$},  {${t}/{\tau} \gg O (1)$},
{$T \gg 0^{\circ} K$} 
 where $L$ is the size of the system, $\Lambda$ is MFP of the heat carries, $\tau$ is the relaxation time. The ratio $\Lambda/L$ is the Knudsen number $Kn$ as in the rarefied gas dynamics. 

Heat waves in the form of the second sound \cite{ss} --- the thermal transport regime where the heat is carried by the temperature waves similarly to the propagation of sound wave in gases --- were predicted by Landau \cite{lan47} (see also \cite{tis47}) who studied the behaviour of the quasiparticles in the superfluid liquid helium II using the two-fluid model  and were observed in the helium II at 1.4 K by V. Peshkov in 1944 with velocity about 19 m/sec that is one order of magnitude less than the speed of sound \cite{pesh}. 

Later the second sound was observed at the cryogenic conditions in other materials \cite{war52,shk67,gur68} --- solid helium-3 \cite{ack66,nar75}, sodium  fluoride (at ca. 10 - 20 K \cite{mcn70,van17}), bismuth (at 1.2 - 4.0 K \cite{nar72}), sapphire, strontium titanate $SrTiO_3$  \cite{kor07,mar18}, in pyrolytic graphite at temperature above 100 K \cite{graph}. Hydrodynamic effects of the phonon transport usually observed at   the low temperatures are significant in the graphitic materials even at the moderate temperature due to both the high Debye temperature and the strong anharmonicity \cite{guo21}.

The existence of the hydrodynamic effects in graphite --- a 3D material ---
above the liquid nitrogen temperature was predicted by {\em ab initio} computations by Ding et al. \cite{ding17} who determined  the phonon distribution function in an infinitely large crystal under a constant temperature gradient. The authors found that at 100 K heat transport in graphite is hydrodynamic --- dominated by drifting phonons.  They studied the effect of vacancies that are treated  as the mass disorder and found that the collective phonon drift motion is destroyed when the vacancy concentration is about 0.01\%, but can still be observed when the concentration is about 0.001\%.
The unusual behaviour of graphite the authors attribute to its strong intralayer sp2 hybrid bonding and to the weak van der Waals interlayer interactions. The authors also noted that the reflection symmetry associated with a single graphene layer is broken in the graphite, which opens up more momentum-conserving phonon-phonon scattering channels and results in stronger hydrodynamic features in graphite than graphene.

In all observations of the second 
sound, the dominance of the momentum conserving phonon scattering
(the Normal processes) with respect to the resistive scattering
(the Umklapp processes, the isotope or the impurity scattering) was critical --- the second sound was observed almost exclusively in the very low temperature regime, with the exception of  the experiment by Huberman et al. in the graphite \cite{graph}. Thus the condition for the  detection of the second sound was found to be $\tau_N < \tau_{exp} < \tau_R$, i.e., "the observation times" (time sensitivity) $\tau_{exp}$
must be larger than the normal scattering times $\tau_N$ to allow the momentum redistribution but smaller than the resistive phonon scattering times $\tau_R$ to avoid the decay of the phonon wave packet into the phonon equilibrium distribution \cite{bea21}.
Recently Beardo et al. \cite{bea21}  demonstrated the existence of the second sound in bulk Ge between 7 K and the room temperature
by studying the phase lag of the thermal response under the harmonic high-frequency external excitation. 
 
The wave nature of the heat propagation or the time lags become dominant and the memory, nonlinear or spatial nonlocal effects significant \cite{z21} 
\begin{itemize}
\item{in the ultrafast heating
(the pulsed laser heating and melting \cite{san95,liu01a,melt}, e.g., the heating of metal films \cite{bro87,wan12a,yud15} or films of solid argon \cite{seb96,sho14}), 
welding of metals, surface annealing, sintering of ceramics \cite{yu14}, micro-machining \cite{hen13},
the rapid solidification \cite{mul97,wan00} (e.g., the solidification rates are up to 70 m/sec in pure Ni and Cu-Ni alloy \cite{sob95}),
the glass transition of supercooled liquids  \cite{hil02,sob95},
the glassy polymers near the glass transition temperature \cite{jou91}, 
the heat pulse experiments at the room temperature,
including "the book experiment" \cite{dre93,van17a,kov19a}, the sudden contact of two liquids such as uranium oxide and sodium \cite{sud},}
\item{ in the heat transfer at nanoscale \cite{cah03,zhang07,cah14,che21} (microelectronic and optoelectronic  devices \cite{ver08,ral08,tor18,fra19}, e.g. the hot spots in nanotransistors \cite{art89,chen04,pop06,pop06a,fra18}\footnote{
The integrated schemes contain billions of transistors (about 9 billions CPUs in 2015 \cite{vol16}) that generate huge heat fluxes in a very small area; these "hot spots" become  the bottleneck of the future development in terms of both performance \cite{zhang07}  and reliability \cite{moo14}. The hotspot within the transistor drain region where the energy transfer from electrons to the lattice is the most intense can be 10 nm thick \cite{sve01} and lead to the increase of the drain series, the source injection electrical resistances and thermal-induced breakdown  \cite{pop06}. Sverdrup et al. measured the ballistic phonon conduction near hotspot  and found that the temperature exceeds by 60\% a value predicted by the Fourier law \cite{sve01}.

The Fourier law for heat conduction dramatically over-predicts the rate of the heat dissipation from sources with dimensions smaller than the dominant MFP \cite{sim10,hoog} that is very important in the thermal management in microelectronics.
	The scaling problem for the thermal management in nanoelectronics is mitigated by the phenomenon discovered  by Hoogeboom-Pot et al. \cite{hoog} --- "collective diffusion": when the separation between nanoscale heat sources is small to compared the phonon MFP, phonons can scatter with phonons originated from the neighbouring heat source increasing the heat transfer efficiency to near the diffusion limit.},
the nanostructured devices for the solid-sate energy conversion, e.g., the thermoelectric/thermoionic refrigeration \cite{ven01,chen05,bou08}, 
the nano-electromechanical systems (NEMS) \cite{luo21},
2D materials \cite{gu17},
the heterostructures \cite{gu17,vaz20},
the layered strongly correlated materials \cite{maz21} such as, for example, transition metal oxides in which the strong short-ranged electronic interactions lead to the failure of the independent-electron approximation and which  exhibit the wave-like  temperature propagation \cite{gan16},
the laser plasma in irradiating small targets \cite{vol08}),
the nano suspensions for the radiative cooling \cite{37} and the volumetric solar thermal energy absorption \cite{38},
the heat transfer  in the DNA during its denaturation ("melting") --- unravelling of the double-stranded structure into two single strands \cite{vel11a}, ultra-high thermal isolation across heterogeneously layered  2D materials \cite{vaz19},}
\item{in heat transfer in a solid state laser medium at     the high pump density or under the ultra-short pulse duration \cite{tao06},} 
\item{in the heat transfer in the granular and porous  materials \cite{lui75,gra2,gran,sil13} including  porous silicon (pSi)\footnote{
Porous silicon was discovered in 1956 \cite{uhl56}. Now it is used in Light Emitting Diodes (LEDs), sensors, thermoelectric devices \cite{lee08}, as insulation for microelectronic device; thermal conductivity of pSi is two to five orders of magnitude smaller than that of the bulk Si \cite{lee07,alv10}.
Nanoscale porous materials are also called {\em Cantor} materials.

},
in fractured geological media \cite{fr1,luch11,fr2},
} 
\item {at the extremely high values of the heat flux \cite{mau73}  --- e.g., in the beam deep penetration welding with the heat flux greater than $10^8 W/cm^2$	  \cite{zha04}, }
\item {in the heat transfer in the biological tissues \cite{liu00,liu08,zhou08,z9,ahm12a,str13,sob17,z17,z17a,zhm18}.}
\end{itemize}

The errors due to the use of the Fourier law  at nanoscale are not always important. E.g., Wilson \& Cahill \cite{win15} listed reasons why the errors in the diamond thermal conduction are not essential  in analysis of the thermal management of the  devices with diamond as the heat spreader:
(1) {the magnitude of the ballistic-diffusive effects in the polycrystalline films grown by the chemical vapour deposition  will be smaller than in the single crystal due to the phonon scattering by grain boundaries \cite{cah02};}
(2) {at least for the GaN devices such as HEMTs the substrate thermal conductivity is a major factor in the performance only when the dimensions of the active region  exceed 1 $\mu m$;}
(3) {for a sufficiently high areal density of devices there is no high in-plane temperature gradient.}

The extensions of the Fourier law are important not only in the heat transfer itself but in the related areas such as, e.g., thermoelasticity \cite{te1,te2,te3} and piezoelectric thermoelasticity (PETE) \cite{pte1,pte3,guo18a} as well.

The relaxation time $\tau$ is the characteristic time needed by the system to return to the steady state  after it has been removed from it.
It is related to the mean collision time $\tau_c$ of the particles responsible for the heat transfer. There is no universal relation between $\tau$ and $\tau_c$: in some cases $\tau$ may correspond to just a few $\tau_c$ while in others the difference could be huge (e.g., in the model of the early universe the relaxation time of the shear viscosity could be orders of magnitude larger than the collision time between photons and electrons \cite{her01}).

The origin of the time lag could be the heterogeneous inner structure or  the existence of several energy carriers \cite{tzou09} (e.g., the relaxation between the electron and phonon subsystems \cite{zim} in  solids --- the heat transfer from free electrons to the lattice \cite{ret02,pol15} in the metal heating by the ultra short laser pulses \cite{ani74,qi94,sch02}:  the time necessary to established an electron temperature is under picosecond, for example, for gold is about 800 fs, while the electron-lattice relaxation time is on the order of a few picoseconds \cite{ho14}). 

The biological tissues contain cells, membranes, organelles, superstructures, liquids, solid/soft elements (sometimes the tissues are referred to as to the {\em mesoporous} structure \cite{mag13}).
The heating or cooling of the living tissues induces a series of chemical, electrical and mechanical processes, e.g. diffusion, electrical potential change and osmosis across the cell  membrane; the cell membranes could store energy \cite{liu00}. Thus, the heat propagation involves the multi-mode energy conversion at the different cellular levels \cite{liu00,lin11,nak13}.

The relaxation time is the macroscopic parameter that integrates a series of microscopic interactions and is associated with the communication time between the particles such as photons, electrons, phonons \cite{tan97}.
 
In common materials it ranges from $10^{-8}$ to $10^{-14}$ sec \cite{kam90,sob91,oz94,gal03} (e.g., it equals  3.5 ps \cite{ver08} for Si, 4,5-6.4 ps for Hg, 5.1-7.3 ps for molten Ga \cite{kha15}) --- the non-Fourier effects in these case are small. Saad \& Didlake \cite{sad97} studied the Stefan problem using the Cattaneo model  and found for the Al (the relaxation time $10^{-10}$ - $10^{-12}$ sec) the non-Fourier effects are significant for times  $10^{-9}$ - $10^{-11}$ sec and in the region within $10^{-4}$ - $10^{-5}$ cm from the phase boundary.    

However, the thermal relaxation time could be as high as 1 sec in the degenerate cores of the aged stars \cite{her01} and  its reported value in the granular and biological objects varies up to 30 sec.  
Kaminski \cite{kam90} reported the relaxation time of 20 sec for sand and 29 sec for $NaHCO_3$,  
Mitra et al. \cite{mi95} found the value of the relaxation time for the processed meat to be c.a. 15 seconds.
However, Grassman \& Peters \cite{gra99} and Herwig \& Beckert \cite{her00,her00a} did not found the evidence for the hyperbolic heat conduction in materials with the nonhomogeneous inner structure.
These discrepancies  were explained by Roetzel  et al. \cite{roe03} as an inconsistency in the early experiments with determination of the thermophysical properties independently from the relaxation time measurements. Roetzel  et al. obtained all parameters simultaneously from the single experiment and confirmed the non-Fourier character of heat transfer but reported smaller values of the relaxation times (2.26 sec for sand, 1.77 sec for meat).
Later Antaki \cite{an05} gave a  value of 2 seconds for the processed meat. 

For observation of the temperature oscillation in the living tissue \cite{liu95} see, e.g., \cite{sco09} and review \cite{xu08a}. The errors in the predicted temperature distribution in the case of the cryosurgery and the cryopreservation can manifest themselves in the thermal stress distribution \cite{ra98,ra00,de03} and estimation of the tissue fracture \cite{shi99,hu01} due to the large volumetric expansion \cite{shi98}.

To analyse the heat transfer in the biological tissue {\em in vivo} one should account for the heat transfer by the arterial and venous blood \cite{av2,zhm18}. 
The continuum models of the microvascular heat transfer are derived with intention to average the effects of a large number of the blood vessels present in the region of interest \cite{b_1,b_2,b_3,b_4}.
The best known model based on the Fourier equation was suggested by H.H. Pennes in 1948 \cite{pennes} and referred to as  the Pennes equation or ''bioheat" equation (also called ''heat sink model" \cite{crezee})
\begin{equation}
\label{pennes}
\varrho c \frac{\partial T}{\partial t} = \nabla \cdot \lambda \nabla T  + Q_p, \quad 
Q_p = c_{\mathrm{b}}\omega_b (T_{\mathrm{a}} - T) + \dot{q}_{\mathrm{met}} + Q^{\mathrm{ext}},
\end{equation}
\noindent where  $T$, $\varrho$, $c$ and $\lambda $ are the temperature,  density,  specific heat  and thermal conductivity of the tissue as homogeneous medium, $\omega_b$  is the blood perfusion rate, $c_b$  is the blood specific heat, $T_{\mathrm{a}}$ is the temperature of the arterial blood, $\dot{q}_{\mathrm{met}}$ and $Q^{ext}$ are the heat sources due to the metabolic reactions (usually could be neglected in the cryobiology problems) and the external source of energy. 

An obvious extension ---  nonlinear ("modified") Pennes equation --- accounts for temperature dependence of the blood perfusion rate $\omega_b = \omega_b^0 + \omega_b^1 T$ \cite{lak10}.

The different models of bioheat equations have been presented where the vascular structures of tissue have been supposed to be uniformly distributed in order to consider the physical model as a uniform porous medium \cite{khaled,nk,yua09,he10,li19a,and19}.
E.g., Xuan \& Roetzel \cite{rh} introduced a two-equation model that considers the heat transfer in porous media called the Local Thermal Non-Equilibrium Equations. They modelled the tissue by dividing it into two regions: the tissue  (extravascular) region  and the blood (vascular) region  and introducing an equivalent effective thermal conductivity in the energy equations of blood and tissue. 
They proposed an interfacial convective heat transfer term instead of perfusion one. Yuan \cite{yua09} found that this  coefficient  is inversely related to the blood vessel diameter.
 The coupled differential equations for the conservation of energy in the tissue and blood are formulated as follows:
\begin{equation}
	(1 - \epsilon) (\rho C_p)_t \frac{\partial T_t}{\partial t} = (1 - \epsilon) \lambda_t \nabla^2 T_t + h a (T_b - T_t) + (1 - \epsilon) Q_t,
\end{equation}	
\begin{equation}
	 \epsilon (\rho C_p)_b  \left(\frac{\partial T_b}{\partial t} + \vec{u}_b \cdot \nabla T_b \right) =  \epsilon \lambda_t \nabla^2 T_b - h a (T_b - T_t) +  \epsilon Q_b
\end{equation}
where $T_t$ and $T_b$ are temperatures averaged over the tissue and blood volumes, $\epsilon$ is the porosity, $h$ is the
heat transfer coefficient, $u_b$ is the blood velocity, $a$ is the transfer area between tissue
and blood, and $Q$ is the absorbed power density.

Later  Roetzel \& Xuan \cite{rh1} developed a 3D equation by dividing the biological tissue into artery, vein and tissue region and investigated  the transient heat transfer process between these regions in a cylinder model.

Nakayama \& Kuwahara \cite{nk}  developed a generalized three-equation bioheat models for vascular and extravascular space in local thermal non-equilibrium condition and incorporated blood
perfusion term within the two sub-volume equations. Then  considered the effect of heat transfer in closely spaced countercurrent artery-vein pair. 
Bazett et al. \cite{baz} found that the axial temperature gradient in the limb artery of human under a low ambient temperature is an order of magnitude higher than that in normal ambient condition due to effect of countercurrent heat exchange in bioheat transfer. The countercurrent heat exchange also reduces heat loss from the extremity to the surroundings \cite{mm}.

The three equations are derived for the arterial blood, the venous blood and the tissue with different temperatures. If the two temperatures are eliminated, the equation for the single (tissue) temperature could be written that contains the derivatives  up to the sixth order in space \cite{tz12,li21}.

Deviations from the Fourier law are also observed for the low-dimension objects (i.e. spatially constrained systems \cite{liv03}) such as thin films, carbon, boron-nitride, bismuth \cite{roh11}, silicon (both crystalline and with amorphous surface \cite{li03a,zou01,mcg11}), silicon-germanium (single crystalline \cite{che09} and core-shell \cite{hu11}) and indium arsenide (both with wurzite and zinc blende phases)  nanotubes \cite{cha08,el09,zho11}, whiskers, graphene nanoribbons \cite{ber10,bal11,sel12g,sel12,pum12,lee15,g23}  (GNRs, a monolayer of $sp^2$-hybridized carbon atoms tightly packed into a 2D honeycomb lattice \cite{geim,gho08}) 
as well as various graphen allotropes  such as, for example, $\alpha-,\; \beta-,\; \delta-$ and $\gamma-$graphynes \cite{cao18}, polymer chains \cite{hen08,hen09}; in nanotubes the violation of the Fourier was observed even when the phonon MFP was much shorter than the sample length as showed experiments by Chang et al. \cite{cha08} with the individual multiwalled carbon and boron-nitride nanotubes.

The low dimensional materials demonstrate the size effect \cite{fra19}\footnote{
The size effect is usually less important in amorphous materials since MFP of phonons is short \cite{che00} and the opposite trend is observed in polyethylene chains because of the reduction of the chain-chain anharmonic scattering \cite{hen10}.
} --- the reducing the thermal conductivity with reducing the size of the sample. E.g., the conductivity of the crystalline nanowires (NWs) is significantly lower than the bulk value and decreases with the wire diameter \cite{alv12} --- thermal conductivity of the Si NW (one of the most promising materials due to the compatibility with the Si-based technology) is about two orders of magnitude smaller that that of bulk Si \cite{yang12} due to the increase of the   roughening \cite{jou_r}. This model combines the incoherent surface scattering of the short-wavelength phonon with the nearly ballistic behaviour of the long-wavelength phonons. 
The thermal conductivity of the thinnest possible Si NWs reaches a superhigh level that is as large as more than an order of magnitude higher than its bulk counterpart and shows the nonmonotonic diameter dependence \cite{zhou17}. This is explained in terms of the dominant normal  process  of the low frequency acoustic phonons that induces the hydrodynamic phonon flow in NWs without being scattered. With diameter increasing, the downward shift of optical phonons triggers strong Umklapp scattering with acoustic phonons. The two competing
mechanisms result in the nonmonotonic diameter dependence of the thermal conductivity with minimum at the critical diameter of 2-3 nm. 

Since the MFP is inversely proportional to a certain power of the phonon  frequency, no matter what the distance scale one has for the temperature variation, there are always low frequency phonons with the greater MFP than this scale,  the nonlocal theory is needed in the analysis of the heat transport by phonons in 
semiconductor and dielectric crystals, especially when the low frequency phonons have a significant contribution to the process \cite{ezz20}.

For energy transport across a thin film or in a multilayer structure the thermal boundary resistance (TBR) become significant; due to the wave-particle duality in some cases the electron wave or the phonon wave effects should be considered \cite{zhang07}.
The thermal conductivity of the superlattices (SLs) is significantly reduced compared to the bulk values of the corresponding alloys \cite{che97,kot17}\footnote{The reduced thermal conductivity of SLs is beneficial for  applications such as
	thermoelectrics \cite{chen11,mu15,qiu15}. Coherent heat conduction has  been confirmed experimentally in
	SL structures. Such travelling coherent phonon waves in SL lead to increase of thermal conductivity as the  number of periods increases. For applications such as
	thermal insulation or thermoelectrics, minimization of phonon coherent effect is desirable. It is found that either aperiodic SLs \cite{hu20} or SLs with rough interfaces can 	disrupt coherent transport.
} and  non-monotonously depends on the SL period thickness \cite{sh152,vaz20}.

As the size further shrinks, a nanowire becomes a  molecular chain and a thin film becomes a molecular sheet. The famous experiment of Fermi, Pasta  \& Ulam started the study of the thermal conductivity in long chains of interacting particles that showed that   conductivity can diverge with the chain length (it scales with a positive power of the system size in 1D and shows a logarithmic divergence in 2D \cite{gra02}) in the case of the integrable systems (e.g., FPU lattice, disordered harmonic chain, diatomic Toda lattice). 
The heat transport through the chain depends on the boundary conditions and variation of these conditions could change the  chain from a heat superconductor to an  insulator \cite{lep08}. 

These  problems as well as their connection to the extremely high thermal conductivity of the carbon and boron nitride nanotubes \cite{cha08,el09} and the graphene  \cite{gho08,bal08,nik12} or silicene \cite{fle12} sheets or hybrid graphene/silicene monolayers \cite{liu14b} as well as the strain effects on the heat transfer in nanostructures \cite{l63,l65} are not considered in the present review --- these topics are examined in detail in the reviews 
 {S. Lepri, R. Livi, A. Politi,  Thermal conduction in classical low-dimensional lattices, Phys. Rep. 2003, v. 377, p. 1-80;}
{S. Liu, X. F. Xu, R. G. Xie, G. Zhang, B. W. Li, 
Anomalous heat conduction and anomalous diffusion in low dimensional nanoscale systems, Eur. Phys. J. B 85, 337 (2012)
and} {a monograph S. Lepri, Thermal Transport in Low Dimensions: from Statistical Physics to Nanoscale Heat Transfer, Lect. Notes Phys. 921, Springer, 2016. }
The present review also does not considers the problems of the heat transport across interfaces that has an essential role in nanodevices and  cooling of electronic circuits and such topics as 
the temperature jump  boundary condition or the thermal boundary resistance \cite{chen15,li15,kot17,ren18} or the trivial deviations from the Fourier law such as  quantum transport or heat transfer via convection or/and radiation \cite{zhm21}.
 
 A number of non-Fourier models have been developed that are based on the modification of the constitutive relation between the heat flux and the temperature gradient.
Most of these models are incorporate the time non-locality for materials with memory \cite{nun71,kol78} (the hereditary materials, including the fractional hereditary materials \cite{fhm}), some are also consider the space non-locality, i.e. for materials with the nonhomogeneous inner structure.

The non-Fourier models must be
compatible with the kinetic theory of gases --- rarefied gases
have the best understood microscopic composition among
continua --- and with the second law of thermodynamics
and the spacetime requirements of nonequilibrium thermodynamics that
are universal, independent of material properties, therefore their consequences are
universal as well \cite{sho21}; moreover, many models could be derived using thermodynamic considerations.

\section{Phase-Lag Models}
 The constitutive relation (\ref{four}) could be formulated in the integral form \cite{col1,col2}
 ("the integrated history of the temperature gradient" \cite{li19})  similar to the non-linear materials (e.g, the viscoelastic materials \cite{col61}) in which the stress at the point depends on the deformation gradients at previous instants \cite{gre1}.
 
 A general expression for the heat flux in materials with memory was written by Gurtin \& Pipkin half a century ago in analogy with the stress-strain relationship of viscoelastic materials  as \cite{gur69} (see also \cite{col67})
  \begin{equation}
  	\label{gp}
 \vec{q} = -  \int\limits_{- \infty}^{t} Q (t - t^{\prime}) \nabla T  t^{\prime}
 \end{equation}
 where $Q (s)$ is a positive decreasing function (called the  relaxation kernel of the Jeffrey's type \cite{tw,hri13,ruk13}, also the {\em memory} kernel)  tending to zero as $s \rightarrow \infty$. 
 
 The condition that  the  relaxation kernel $Q (t - t^{\prime})$
 is the decreasing function means the decrease of the relevance of the older temperature gradients in comparison with the newer ones, thus the term "fading memory" \cite{col74} is used\footnote{
 Herrera \cite{her19} has recently considered the heat 
  conduction that contravenes the "fading memory" paradigm with the  relaxation kernel of the form  
    \begin{equation}
  Q (t - t^{\prime}) = c \left[ 1 -	e^{- \frac{t - t^{\prime}}{\tau}} \right].
  \end{equation}
Evidently, in this case the older temperature gradients have more influence on the heat flux than the present ones and the most impotant contribution to $\vec{q}$ is from the remote past ($t^{\prime} \rightarrow \ - \infty$).
The author considered two problems where such kernel is relevant:
\begin{enumerate}
	\item {The thermohaline convection in a layer of warm salt water that is above a layer of fresh cold water and due to the cooling of the warm salt water the "salt fingers" appear.}
	\item {The secular stability of the nuclear burning when the hydrogen falling onto the surface of a neutron star in a close binary system undergoes nuclear fusion, independently on how low the temperature may be. The nuclear instability  appears whenever the characteristic time for the increasing of the thermal energy generated by the nuclear burning is smaller than the time required for the removal of this energy. 
	
}
\end{enumerate}
}.  
 
 Note that selection of the integral domain $(\infty,t]$ instead of $[0, t]$ means that the initial effects are implicitly neglected \cite{her19}.

 For $Q(s) = \lambda \delta (s)$ where $\delta (s)$ is the one-sided Dirac delta function \cite{tw}
 we get the Fourier law (\ref{four}) that corresponds to the "zero-memory" material.
 
 Different choices of the constitutive relation leads to numerous "lagging"{\ } models (see, for example,  \cite{liu00,tzou09,akb14,liu15} and references therein). 
 
\subsection{Maxwell-Cattaneo-Vernotte Equation}
A constitutive heat-flux equation of the Jeffreys type is  \cite{tw,tw1}
\begin{equation}
\label{jeff}
\tau \frac{\partial \vec{q}}{\partial t} + \vec{q} = - \lambda \nabla T - \tau \lambda_1 \frac{\partial}{\partial t} \nabla T.
\end{equation}

For the case $\lambda_1 = 0$ the equation (\ref{jeff}) reduces to the well known the Cattaneo (also called Maxwell-Cattaneo-Vernotte (MCV)  \cite{sho21}) constitutive relation  that was independently formulated by P.M. Morse \& H. Feshbach in their famous  'Methods of Theoretical Physics" \cite{mors}, by H. Grad (1958), by Carlo Cattaneo (1958) \cite{c58} and by Pierre Vernotte (1958) \cite{v58,v61} (see also \cite{bak03,sob18})
\begin{equation}
\label{catt}
\tau \frac{\partial \vec{q}}{\partial t} + \vec{q} = - \lambda {}\nabla T 
\end{equation}
If additionally $\tau = 0$ the equation (\ref{jeff}) reduces to the Fourier law. The term $\tau {\partial \vec{q}}/{\partial t}$
is sometimes reffered to as "thermal inertia" \cite{zub18}.

The Cattaneo relation (sometimes called "modification of the Fourier law" (MFL) \cite{sob18}) is obtained if in the general relation (\ref{gp}) the kernel is chosen as
\begin{equation}
Q(s) = {\displaystyle	\frac{\kappa}{\tau} e^{- \frac{t - t^{\prime}}{\tau}}}
\end{equation}
where $\kappa = \lambda/(\rho C)$ is the thermal diffusivity.

Thus the Cattaneo equation  (for constant properties) is \cite{cie16,cho16} 
\begin{equation}
	\label{ce}
 \tau \frac{\partial^2 T}{\partial t^2} + 	\frac{\partial T}{\partial t}  = \kappa \nabla^2 T.
\end{equation}

The equation (\ref{ce}) (a particular case of telegrapher's equation \cite{zhu18}) describes the crossover between  ballistic motion and the diffusion --- the transition from the reversible to the irreversible behaviour --- at the characteristic time $\tau$:
\begin{itemize}
	\item {for small times $t \ll \tau$ the first term in the equation (\ref{ce}) is dominant and it reduces to the wave equation that describes the reversible process since it is invariant with respect to the time inversion
\begin{equation}
	\tau \frac{\partial^2 T}{\partial t^2}  = \kappa \nabla^2 T.
\end{equation}	

}
\item {for large time $t \gg \tau$ the first term could be neglected and one get the parabolic equation of the irreversible process - the heat diffusion.}
\end{itemize}

Thus, the hyperbolic nature of the equation (\ref{ce}) is the most significant  at short times $t \approx \tau$. This corresponds to an initial condition when all particles move in the same direction; after the time $\tau$ due to the  randomization of the particle motion  the equation  (\ref{ce}) reduces to the Fourier equation \cite{sob18}.
The velocity of propagation of waves corresponding to the equation (\ref{ce}) \cite{leb14} is  
\begin{equation}
{\displaystyle
	v_{ph} = \frac{\sqrt{2 \kappa \omega}}{\sqrt{\tau \omega + \sqrt{1 + \tau^2} \omega^2}}
}
\end{equation}
In the high-frequency limit the phase velocity is finite for non-zero values of the relaxation time $\tau$
$v_{ph} = U = \sqrt{{\kappa}/{\tau}}$.	
The velocity $U$ is called the {\em second sound}, it is a damped temperature wave while the first sound is the pressure wave \cite{leb14}. 

The Cattaneo equation (\ref{ce}) can be derived in different ways:
{from the BTE under the relaxation time approximation (RTA)} or 
{in the framework of the extended irreversible thermodynamics (EIT).}

The basic assumption of the RTA (in 1D case)
\begin{equation}
	\label{rta}
	\frac{\partial f}{\partial t} + v_x \frac{\partial f}{\partial x} = \frac{f_0 - f}{\tau}
\end{equation}
 is that the distribution function $f$ is close to the equilibrium distribution function $f_0$, i.e. one can assume that
 \begin{equation}
 	\frac{\partial f}{\partial x} \approx \frac{\partial f_0}{\partial x} = \frac{\partial f_0}{\partial T} \frac{\partial T}{\partial x}.
 \end{equation}
 
 Multiplying the equation (\ref{rta}) by $\tau \epsilon v_x$  ($\epsilon$ is the particle energy) and integrating over the momentum space gives equation (\ref{ce}).

In EIT  the dissipative fluxes such as the heat flux are considered as the basic independent variables \cite{leb11,ser13}. Thus, an entropy depends on the internal energy and the heat flux $s = s(u,q)$  obeys the evolution equation 
\begin{equation}
\frac{\partial s}{\partial t} + \nabla \cdot \vec{J}^s = \sigma^s \ge 0,
\end{equation}
\noindent where $\vec{J}^s$ is the entropy flux, $\sigma^s$ is the entropy production rate.

Definition of the non-equilibrium temperature as $T^{-1} = \partial s/\partial u$ and the assumption that $\partial s/\partial q = - \alpha T$, where $\alpha$ is a material coefficient, lead (with the energy balance for the rigid conductor at rest) to the equation   
\begin{equation}
\frac{\partial s}{\partial t} = - \nabla \cdot \frac{\vec{q}}{T} + \vec{q} \cdot \left(\nabla T^{-1} - \alpha \frac{\partial \vec{q}}{\partial t} \right), 
\end{equation}
\noindent and thus
\begin{equation}
\sigma^s = \vec{q} \cdot \left(\nabla T^{-1} - \alpha \frac{\partial \vec{q}}{\partial t} \right).
\end{equation}
The entropy production rate $\sigma^s$ is positive if the heat flux linearly depends on the thermodynamic force in the parentheses
\begin{equation}
\nabla T^{-1} - \alpha \frac{\partial \vec{q}}{\partial t} = \mu \vec{q},
\end{equation}
\noindent where $\mu$ is a positive coefficient.
By introduction of the notation 
${\alpha}/{\mu} = \tau$, ${1}/{\mu T^2} = \lambda$
 one recovers the Cattaneo constitutive relation (\ref{catt}).

Jou \& Casas-Vazquez \cite{jou90} have shown that in a similar way it is possible to include the nonlocal term into the Cattaneo equation assuming that the generalized entropy, entropy flux and entropy production explicitly depend on the the flux of the heat flux, the tensor $\hat{Q}$, to get
\begin{equation}
\tau \frac{\partial \vec{q}}{\partial t} + \vec{q} = - \lambda \nabla T + l^2 \nabla^2 \vec{q}.
\end{equation}
This equation differs from the Guyer-Krumhansl equation (see section \ref{gk}) by the absence of the term of the form $\nabla \nabla \cdot \vec{q}$.
The thermal disturbances in this model propagate with a finite speed $s = \sqrt{{\lambda}/{\rho C \tau}}$.

Estimates for the relaxation time in solids and rarefied gas could be written, respectively, as \cite{kha15}
$\tau_s = {3 \lambda}/{\bar{c}^2}, \tau_g = {3 \nu}/{\bar{c}^2}$,
where $\bar{c}$ is the phonon velocity in the solid or the mean molecular velocity in the gas
$\bar{c} = \sqrt{{8k_B T}/{\pi m}}$,  $\nu$ is the gas kinematic viscosity, $m$ is the mass of molecule. The 
relaxation time for metals at low temperatures could be considered constant \cite{wil75}.

The Cattaneo constitutive relation (\ref{catt}) could be considered as Taylor's  expansion of the relation with a time lag
\begin{equation}
\label{spl}	
\vec{q} (\vec{r}, t + \tau) = - \lambda \nabla T (\vec{r}, t)
\end{equation}
\noindent which is called sometimes as "improved" {\ } Cattaneo model or single-phase lag (SPL) model \cite{yud15}.
SPL model (\ref{spl}) is Galilean invariant \cite{xu11b}. 
 
Chen et al. \cite{che08} have derived the SPL model equation from the BTE using for the  time derivative approximation \cite{xu11b} 
\begin{equation}
\frac{\partial f}{\partial t} \approx \frac{f (t + \Delta t)}{\Delta t} = \frac{f (t + \tau)}{\tau}.
\end{equation}

Recently Li \& Cao \cite{li16} noted that the Cattaneo model should not be considered as a consequence of SPL model since the accuracy of this approximation is uncertain --- the remaining higher order terms could be very large even if the relaxation time is very small and predictions of these models could be different.

The relation (\ref{catt}) could be  written as an integral over the history of the temperature gradient
\begin{equation}
\vec{q} = - \frac{\lambda}{\tau} \int\limits_{- \infty}^{t} \exp \left(- \frac{t - t^{\prime}}{\tau} \right) \nabla T d  t^{\prime}.
\end{equation}

Frankel et al. \cite{fra85} noted that alternative formulation --- in terms of the heat flux (scalar equations for three components in general case) could be useful for problems involving the heat flux in the boundary conditions. The temperature distribution is recovered by integration of the  energy balance equation over time 
\begin{equation}
T(t) = T(0) + \int\limits_{t^{\prime} = 0}^t \frac{1}{\rho c} [- \nabla \cdot q(t^{\prime}) + Q] dt^{\prime}
\end{equation} 

Nir \& Cao \cite{nie19} have compared numerical simulations (using the alternating directions implicit, ADI, finite difference scheme) based on the three representations: temperature, flux and hybrid and found the last to be preferable.

Sometimes the Cattaneo model is called a damped version of Fourier law (Damped Wave Equation, DWE) \cite{mar93,ber18}.

Cattaneo law removes the paradox of the infinite speed of propagation of disturbances but causes another one for the  heat transfer in the moving body: the Cattaneo law is not Galilean invariant.
The heat conduction in the moving medium is governed by the equation (in dimensionless variables) \cite{chr05}
\begin{equation}
\label{chr8}
\frac{\partial^2 T}{\partial t^2} + \frac{\partial (\vec{u} \cdot \nabla T)}{\partial t} + \frac{\partial T}{\partial t} + \vec{u} \cdot \nabla T = \Delta T.	
\end{equation}
 
In the case of the  constant velocity $u (x, t) = U$  (\ref{chr8}) in 1D reduces to
\begin{equation}
	\frac{\partial^2 T}{\partial t^2} + U \frac{\partial^2 T}{\partial x \partial t} + \frac{\partial T}{\partial t} + U \frac{\partial T}{\partial x} = \Delta T.	\end{equation}

The speeds of disturbances in the coordinate system moving with a velocity $U$ are nonlinear function of $U$ \cite{chr05}:
\begin{equation}
c_{1,2} = \frac{1}{2} \left[ U \pm \sqrt{U^2 + 4} \right]. 
\end{equation}

\noindent Evidently, $c_{1,2} \ne U \pm 1$, i.e.,  the sum or the difference of the dimensionless frame velocity and the thermal wave speed.

This paradox is removed when instead of the partial time derivative a material derivative is used \cite{chr05}. According to the Galilean principle of relativity the equation should be the same in the inertial moving frame, i.e. do not change under Galilean transformation $\xi = x - U t, \quad \tau = t$.

In 3D case the Cattaneo relation in the material framework is written as
\begin{equation}
	\label{chr20}
	\vec{q} + \tau \left( \frac{\partial}{\partial t} + \vec{u} \cdot \nabla \right) \vec{q} = \kappa \nabla T.
\end{equation}

The equation (\ref{chr20}) cannot be resolved with respect to the heat flux and thus a single equation for the temperature  cannot be derived \cite{chr05}.

Later Christov \cite{chr09} suggested to use the frame-indifferent upper-convected \cite{kha11} Oldroyd derivative  and to write the Cattaneo equation (\ref{catt}) as 
\begin{equation}
\tau \left[ \frac{\partial \vec{q}}{\partial t} + \vec{v} \cdot \nabla \vec{q} + \vec{q} \cdot \nabla \vec{v} + (\nabla \cdot \vec{v}) \vec{q} \right] + \vec{q} = - \kappa \nabla T
\end{equation} 
that allows elimination of $\vec{q}$ to get for the temperature the equation \cite{chr09}
\begin{multline}
	\label{chr16}
	\tau \left [\frac{\partial^2 T}{\partial t^2} + 2 \vec{u} \cdot \nabla \frac{\partial T}{\partial t} + \frac{\partial \vec{u}}{\partial t} \cdot \nabla T + \left( \frac{\partial T}{\partial t} + \vec{u} \cdot \nabla T \right) (\nabla \cdot \vec{u}) + \vec{u} \cdot \nabla (\vec{u} \cdot \nabla T)
	 \right] \\ + \frac{\partial T}{\partial t} + \vec{u} \cdot \nabla T = \nabla \cdot (\kappa \nabla T).
\end{multline}
Christov showed that equation (\ref{chr16}) retains its form under change of variables corresponding to frame that moves with a constant velocity.

Thermal convection with Cattaneo-Christov relation  was studied by
Straughan \cite{hay14}.  Stability
for Cattaneo-Christov equation has been investigated by Ciarletta
\& Straughan \cite{hay15}. Haddad \cite{hay16} studied thermal stability in the Brinkman permeable space  of improved heat conduction Cattaneo-Christov  model. Fluid flow and heat transfer characteristics of Maxwell material over a stretched sheet by employing this model is addressed by Han et al. \cite{hay17}. Hayat et al. \cite{hay18} simulated 3D flow of Prandtl
fluid. Zhang et al. \cite{zha19} studied the heat transfer of blood vessels subject to the transient laser irradiation.

Joseph \& Preziosi \cite{tw} suggested to use the relaxation kernel in the form $R_{JP} = \lambda_1 \delta(s)  + (\lambda_2/\tau) exp (-s/\tau)$ where $\lambda_1$ is the effective thermal conductivity and $\lambda_2$ is the "elastic" conductivity, In this case the heat flux is equals  to
\begin{equation}
q = -\lambda_1 \frac{\partial T}{\partial x} - \frac{\lambda_2}{\tau} \int\limits_{-\infty}^t exp \left(\frac{t - s}{\tau} \right) \frac{\partial T}{\partial x} d s . 
\end{equation}

The Cattaneo (Cattaneo-Vernotte) model may give unphysical predictions such as the negative temperature when two cooling waves meet \cite{wan11}.

K\"orner \& Bergmann \cite{kor98} and Barletta \& Zancini \cite{bar97,bar97a,zan99} analysed the compatibility of the Cattaneo law with the second law. Barletta \& Zancini  also considered  the Taitel's paradox (a temperature exceeding the difference of the boundary values in a slab whose sides are kept at the different temperatures).
In the frame of the CIT the entropy production could be written as \cite{leb14}
\begin{equation}
	\sigma^s = \lambda \frac{(\nabla T)^2}{T^2} + \frac{\tau}{T^2} \frac{\partial \vec{q}}{\partial t} + \cdot \nabla T.
\end{equation}

Barletta \& Zancini  found that the production of the entropy  could be negative in the regions where the heat flux decreases so steeply that $|\partial \vec{q}/\partial t| > |\vec{q}| / \tau$. 

Torii \& Yang \cite{tor05} 
studied the heat transfer in the thin film under the continuous-operated and pulse laser heat sources
and found that CV model could lead to {\em overshooting} phenomenon in the propagation of the thermal wave that  
 seemingly violates the second law of thermodynamics.

However, 
these results does not contradicts the second law since 
classical thermodynamics is based on the local equilibrium hypothesis that is not fulfilled in this case \cite{con10} and the concept of "temperature" in  the hyperbolic heat transfer equation cannot be  interpreted in the conventional sense \cite{leb14}.
Cattaneo law is compatible with the second law in the EIT  \cite{jou,jou1}. 

Li \& Cao \cite{li16a} studied the thermodynamics problems of the SPL model. Using the expression of the entropy production rate
\begin{equation}
S = - \frac{\vec{q} \cdot \nabla T}{T^2}
\end{equation}
\noindent from the CIT the authors got for the Fourier law and for SPL model
\begin{equation}
S_F = - \frac{q^2}{\lambda T^2} \ge 0, \qquad
S_{SPL}^{CIT} = - \frac{\vec{q}(t) \cdot \vec{q} (t + \tau|)}{\lambda T^2}.
\end{equation}

The entropy production rate for the SPL model is not necessary positive or zero.
The second law of thermodynamics is satisfied in the EIT \cite{jou} where the entropy production rate is expresses as
\begin{equation}
S_{SPL}^{EIT} = - \frac{\vec{q}(t + \tau) \cdot \vec{q} (t + \tau|)}{\lambda T^2}.
\end{equation}

Still, SPL model could violate the second law of thermodynamics by the breaking the equilibrium spontaneously in special circumstances. 
 Li \& Cao \cite{li16a} constructed a simple example of such behaviour. They considered the case 
 ${2 \tau n^2 \pi}/{\rho C_V l^2} = 1$
 and obtained the solution
\begin{equation}
\label{sol}
T (x, t) = C_1 \sin \frac{n \pi x}{l} \sin \frac{\pi t}{2 \tau} + C_0 	
\end{equation}
of the SPL model equations with the boundary conditions 
$T (0, t) = C_0, \quad T (l, 0) = C_0$ and the initial condition $T (x, 0) = C_0$.

The coefficient $C_1$ in the solution (\ref{sol}) is arbitrary and the initial equilibrium $T = C_0$ could be broken spontaneously. Note that both Fourier and Cattaneo models preserve the equilibrium \cite{li16a}:
\begin{equation}
	\frac{d E^F}{d t} = - \int  \frac{\lambda}{\rho C_v} (\nabla T)^2 d V \le 0,\qquad \frac{d E^{CV}}{d t} = - \frac{2}{\tau} \int \left(\frac{\partial T}{\partial t} \right)^2 d V \le 0.
\end{equation}

In biological problems the Cattaneo equation is called Thermal Wave model or Thermal Wave Model Bioheat Transfer (TWMBT) \cite{liu99a,zh13,mal15}, sometimes the terms "heat wave" or "temperature wave" are also used \cite{sal06,zhang07}.

Conejero et al. \cite{con10} studied the chaotic asymptotic behaviour of the solutions of the Cauchy problem for the heat transfer equation (\ref{ce}) 
\begin{equation}
\begin{cases}
	T (0, x) = \phi_1 (x), \quad x \in \mathbb{R}\\
	\frac{\partial T}{\partial t} (0, x) = \phi_2 (x), \quad x \in \mathbb{R}
\end{cases}
\end{equation}
where $\phi_1 (x)$ and $\phi_2 (x)$ are the initial temperature and  temperature variation.
 
The authors expressed the hyperbolic heat transfer equation as a first-order equation and represented the solutions as a $C_0$-semigroup\footnote{
A family $\{T_t\}_{t \ge 0}$ of linear continuous operators
on the Banach space $\mathbb{X}$ is called $C_0$-semigroup if
 {$T_t T_s = T_{t +s}$ for all $t,s \ge 0$ and}
 {$\lim_{t \rightarrow s} T_t = T_s$ for all $x \in \mathbb{X}$  and $s \ge 0$.}
} on the product of certain function space $\mathbb{X}$ with itself \cite{how95} and showed that this semigroup is hypercyclic and chaotic in the sense of Devaney \cite{con10}.

Podio-Guidugli \cite{pod12} (see also \cite{bar14}) derived the  MCV-type equations using the splitting the entropy into the standard and dissipative parts.

\subsection{Dual-Phase-Lag Model}

To include effects of both  the relaxation and the microstructure, Tzou \cite{ts93,tz95,ts96} introduced the dual phase lag (DPL) model 
\begin{equation}
\label{dpl}
\vec{q} (\vec{r}, t + \tau_q) = - \lambda \nabla T (\vec{r}, t + \tau_T) 
\end{equation}
where $\tau_q$ and $\tau_T$ are the phase lags for the heat flux vector and for the temperature gradient, respectively,  arising from "thermal inertia" and "microstructural interaction" \cite{xu09} that are the intrinsic properties of the material. The phase-lag of heat flux can be interpreted as the time delay due to the fast transient effect of thermal inertia, while the phase-lag of the temperature gradient represents the effect of phonon–electron interactions and phonon scattering  \cite{akb17}.

Tang \& Araki \cite{tan00} (see also the paper by Shen \& Zhang \cite{she08}) distinguish four heat propagation modes of the DPL model:
{wave mode ($\tau_T = 0$);}
{wavelike mode ($0 < \tau_T < \tau_q$);}
{diffusion mode ($\tau_T  = \tau_q$);}
{over-diffusion mode ($\tau_T  > \tau_q$).}

Zhang et al. \cite{zha14} considered the damping of thermal waves measured as the relative decrease of the temperature and the heat flux for the Cattaneo and DPL models (and for the thermomass model, see below ) and introduced the dimensionless "damping factor" $\xi = L/\sqrt{\kappa \tau}$. In the case of the heat conduction by the photon transport this factor is shown
to be inversely proportional to the Knudsen number ($Kn = l/L$, $l$ is MFP of phonons) $\xi = \sqrt{3} / Kn$ --- the decrease of the Knudsen number reflects stronger collisions between phonons and thus quicker damping of the energy transported by the thermal wave.

Both relaxation times are very small for common materials. For example, the phase lags $\tau_q$ and $\tau_T$ for gold are 8.5 ps and 90 ps, respectively \cite{da00}. The relaxation time $\tau_T$ was found by the molecular dynamics (MD) computations to be smaller than $\tau_q$ and both times being in the range from a few picoseconds to tens of picoseconds
in the solid argon films \cite{liu08a}.
Goicochea et al. \cite{goi10} found for the bulk silicon from the MD computations that the relaxation time $\tau_q$ is inversely proportional to the cube of the temperature $\tau_q \propto T^{-3}$ and to the temperature $\tau_q \propto T^{-1}$ when the temperature is higher than the Debye temperature. Thus, the thermal wave travels faster at higher temperature \cite{zha13a}.

There is a lot of controversies in the literature concerning the values of relaxation times for the biological tissues. For the processed  meat these values are estimated as 14-16 s and 0.043 - 0.056 s\cite{an05}, experiments with the muscle tissue from cow give values 7.36-8.43 s and 14.54-21.03 s \cite{liu10}. Zhang \cite{zha9a} estimated the relaxation times in term of the blood and tissue properties and found they to be close to each other.  

The DPL model reduces to the SPL model when $\tau_T = 0$ and to the Fourier law if $\tau_T = \tau_q =0$.
Equation (\ref{dpl}) could be re-written as
\begin{equation}
\label{dpl1}
\vec{q} (\vec{r}, t) = - \lambda \nabla T (\vec{r}, t + (\tau_T - \tau_q)) 
\end{equation}

Thus, the solution of the DPL model does not depend on the relaxation times $\tau_T$ and $\tau_q$ separately but only on their difference \cite{kul04} and the SPL and the DPL models are mathematically equivalent \cite{ord11,fab14a,van17b}.
 
Fabrizio et al. \cite{fab14a,fab17} showed  that
 there are mathematical conditions beyond the physical ones to obtain an exponentially  stable equilibrium solution for DPL equation. Such condition requires  negative time delay $\tau_d$ (called as retarded effect) between the heat flux and temperature  gradient $\tau_d = \tau_q - \tau_T \le 0 $.
  The opposite case, i.e. $\tau_d > 0$ is ill-posed which enlightens the validity of MCV equation
 but excludes equations based on the arbitrary Taylor series expansion.

The DPL model is closely related \cite{xu11} to the hyperbolic models that describe the energy exchange between electrons and phonons by a pair of coupled nonlinear equations governing the effective temperatures of electrons and phonons  \cite{kag57,ani74,qiu93,sob93,sob94,che06,li14a,yud15} (two temperature models, TTM).

It is assumed that the electron and phonon subsystems are at their local equilibrium, heat conduction by phonon is neglected and the lattice temperature is near or above the Debye temperature (thus the electron-electron and electron-defect scatterings are insignificant compared with the electron-phonon scattering \cite{zhang07}). The equations of the TTM  for the 1D could be written as
\begin{equation}
c_e \frac{\partial T_e}{\partial t} = - \frac{\partial q}{\partial x} - G (T_e - T_l) + Q, \quad
c_l \frac{\partial T_l}{\partial t} =  G (T_e - T_l), \quad
\tau \frac{\partial q}{\partial t}  + q = - \lambda \frac{\partial T_e}{\partial x}
\end{equation}

\noindent where $T_e$ is the temperature of the electron gas, $T_l$ is the temperature of the lattice, $c_e$ and $c_l$ are the heat capacity of the electron gas and the lattice, respectively, $G$ is the electron-phonon coupling factor that is estimated as \cite{jia07}
$G = \pi^2 m_e n_e c_s^2 /{\tau_e (T_e) T_e}$
where $m_e$ is the mass of electron, $n_e$ is the density of the free electrons, $\tau_e$ is the electron relaxation time, $c_s$ is the speed of sound in bulk material $c_s = \sqrt{B/\rho_m}$, $B$ is the bulk modulus, $\rho_m$ is the density.

Qiu \& Tien \cite{qi94,wan11a} suggested a similar model named a hyperbolic two-step (HTS) model that accounts for the lattice conductivity
\begin{equation}
	c_e \frac{\partial T_e}{\partial t} = - \frac{\partial q_e}{\partial x} - G (T_e - T_l) + Q, \quad
	c_l \frac{\partial T_l}{\partial t} = - \frac{\partial q_e}{\partial x} + G (T_e - T_l),
\end{equation}	
\begin{equation}	
	\tau_e \frac{\partial q_e}{\partial t}  + q_e = - \lambda \frac{\partial T_e}{\partial x}, \quad
\tau_i \frac{\partial q_i}{\partial t}  + q_i = - \lambda \frac{\partial T_i}{\partial x}.
\end{equation}

For pure metals heat conduction in the lattice is small compared to that in electrons and can be neglected as well as the electron relaxation time, thus the dual-hyperbolic models reduce to the parabolic two temperature model \cite{ho14}.

Tzou \cite{tz95} had estimated the relaxation times of the DPL model in terms of the parameters $G,\quad c_e,\quad c_l$ of this model and obtained for copper, silver, gold and lead typical values of $\tau_T$ and $\tau_q$ about $10^{-11}$ and $10^{-13}$, respectively.

Zhang \cite{zha9a} proposed explicit estimates for the  biological tissues
\begin{equation}
{\displaystyle
\tau_q = \frac{\epsilon (1 - \epsilon)}{\frac{\epsilon}{C_{tb}} + (1 - \epsilon)} \frac{\rho_b C_b}{G}, \quad 
\tau_T = \frac{\epsilon (1 - \epsilon)}{\frac{\epsilon}{K_{tb}} + (1 - \epsilon)} \frac{\rho_b C_b}{G}
}
\end{equation}
\noindent where $C_{tb} = \rho_t C_t / (\rho_b C_b)$ is the ratio of heat capacities of tissue and blood, $K_{tb} = \lambda_t / \lambda_b$ is the ratio of thermal conductivities, $\epsilon$ is the porosity of the tissue, $G$ is "lumped convection-perfusion parameter".  

Tzou \& Dai \cite{tzou09} considered lagging in the  multi-carrier systems. The equations for the N-carrier system are written as 
\begin{equation}
C_1 \frac{\partial T_1}{\partial t} = \lambda_1 \nabla^2 T_1 - \sum_{i = 2}^N G_{1i} (T_1 - T_i), 
\end{equation}
\begin{equation}
C_m \frac{\partial T_m}{\partial t} = \lambda_m \nabla^2 T_m + \sum_{j = 1}^{m - 1} G_{jm} (T_j - T_m) -\sum_{i = m + 1}^N G_{mi} (T_m - T_i), m = 2,...(N - 1),
\end{equation}
\begin{equation}
C_N \frac{\partial T_N}{\partial t} = \lambda_N \nabla^2 T_N + \sum_{i = 1}^{N - 1} G_{iN} (T_i - T_N).
\end{equation}

Deriving an equation for the single temperature in the three-carrier system (for example, the composite with three constituents or the polar semiconductor where the heat could be transported by electrons, holes and phonons) Tzou \& Dai found that the nonlinear effects are related to the $\tau_q^2$  and $\tau_T^2$.

The Taylor series expansion can be made to provide a series of DPL
heat conduction models \cite{xu08a,li21}.
Using the first order expansion for both $\tau_q$ and $\tau_T$ in the  equation (\ref{dpl}) gives the constitutive relation 
\begin{equation}
\label{dpl_1}
\vec{q} + \tau_q \frac{\partial \vec{q}}{\partial t} = -\lambda \left[\nabla T + \tau_T \frac{\partial \nabla  T}{\partial t} \right].
\end{equation}
Inserting this relation in the energy conservation law results in  the so called type I \cite{xu08a,liu11,van17b} (also linear \cite{liu14a} or first-order \cite{li21}) DPL model.

The first order DPL bioheat transfer equation is written as
\begin{equation}
	\label{1_dpl}
\frac{\varrho c}{\lambda} \left( \frac{\partial T}{\partial t} + \tau_q \frac{\partial^2 T}{\partial t^2} \right) + \frac{\rho_b c_b  w_b}{\lambda} \left[(T - T_b) + \tau_q  \frac{\partial T}{\partial t}  \right]= \nabla^2 T + \tau_T \frac{\partial}{\partial t} (\nabla^2 T), 	
\end{equation}
in the source term the metabolic reactions and the source of energy are omitted.

The equations of this model could be re-written in terms of the heat flux instead of the temperature \cite{gha15} 
\begin{equation}
\nabla (\nabla \vec{q}) + \tau_T \nabla (\nabla \vec{q}) = 	\frac{1}{\kappa} \frac{\partial \vec{q}}{\partial t} + \frac{\tau_q}{\kappa} \frac{\partial^2 \vec{q}}{\partial t^2}
\end{equation}
and even in terms of the heat flux potential $\phi$  defined by  $\vec{q} = \nabla \phi $ \cite{wan01}.

Wang et al. \cite{wan01} proved the well-posedness of the DPL model on a finite 1D region under the uniform Dirichlet, Neumann or Robin conditions. Later Wang \& Xu \cite{wan02} extended this result to the $n$-dimensional case.

Type II DPL model \cite{van17b} is obtained when the first-order and the second-order expansions are used for the heat flux $q$ and the temperature $T$, respectively
\begin{equation}
q + \tau_q \frac{\partial q}{\partial t} = -\lambda \left[\nabla T + \tau_T \frac{\partial \nabla  T}{\partial t} + \frac{\tau_T}{2} \frac{\partial^2 \nabla T}{\partial t^2}\right].
\end{equation}

\noindent and Type III (second-order \cite{li21}) DPL model \cite{liu11a,van17b}) - when the second-order Taylor series expansions are used for both $q$ and $T$
\begin{equation}
q + \tau_q \frac{\partial q}{\partial t} + \frac{\tau_q}{2} \frac{\partial^2 q}{\partial t^2} = -\lambda \left[\nabla T + \tau_T \frac{\partial \nabla  T}{\partial t} + \frac{\tau_T}{2} \frac{\partial^2 \nabla T}{\partial t^2}\right].
\end{equation}

The second-order DPL bioheat transfer equation is written as follows
\begin{multline}
\frac{\varrho c}{\lambda} \left( \frac{\partial T}{\partial t} + \tau_q \frac{\partial^2 T}{\partial t^2} + \frac{\tau_q^2}{2} \frac{\partial^3 T}{\partial t^3} \right) 
+ \frac{\rho_b c_b  w_b}{\lambda} \left[(T - T_b) + \tau_q  \frac{\partial T}{\partial t} + \frac{\tau_q^2}{2} \frac{\partial^2 T}{\partial t^2} \right]\\
= \nabla^2 T + \tau_T \frac{\partial}{\partial t} (\nabla^2 T)	
+ \frac{\tau_T^2}{2} \frac{\partial^2}{\partial t^2} (\nabla^2 T).
\end{multline}

Sometimes the other notation is used to distinguish the DPL models, indicating the orders of the Taylor expansions, for example, DPL (2,1) \cite{esc11}.

Rukolaine \cite{ruk14} found the solutions of the DPL model to be unstable. An analytical solution for
DPL equation with a Gaussian initial condition gives an unphysical behaviour of temperature history, it
goes into the negative domain.
Later he confirmed his conclusion for the type III DPL model \cite{ruk17}.

Quintanilla \& Racke \cite{qui06} (see also \cite{qui02})  analysed the stability of the solution of the different DPL versions.  Quintanilla introduced the ratio of the two relaxation times of the DPL model 
$\xi = \frac{\tau_T}{\tau_q}$ 
as a parameter that controls the stability of the DPL model.
The author considered the characteristic polynomial associated to the Laplace operator for the case of the Dirichlet boundary conditions in a bounded domain. Stability of the solution is determined by the real part of the eigenvalues.
The results of the study could be summarized as
\begin{itemize}
	\item {When approximation is up to first order in $\tau_q$ and up to first or  second order in $\tau_T$ is used, the system is always stable.}
\item {When approximation is up to second order in $\tau_q$ and up to first order in $\tau_T$ is used, the system is stable if $\xi > 1/2$ and unstable if $\xi < 1/2$.}
\item {When approximation is up to second order in both $\tau_q$ and $\tau_T$ is used, the system is stable if $\xi > 2 - \sqrt{3}$ and unstable if $\xi < 2 - \sqrt{3}$.}
\item {Whenever $\xi > 1/2$, the several models predict the same behaviour }
\end{itemize}
 
 Restrictions on the ratio $\xi = {\tau_T}/{\tau_q}$ were also derived from the second law of thermodynamics by Fabrizio \& Lazzari \cite{fab14}.
 
The compatibility of the DPL model with the second law of the Extended Irreversible Thermodynamics was proved by Xu \cite{xu11a}.  
 
Both TWMBT and DPL models are used to study the heat transfer in biological objects.
 E. g., interaction of laser radiation with the tissue, including the laser interstitial thermal therapy (LITT) \cite{sho21}, have been considered by  Zhou et al. \cite{zhou09},  Afrin et al. \cite{afr12},  Sahoo et al. \cite{sah14},  Hooshmand et al. \cite{hoo15}, Kumar \& Srivastava \cite{kum15a}, Mohajer et al. \cite{sh223}, Jasinsky et al. \cite{jas16}, Liu \& Chen \cite{liu16a}, Zhng et al. \cite{zha17}. 
  Singh \& Melnik \cite{sin19} used the DPL model incorporating
 both the tissue contraction and expansion during the procedure and consider DPL along  with Helmholtz harmonic wave equation, the modified  stress-strain and the thermoelastic wave equations to simulate radiofrequency ablation.
  High intensity focused ultrasound ablation was investigated by Li et al. \cite{li18}, Namakshenas \& Mojra \cite{sh220} and Singh et al. \cite{sh222}.
  Kumar et al. used the finite element wavelet Galerkin method to study the hyperthermia \cite{kum15b}. Ho et al. exploited the lattice Boltzmann method  for solution of the DPL model for heat transfer in the two-layer system \cite{ho13}. 
Moradi \& Ahmadikia \cite{mor12} used the DPL model to study the heat transfer during the fast freezing of the biological tissues (the cooling rate about $1000^{\circ}/s$ \cite{ahm12b}) when the frozen region tends to form the amorphous ice \cite{z9}.
Liu \& Chen \cite{sh215}, Liu \& Cheng \cite{sh216}, Liu \& Yang \cite{sh217}, Raouf et al. \cite{sh218} and Kumar et al. \cite{sh213,kum19a} used the DPL model to study the magnetic fluid hyperthermia. The effect of the lagging times only appears in the
tumour: the temperature increases with
the increment  of $\tau_q$ and decreases with the augmentation of the value of the $\tau_T$ in the tumour zone \cite{sho21}.
 
 Li et al. \cite{li20} applied DPL to investigation of the thermo-viscoelastic behaviour of the biological tissue 
 subjected to the hyperthermia treatment.
 
 Liu et al. \cite{liu13} applied the second-order Taylor expansion of dual-phase lag model for analysis of
 the thermal behaviour in the laser-irradiated layered tissue, which was stratified into skin, fat, and muscle.
 
Gandolpi et al. \cite{gan20} used the first-order expansion of DPL (\ref{dpl_1}) that results in the temperature equation
\begin{equation}
\label{gan}
\left(\frac{\tau_q}{\kappa}\right) \frac{\partial^2 T}{\partial t^2} - \frac{\partial^2 T}{\partial x^2} + \frac{1}{\kappa} \frac{\partial T}{\partial t} - \tau_T \frac{\partial^3 T}{\partial t \partial x^2} = 0 
\end{equation}
as the basis for the design of the temperonic crystal (TC) --- a periodic structure with a unit cell made of two slabs sustaining temperature
wavelike oscillations on short timescales. The temperonic crystal
is the similar to  electronic,
phononic, and photonic SL. The complex-valued
dispersion relation for the temperature scalar field in TCs  discloses frequency gaps, tunable upon
varying the slabs thermal properties and dimensions,
serving, for instance, as a frequency filter for a temperature
pulse \cite{gan20}.

  For analysis of the equation (\ref{gan}) Gandolpi et al. \cite{gan19} used dimensionless variables ($T_{eq}$ is the equilibrium temperature)
  \begin{equation}
  \beta = \frac{t}{\tau_q}, \qquad \xi = \frac{x}{\sqrt{\kappa \tau_q}}, \qquad Z = \frac{\tau_T}{\tau_q}, \qquad \theta = \frac{T}{T_{eq}}.
  \end{equation}
The equation (\ref{gan}) is re-written as
\begin{equation}
\label{gana}
 \frac{\partial^2 \theta}{\partial t^2} - \frac{\partial^2 \theta}{\partial x^2} + \frac{1}{\kappa} \frac{\partial \theta}{\partial t} - Z \frac{\partial^3 \theta}{\partial t \partial x^2} = 0 
\end{equation}
 
 The equation (\ref{gan}) is parabolic but its
 solution may bear wavelike characteristics if \cite{gan19}
 $Z  < {1}/{2}$ 
 since the first two terms constitute the homogeneous  wave equation, while the last two are responsible for  damping effects; $\sqrt{\kappa/\tau_q}$ is the speed of propagation and $\sqrt{\kappa \tau_q}$ is the diffusion
 length \cite{gan16}.
 
 The authors sought the solution of the equation  (\ref{gana}) in the form
 \begin{equation}
 \theta (\xi, \beta) = \theta_0 e^{i (\tilde{k} \xi + \tilde{\omega} \beta)}
 \end{equation}
 where
 the complex-valued dimensionless wave vector $\tilde{k}$ and angular frequency $\tilde{\omega}$  are
 linked to their dimensional counterparts $\tilde{k} = \sqrt{\kappa \tau_q} k$ and $\tilde{\omega} = \tau_q \omega$.
 
 The equation (\ref{gana}) gives the dispersion relation
 \begin{equation}
 \tilde{k}^2 \left(1 + i Z \tilde{\omega} \right) = \tilde{\omega}^2 \left(1 - \frac{1}{\tilde{\omega}} \right).
 \end{equation}

 Numerous researches used the DPL model to study the heat transfer in nanoelectronic devices, see, for example, papers \cite{sh200,sh202,sh203}.
  
\paragraph{Nonlocal Dual-Phase-Lag Model}  
 To accommodate the effect of thermomass (see section \ref{tm}), the
 distinctive mass of heat, of dielectric lattices in the heat
 conduction and size-dependency of thermophysical properties,
 Tzou \& Guo \cite{tz10} and Tzou \cite{tz11} have introduced the nonlocal (NL) behavior of heat transport
 in space in addition to the lagging of temperature
 gradient and heat flux, in time. The nonlocal dual-phase-lag (NL
 DPL) heat conduction can be written as: 
 \begin{equation}
 	\label{ndpl}
 	\vec{q} (\vec{r} + \vec{\lambda}_q, t + \tau_q) = - \lambda \nabla T (\vec{r}, t + \tau_T) 
 \end{equation}
 where $\vec{\lambda}_q$ is the correlating length of nonlocal heat flux that  has the same form of  the length parameter in the thermomass model \cite{akb17}.
  The natural extension of the NL DPL (\ref{ndpl}) is the introduction of another local space scale $\vec{\lambda}_T$
 \begin{equation}
 	\label{ndpl1}
 	\vec{q} (\vec{r} + \vec{\lambda}_q, t + \tau_q) = - \lambda \nabla T (\vec{r} + \vec{\lambda}_T, t + \tau_T). 
 \end{equation}
 
 Using the first-order Taylor series expansion of equations (\ref{ndpl},\ref{ndpl1})  with respect to  both correlating lengths and phase-lags, Tzou has developed  the nonlocal  DPL  (NL DPL) heat conduction models.
 
 The first-order expansion in lagging time and second-order
 effect in nonlocal parameters of equations (\ref{ndpl}, \ref{ndpl1}) gives in 1D case \cite{li21}
 \begin{equation}
 	q (x,t) - \frac{\lambda_q^2}{2} \frac{\partial^2 q (x,t)}{\partial x^2} + \tau_q \frac{\partial q (x,t)}{\partial t} = - \lambda \frac{\partial T (x,t)}{\partial x}
 \end{equation}
  and
  \begin{equation}
  	q (x,t) -  \frac{\lambda_q^2}{2} \frac{\partial^2 q (x,t)}{\partial x^2} + \tau_q \frac{\partial q (x,t)}{\partial t} = - \left[ \lambda \frac{\partial T (x,t)}{\partial x} -  \frac{\lambda_T^2}{2} \frac{\partial^3 q (x,t)}{\partial x^3} \right].
  \end{equation}
 
  These expressions are used to derive nonlocal DPL bioheat equations \cite{li21}.

  Li et al. \cite{li21} considered the introduction of the nonlocal parameter $\lambda_T$ into the 1D nonlocal  DPL model
  \begin{equation}
  	\label{l17}
  	q (x,t + \tau_q ) = -\lambda \frac{\partial T (x + \lambda_T, t + \tau_T)}{\partial x}
  \end{equation}
  whose $1^{st}$ order expansion in the lagging time and $2^{d}$ order in the nonlocal parameter yields
  \begin{equation}
  	q (x, t) + \tau_q \frac{\partial q (x, t)}{\partial t} = -\lambda \left[ \frac{\partial T (x, t)}{\partial t}  - \frac{\lambda_T^2}{2} \frac{\partial^3 T (x, t)}{\partial t^3} + \tau_T \frac{\partial^2 T (x, t)}{\partial t \partial x} \right]
  \end{equation}
  and the corresponding bioheat transfer equation
  \begin{multline*}
  	\frac{\varrho c}{\lambda} \left( \frac{\partial T}{\partial t} + \tau_q \frac{\partial^2 T}{\partial t^2} \right) 
  	+ \frac{\rho_b c_b  w_b}{\lambda} \left[(T - T_b) + \tau_q  \frac{\partial T}{\partial t}  \right]
  	= \frac{\partial^2 T}{\partial x^2} + \tau_T \frac{\partial^3 T}{\partial t \partial x^2} 
  	- \frac{\lambda_T^2}{2} \frac{\partial^4 T}{\partial x^4}.
  \end{multline*} 
  
  Li et al. \cite{li21} also investigated the fourth-order expansion in $\lambda_T$ and   the second-order expansion in $\tau_q$ and $\tau_T$ in equation (\ref{l17})
  \begin{multline*}
  	q + \tau_q \frac{\partial q}{\partial t} + \frac{\tau_q^2}{2} \frac{\partial^2 q}{\partial t^2} =\\ -\lambda \left[ \frac{\partial T}{\partial t} + \tau_T \frac{\partial^2 T}{\partial t \partial x} + \frac{\tau_T^2}{2} \frac{\partial^3 T}{\partial t^2 \partial x}
  	 + \frac{\lambda_T^4}{4!} \frac{\partial^5 T}{\partial x^5} - \frac{\lambda_T^2}{2} \frac{\partial^3 T}{\partial x^3} - \frac {\lambda_T^2 \tau_T}{2} \frac{\partial^4 T}{\partial t \partial x^3} 	\right].	
  \end{multline*}
  The corresponding bioheat transfer equation could be found in \cite{li21}.
  
\subsection{Triple-Phase-Lag Model}
Triple-phase-model is obtained by Choudhuri \cite{cho07}  
(who extends the thermoelastic model suggested  by Green \& Naghdi \cite{gr2})
by introduction to DPL additionally to the relaxation times for the heat flux and the temperature gradient the relaxation time for the thermal displacement\footnote{
Thermal displacement was introduced by H. von Helmholtz \cite{pod16,akb17}. 
It  satisfies the definition $\dot{v} = T$. This quantity was used in models by Green \& Naghdi \cite{gr1,gr2} as " a scalar history variable"
\begin{equation}
v = \int\limits_{t_0}^t T(\tau) d \tau + v_0.
\end{equation}

Bargmann \& Favata \cite{bar14} used the thermal displacement in their thermomechanical analysis of the laser-pulsed heating in polycrystals including time derivative of the thermal displacement $\dot{v}$ (i.e. the temperature), gradients of the  thermal displacement $\nabla v$ and its time derivative $\nabla \dot{v}$ as the state space variables for all quantities of the coupled equations --- free energy, entropy, first Piola-Kirchhoff stress tensor, chemical potential, defect flux, heat flux.
}
 gradient \cite{cho07,akb14,akb14a,kum19}, sometimes simply the "phase lag of the thermal gradient" \cite{tiw} or the "phase lag of the thermal displacement" \cite{ez12}.    
\begin{equation}
	\label{tpl}
\vec{q} (\vec{r}, t + \tau_q) = - [\lambda \nabla T (\vec{r}, t + \tau_T) + \lambda^{\star} \nabla v (\vec{r}, t + \tau_v)], 
\end{equation}
\noindent $\lambda^{\star}$ is a positive
material constant characteristics of the TPL theory. 

Falahatkar et al. \cite{fal18} used the TPL model to study heat conduction in a laser irradiated tooth. The human tooth is composed of enamel, dentin, and pulp with unstructured shape, uneven boundaries. Earlier Falahatkar et al. \cite{fal17} studied this problem using the DPL model and found that the heat flux phase lag has significant effect on the temperature
profile at early stages while the temperature gradient
phase lag is more important at later stages.

Akbarzadeh et al. \cite{akb14a} used the TPL model to investigate the heat transfer in the functionally graded hollow cylinder (earlier these authors exploited the DPL model to study heat conduction in 1D functionally graded media \cite{akb13}).

\paragraph{Nonlocal Triple-Phase-Lag Model}
Akbarzadeh et al. \cite{akb17} derived 
a non-local three-phase-lag (NL TPL) constitutive
equation for the non-Fourier heat conduction   including the effects of thermal
displacement (a scalar "history variable" used  by Green \&
Naghdi \cite{gr1,gr2}) and its nonlocal length in the NL DPL
thermal analysis:
\begin{equation}
	\label{ntpl}
	\vec{q} (\vec{r} + \vec{\lambda}_q, t + \tau_q) = - - [\lambda \nabla T (\vec{r} + \vec{\lambda}_T, t + \tau_T) + \lambda^{\star} \nabla v (\vec{r} + \vec{\lambda}_v, t + \tau_v)]
\end{equation}
where $\lambda_q$, $\lambda_T$, and $\lambda_v$ represent correlating nonlocal lengths of
heat flux, temperature gradient, and thermal displacement
gradient. The Taylor series expansion of constitutive equation (\ref{ntpl}) 
with respect to either nonlocal lengths and/or phase-lags leads
to alternative nonlocal phase-lag heat conduction models.

\section{Phonon Models}
Phonons are the quantized lattice vibrations (the elastic waves that can exist only at discrete energies); phonons serve as heat carriers in dielectric and semiconductor crystals that undergo scattering in the course of propagation \cite{li12a}. The electronic thermal conductivity could be significant in some cases, e.g., in the channel layer of the AlGaN/GaN high electron mobility transistors (HEMTs) due to the formation of the high density electron 2D gas \cite{liu05}.

Phonons are eigen states of the atomic system and  can propagate without dissipation in purely harmonic solid that should have the infinite thermal conductivity \cite {cher}. 
Peierls \cite{pei} stated that  the origin
of thermal resistance is the combination of anharmonicity
and the discrete nature of  lattice. 
However, the anharmonicity alone cannot induce resistance --- an infinite thermal conductivity is expected if the phonon scattering processes conserve momentum \cite{pei}.

Dissipation occurs due to the phonon scattering.
One should consider different mechanisms of the phonon scattering: the three-phonon inelastic scattering such as normal \cite{hol63,arm85} and Umklapp (both transverse and longitudinal \cite{che05a}), defect scattering (one should distinguish the scattering on the impurity and on the isotope atom \cite{qua17}), scattering at the boundaries of the sample \cite{kli88,pap08} that in general case should account for 
the root-mean-square roughness height \cite{mar09} and 
the dependence of the phonon specularity probability on the phonon frequency, incidence angle and the surface roughness \cite{kaz10}, phonon-electorn scattering \cite{hua06};
the four-phonon processes could be important above the Debye temperature \cite{ec77,mur05}.  Umklapp scattering, isotope scattering and impurity scattering are referred to as the momentum-destroying phonon scattering processes (R-scattering) \cite{ding17}. Phonons, which are the main heat carriers in non-metallic systems, can undergo not
only a diffusive regime that can be described by means of the Fourier law, but also a hydrodynamic regime (Poiseuille-like) and a ballistic one \cite{sel15}. Division of the phonon processes into the normal and Umklapp ones is also applicable to the amorphous materials \cite{kle51}, for example, polymers (except crystalline one) where the structural scattering can occur that reduces the phonon MFP to a few monomer lengths \cite{cho77}.

Recent studies show that  Umklapp scatterings are not necessarily
resistive --- no thermal resistance is induced if the projected momentum is conserved in the direction of heat flow \cite{din18}.
This feature is especially important in anisotropic materials such as graphite where phonons dispersion along one direction (denoted as “soft” axis) is much softer than the other directions (“stiff” directions). Ding et al. proposed that a classification of N scattering and U scattering should be based on the projected phonon momentum in the heat flow direction and the condition of the quasimomentum conservation (\ref{qm}) should be modified to
$(\vec{k}_1)_j + (\vec{k}_2)_j = (\vec{k}_3)_j + \vec{b}_j$ 
where $j$ is the heat flux direction. Thus the scattering event is the normal  scattering as long as $\vec{b}_j = 0$, which holds
when $\vec{b} = 0$ or $\vec{b}$ is a reciprocal lattice vector orthorgonal to $\vec{j}$ \cite{din18}. By comparison
with the exact solutions of the phonon BTE, Ding et al. demonstrated that the new classification of N-scattering and U-scattering processes leads to the more accurate predictions
of the thermal conductivity using the Callaway model.

The semi-classical BTE (also called the Peierls-Boltzmann equation) for the phonon distribution function  $f$ is 
\cite{kav} 

\begin{equation}
	\label{bte}
\frac{\partial f}{\partial t} + \vec{v} \cdot \nabla f  + \vec{F} \frac{\partial f}{\partial \vec{P}} = \left( \frac{\partial f}{\partial t} \right)_{scatt}
\end{equation} 
\noindent where $\vec{P}$ is the momentum, $\vec{F}$ is the external force.

In disordered systems the typical phonon MFP may be so short that the quasi-particle picture  is invalid and the BTE is not applicable \cite{isa19}.

To solve the BTE one needs to know the phonon dispersion relation $\omega (\vec{k})$, the group velocity $\vec{v}_g = \partial \omega/\partial \vec{k}$  and the rate of collisions.

There are two strategies to solve in the BTE:
{the full integral scattering term} and  {RTAs.}
In the first one, the  solution is obtained using the empirical  scattering term \cite{gz14,lee15,gz31,gz32,gz33} or via the  Monte Carlo (MC) scheme; however, this approach is very expensive \cite{g19,g20}. The RTAs is more suited for engineering applications; several assumptions are made: 
{single mean relaxation time;}	
{local near-to-equilibrium;}
{local occupation number;} 
{ad hoc or fitted boundary scattering rates;}
{neglecting the cross mode correlations.}
 The RTA is
\begin{equation}
	\left( \frac{\partial f}{\partial t} \right)_{scatt} = - \frac{f - f_0}{\tau}
\end{equation} 
\noindent where $f_0$ is the equilibrium distribution function.

It is also called single-mode relaxation time approximation (SMRTA) \cite{gu17}), the Callaway model or the gray relaxation-time approximation \cite{rez,rez1}. This approximation 
 is known to underestimate the thermal conductivity \cite{cher,cher11}, especially when strong normal scattering is present \cite{war09,din18}. 
 
 The RTA is useful if  the size of the samples is large and the experiments are performed under slow heating conditions. The  new electronic devices reduce the sample size down to
few nanometers and the heating times can be of the order of picoseconds. In these cases  the RTA solution is often far from the real solution, thus more accurate solutions to the linearized BTE
are needed such as the direct solution derived, the iterative
solution or  the kinetic collective model 
approach. Recently Torres et al. \cite{tor19} studied the transition metal dichalcogenides  MX2 (M = Mo, W; X = S, Se) from
first-principles  by solving the BTE and found that RTA  can result in underprediction of the conductivity up to the 50\%.

 The common formulas for the bulk thermal conductivity
 are based on the kinetic theory and  can be derived
 from the BTE in the RTA by summation of the polarization dependence and
 integration of the frequency dependence \cite{yan13a}:
 \begin{equation}
 	\label{y1}
 	\kappa_{bulk} = \sum_s \int_0^{\infty} C v \Lambda_{bulk} d\omega
 \end{equation}
 where $C$ is the specific heat capacity per unit
 frequency, $v$ is the group velocity,  $\Lambda_{bulk}$ is the bulk MFP, $\omega$  is the frequency, and $s$ indexes the polarizations.
 
 It is assumed that the dispersion
 relation and bulk MFPs are isotropic that  is exact for  gases  and  amorphous materials. In crystalline materials the dispersion relation depends on the direction and a more general form of the equation (\ref{y1}) is needed \cite{bro07} (the thermal conductivity $\kappa_{bulk}$ itself  still can be is isotropic, e.g.,  in crystals with  cubic symmetries). The phonon dispersion relation could be determined experimentally or calculated using the elastic wave theory \cite{che05a}.

Yang \& Dames \cite{yan13a} suggested to change the integration variable in the equation (\ref{y1}) from the frequency $\omega$ to the bulk MFP $\Lambda_{bulk}$ to obtain 
 \begin{equation}
 	\label{y2}
 	\kappa_{bulk} = - \sum_s \int_0^{\infty} \frac{1}{3} C v \Lambda_{bulk} \left( \frac{d\Lambda_{bulk}}{d\omega} \right)^{-1}  d\Lambda_{bulk}.
 \end{equation}
 
 The authors interpreted the change of variables  as the change of the labelling scheme for  the energy carriers from $(\vec{q},s)$ to $(\Lambda_{bulk},s)$ and then applied Fubini’s theorem to change the orders of summation
 and integration to obtain  
 \begin{equation}
 	\label{y4}
 	\kappa_{bulk} =  \int_0^{\infty} K_{\Lambda} d\Lambda_{bulk}, \quad
  	K_{\Lambda}	 = - \sum_s  \frac{1}{3} C v \Lambda_{bulk} \left( \frac{d\Lambda_{bulk}}{d\omega} \right)^{-1}
 \end{equation}
 $K_{\Lambda}$  is the thermal conductivity per MFP.
 This function is known as the MFP distribution or MFP
 spectrum for the bulk thermal conductivity.
 
 Yang \& Dames introduced {\em the thermal conductivity
 	accumulation function}
 \begin{equation}
 	\label{y5}
 \alpha(\Lambda_{\alpha})	= \frac{1}{\kappa_{bulk}}   \int_0^{\Lambda_{\alpha}} K_{\Lambda} d\Lambda_{bulk}.
 \end{equation}
 
 This function  represents the fraction of the total thermal conductivity  due to carriers with MFPs less than $\Lambda_{\alpha}$.
  Equations (\ref{y4}) and (\ref{y5}) determine the range of MFPs that contribute  to heat conduction flux. Frequently this range is described by a single "effective"  (or gray) MFP
 \begin{equation}
 	\label{y6}
 	\Lambda_{gray} = \kappa_{bulk} \left( \sum_s \int_0^{\infty} 
 	\frac{1}{3} C v  d\omega \right)^{-1}.
 \end{equation}
 
 The equation (\ref{y6})  corresponds to  a MFP distribution that given by  a Dirac $\delta$  function with weight $\kappa_{bulk}$ centered on $\Lambda_{gray}$ that is a good approximation in systems where the
 real MFP distribution is narrow (e.g., including ideal gases or 
 free electron gases). However, in systems with strongly
 frequency-dependent scattering (e.g., semiconductor crystals), the distributions can be  broad \cite{yan13a}. 
 
 As noted by Yang \& Dames \cite{yan13a}, as long as
 the structure’s characteristic length $L_c$  is much larger than
 the thermal wavelengths of the energy carriers and
 there is no coherence effects, the group velocity and spectral heat
 capacity in the nanostructure are identical to those in bulk,
 so the only effect of the nanostructuring is to reduce the
 effective MFP by scattering at boundaries and interfaces.
 Thus, the nanostructure thermal conductivity could be
 written as
\begin{equation}
 	\label{y7}
 	\kappa_{nano,t} = \sum_s \int_0^{\infty} C v \Lambda_{nano} d\omega
 \end{equation}
 where $\Lambda_{nano} (\omega,s) < \Lambda_{bulk} (\omega,s)$ and the subscript $t$ indicates the “type” of geometry, e.g., "wire" or "film".
After  change variables one gets
 \begin{equation}
 	\label{y8}
 	\kappa_{nano,t} = \int_0^{\infty} K_{\Lambda}  \frac{\Lambda_{nano}}{\Lambda_{bulk}} d\Lambda_{bulk}.
 \end{equation}
 
 For a wide variety  of the geometries  the ratio $\Lambda_{nano}/\Lambda_{bulk}$ 
 depends only on the Knudsen number $Kn = \Lambda_{bulk}/L_c$   \cite{yan13a}. 
 
 Often the Callaway {\em dual} relaxation approximation \cite{call} is used \cite{tam98,g23}
 \begin{equation}
 \left( \frac{\partial f}{\partial t} \right)_{scatt} = - \frac{f - f_{\lambda}}{\tau_N} -  \frac{f - f_0}{\tau_R},
 \end{equation}
  \noindent where $f_{\lambda}$ is the distribution function of the uniformly drifting phonon gas, $\tau_N$ is the relaxation time for the normal scattering and  
 $\tau_R$ is the relaxation time for the Umklapp process. This  approximation allows a separation of N-processes and U-processes and  provides better estimation for the 2D materials than the SMRTA \cite{ma14,li19}.
  Although the Callaway’s model has been widely used in modelling  hydrodynamic phonon transport \cite{g19,g23,gz38,ding17}, the direct numerical solution of the BTE using the Callaway’s scattering term has been advanced only recently \cite{g18a,g24}.

 Guo et al. \cite{guo21} studied the phonon vortices in the graphene ribbon using the deterministic  discrete-ordinate method  \cite{g18a} for the solution  of the phonon BTE under the Callaway’s dual relaxation model. The phonon scattering rates of normal and resistive processes were acquired from ab initio calculations.

 In recent years there is increasing interest in modelling hydrodynamic phonon transport in graphene ribbon. First works were focused on the temperature profile and thermal conductivity  in-plane \cite{g18a,g19} and cross-plane \cite{g21} in rectangular geometries and considered the steady-state  phonon transport; the transient transport was studied only in few works \cite{g23,g24}. 
  The phonon transport in the hydrodynamic regime could be described by the the Guyer-Krumhansl equation (\ref{krum}) where 
 the contribution of the heat flux may be  neglected with respect to its spatial derivatives \cite{sel15} ---
 a signature of hydrodynamic phonon transport is the collective drift motion of phonons that  manifests itself in the distribution function \cite{ding17} and  is similar to viscous fluid flow \cite{sim20}.

  The phonon vortex has attracted less attention than investigation of electron hydrodynamic flow (e.g., experiments by Moll et al. \cite{moll16} on the ultra-pure metal $PdCoO_2$ showed a large viscous contribution and yielded an estimate of the electronic viscosity
   compared to the water viscosity at room temperature)    
   and, in particularity, vortex in graphene ribbons \cite{g27,g28,g29,chan19}.

   There is a demonstration of heat vortex in a rectangular graphene ribbon based on macroscopic phonon hydrodynamic equation in a recent paper \cite{g30}.
 
 Guo et al. \cite{guo21,guo21a} wrote the phonon BTE (\ref{bte}) under the Callaway’s dual relaxation model as
 \begin{equation}
 	\label{g_bte}
 	\frac{\partial f}{\partial t} + \vec{v}_g \cdot \nabla f   = \frac{f_R^{eq} (T_{l,R}) - f}{\tau_R (\omega, p, T)} + \frac{f_N^{eq} (T_{l,N}, \vec{u}) - f}{\tau_N (\omega, p, T)}
 \end{equation} 
 \noindent where the phonon distribution function  $f =f (\vec{x},p, \omega, \vec{s}, t) = f (\vec{x},p, \vec{K}, t),
 \vec{x}$ is the spatial position, $\omega$ is the phonon angular frequency, $p$ is the phonon polarization, $\vec{K}$ is the 2D wave vector that is assumed to be isotropic, i.e., $\vec{K} = |K| \vec{s}$ where $\vec{s}$ is the unit directional vector in 2D coordinate system, $t$ is the time. $\vec{v} = \nabla_{\vec{K}} \omega$
 is the group velocity calculated by the phonon dispersion.

 The relaxation times of the scattering processes are denoted by $\tau_R (\omega, p, T)$ and $\tau_N (\omega, p, T)$, with their local equilibrium distribution functions defined as:
 \begin{equation}
 	f_R^{eq} (T_{l,R}) = \frac{1}{\exp \left( {\displaystyle \frac{\hbar \omega}{k_B T_{l,R}}} \right) - 1}, \quad 
 	f_N^{eq} (T_{l,N}, \vec{u}) =  \frac{1}{\exp \left( {\displaystyle  \frac{\hbar (\omega_{\mu} - \vec{k} \cdot \vec{u})}{k_B T}} \right) - 1}.
\end{equation}
 Here $T_{l,R}$ and $T_{l,N}$ are the local pseudo-temperatures and 
 $\vec{u}$ is  the  drift velocity, which are determined by the energy and momentum conservation conditions of scattering processes. The two temperatures are introduced as mediate mathematical quantities to ensure the energy conservation of  scattering term \cite{gz40}, the local temperature $T$ defined from the local energy density represents the physical temperature of the phonon system, $\vec{k}$ is the  wave vector.
 
 The phonon BTE equation (\ref{g_bte})  is rewritten in terms of a deviational energy 
  \begin{equation}
	\label{d_bte}
	\frac{\partial e}{\partial t} + \vec{v}_g \cdot \nabla e   = \frac{e_R^{eq} (T_{l,R}) - e}{\tau_R (\omega, p, T)} + \frac{e_N^{eq} (T_{l,N}, \vec{u}) - e}{\tau_N (\omega, p, T)}
\end{equation} 
where the deviational distribution functions of energy density are \cite{g18a,g24}
\begin{equation}
	e = \frac{\hbar \omega D(f - f_R^{eq} (T_0))}{2 \pi},
	e_R^{eq} = \frac{\hbar \omega D(f_R^{eq} - f_R^{eq} (T_0))}{2 \pi}, 
	e_N^{eq} = \frac{\hbar \omega D(f_N^{eq} - f_R^{eq} (T_0))}{2 \pi}
\end{equation}
 where $D = |\vec{K} | / ( 2 \pi |\vec{v} | )$ is the density of states, $T_0$ is the reference temperature. The density of states $D (\omega)$ is proportional to $\omega^2$ if the dispersion is linear \cite{zhang07}.

For a small temperature difference $\Delta T \ll T_0$ and a small drift velocity $\vec{K} \cdot \vec{u} \ll \omega$ the equations for deviational distribution functions could be linearized
\begin{equation}
e_R^{eq} = \frac{T - T_0}{2 \pi},\quad e_N^{eq} = \frac{T - T_0}{2 \pi} + C T \frac{\vec{K} \cdot \vec{u}}{2 \pi \omega}
\end{equation}
where 
\begin{equation}
	C (\omega, p, T_0) = 2 \pi \frac{\partial e_R^{eq}}{\partial T} |_{T = T_0}
\end{equation}
is the mode specific heat at $T_0$. 

The equation (\ref{d_bte}) in stationary case is re-written in dimensionless form as 
\begin{equation}
	\frac{\vec{v}}{|\vec{v}|} \cdot \nabla_{x^{\star}} e^{\star} = \beta_R (e_R^{eq, \star} - e^{\star}) + \beta_N (e_N^{eq, \star} - e^{\star})
\end{equation}
where ($L$ is the system size)
\begin{equation}
	x^{\star} = \frac{x}{L},\;  e^{\star} = \frac{e}{C T_0}, \;  e_R^{eq,\star} = \frac{e_R^{eq}}{C T_0},\;  e_N^{eq,\star} = \frac{e_N^{eq}}{C T_0},\; \beta_N = \frac{L}{|\vec{v}| \tau_N},\; \beta_N = \frac{L}{|\vec{v}| \tau_N}, 
\end{equation}
parameters $\beta_N$ and $\beta_R$ represent the ratio of the  system size L to the phonon MFP of the normal  ($l_N = |\vec{v}| \tau_N$) and resistive ($l_R = |\vec{v}| \tau_N$) scattering.

 Guo et al. distinguished three heat transfer regimes:
 {diffusive: $\beta_N = 0$ and $\beta_R \rightarrow \infty$,}
 {ballistic: $\beta_N = \beta_R = 0 $,}
 {hydrodynamic: $\beta_N \rightarrow \infty $ and $\beta_R = 0$.}

The authors used the gray model and Debye approximation, i.e. $\omega = \vec{v}|  \vec{K}|$ \cite{gu1}  and compute the $0^{th}$ and the $1^{st}$ order moments of BTE (\ref{d_bte}) to get 
\begin{equation}
	\frac{\partial E}{\partial t} + \nabla \cdot \vec{q} = 0,\qquad
	\frac{\partial \vec{q}}{\partial t} + \nabla \cdot \hat{Q} = - \frac{\vec{q}}{\tau_R}
\end{equation}
where
\begin{equation}
	E = \sum_p \int \int e d\Omega d\omega,\quad 
	\vec{q} = \sum_p \int \int \vec{v} e d\Omega d\omega,\quad 
	\hat{Q} = \sum_p \int \int \vec{v} \vec{v} e d\Omega d\omega,
\end{equation}
the integral is over the whole 2D solid angle space and the frequency space.

In the diffusive limit the resistivity scattering dominates the phonon transport and causes heat dissipations; at the steady state, the distribution function can be approximated as \cite{guo21} $e \approx e_R^{eq} - \tau_R \vec{v} \cdot \nabla e_R^{eq}$ and the thermal conductivity in Fourier law $\vec{q} = - \kappa_d \nabla T$ in the diffusive limit $	\kappa_d = {1}/{2} C |\vec{v}|^2 \tau_R$.

The authors introduce the vorticity that is defined as the curl of heat flux 
\begin{equation}
\int_l \vec{q} \cdot d\vec{r} 
\end{equation}
where $l$ is  an arbitrary closed curve inside the thermal system,
$d\vec{r}$ is
  the unit tangent vector of the curve $l$ in a clock-wise direction. 
In the diffusive regime  
\begin{equation}
	\int_l \vec{q} \cdot d\vec{r} = - \kappa_d \int_l \frac{dT}{d\vec{r}} \cdot d\vec{r} = 0.
\end{equation}
Thus, there are no heat vortices inside the
system in the diffusive regime.

In both hydrodynamic and ballistic regimes, the heat flux  is conserved in the interior domain \cite{gz14,lee15,gu1} and there is no heat dissipation
\begin{equation}
	\label{g29}
	\frac{\partial \vec{q}}{\partial t} + \nabla \cdot \hat{Q} = 0.
\end{equation}

Although equation (\ref{g29}) is valid  in the ballistic and hydrodynamic limits, these regimes are different. In the ballistic regime \cite{g20}  the momentum conservation is satisfied due to the rare phonon-phonon scattering and the phonon MFP is much larger than the system characteristic length: at the steady state,  the phonon distribution function does not vary with the spatial position until scattering with the boundaries and the appearance of heat vortices depends on the geometry settings or/and boundary conditions.

It is hard to clearly distinguish the (quasi) ballistic or hydrodynamic phonon transport only by the wave like propagation of heat  \cite{zhang21}. 
Up to now there is no general consent on the exact definition of the terms “hydrodynamic phonon transport” or “phonon hydrodynamic”, in particularity, whether  the strong N scattering of phonons is required --- see discussion in the paper by Zhang \& Guo \cite{zhang21}. These authors themselves assumed that the hydrodynamic phonon transport occurs when the N scattering is much stronger than the R scattering.

Zhang \& Guo  suggested to use a transient heat conduction to distinguish the hydrodynamic and  ballistic transport regimes. The authors investigated the transient heat propagation in the homogeneous thermal system using the phonon BTE under the Callaway dual approximation.
The authors considered  the 3D materials; the phonon dispersion was not accounted and the Debye model was used.
The temperature difference in the domain was assumed to be small |compared to the reference temperature $T_0$, i.e., $|T - T_0| \ll T_0$. The BTE for the deviational phonon distribution function of the energy density $e = e (\vec{x}, \vec{s}, t)$ which depends on  position $\vec{x}$ , unit directional vector $\vec{s}$ and time $t$ is 
\begin{equation}
	\label{dim}
\frac{\partial e}{\partial t} + \vec{v}_g \vec{s} \cdot \nabla_{\vec{x}} e = \frac{e_R^{eq} - e}{\tau_R}	+ \frac{e_N^{eq} - e}{\tau_N}
\end{equation}
where $e^{eq}_R$ and $e^{eq}_N$ are the equilibrium distribution function of the R scattering and N scattering linearized by the specific heat $C$
\begin{equation}
e_R^{eq} (T) = \frac{C (T - T_0)}{4 \pi}, \quad e_N^{eq} (T) = \frac{C (T - T_0)}{4 \pi} + \frac{C T}{4 \pi} \frac{\vec{s} \cdot \vec{u}}{v_g} ,	
\end{equation}
$\tau_R$ and $\tau_N$ are the relaxation time of the R scattering and N scattering, respectively, $v_g$ is the value of group velocity, $\vec{u}$ is the drift velocity.

To close the phonon BTE Wang \& Guo used the energy conservation for both N scattering and R scattering
\begin{equation}
\int \frac{e_N^{eq} - e}{\tau_N} d\Omega = \int \frac{e_R^{eq} - e}{\tau_R} d\Omega = 	0,
\end{equation}
where the integral is over the whole solid angle space and used 
momentum conservation for the N scattering
\begin{equation}
	\int \vec{s} \frac{e_N^{eq} - e}{\tau_N} d\Omega = 0.
\end{equation}

The macroscopic spatio-temporal distributions such as the local energy $E (\vec{x}, t)$ or the heat flux $\vec{q} (\vec{x}, t)$  are obtained by taking the moments of the phonon distribution function \cite{g24,g18a,g20,zhang21}
\begin{equation}
	E (\vec{x}, t) = \int e d\Omega, \quad \vec{q} (\vec{x}, t) =  \int v_g \vec{s} e d\Omega.
\end{equation}
The drift velocity is given by \cite{gu1} $\vec{u} = 3 \vec {q}/C T$. 

To wrote the dimensionless version of the BTE equation (\ref{dim}) the authors related the spatial scale to $L$, temporal to $t_{ref} = L/v_g$, distribution functions to $e_{ref} = C \Delta T/4 \pi$ ($\Delta T$ is the temperature difference in the domain) and introduced two Knudsen numbers which represent the ratio between the phonon MFP of the R scattering and N scattering to the characteristic length
\begin{equation}
	Kn_R = \frac{v_g \tau_R}{L}, \quad Kn_N = \frac{v_g \tau_N}{L}
\end{equation}
and mainly determine the transient heat propagation to get
\begin{equation}
	\label{dim1}
	\frac{\partial e^{\star}}{\partial t^{\star}} +  \vec{s} \cdot \nabla_{\vec{x}^{\star}} e^{\star} = \frac{e_R^{eq,\star} - e^{\star}}{Kn_R}	+ \frac{e_N^{eq,\star} - e^{\star}}{Kn_N}.
\end{equation}

In the diffusive limit ($Kn_R \rightarrow 0, Kn_N = \infty$) the Fourier law is valid; in the ballistic limit ($Kn_R = \infty, Kn_N = \infty$) the BTE equation (\ref{dim}) simplifies to 
\begin{equation}
	\frac{\partial e}{\partial t} + \vec{v}_g \vec{s} \cdot \nabla_{\vec{x}} e = 0.
\end{equation}

When the N scattering dominates heat conduction and the R scattering could be neglected,  the phonon  BTE equation (\ref{dim}) becomes 
\begin{equation}
	\label{dim2}
	\frac{\partial e}{\partial t} + \vec{v}_g \vec{s} \cdot \nabla_{\vec{x}} e = \frac{e_N^{eq} - e}{\tau_N}.
\end{equation}

Taking the zeroth- and first- orders moments of  (\ref{dim2}) leads to \cite{zhang21}
\begin{equation}
	\frac{\partial E}{\partial t} + \nabla_{\vec{x}} \cdot \vec{q} = 0,\quad 
	\frac{\partial \vec{q}}{\partial t} + \nabla_{\vec{x}} \cdot \hat{Q} = 0, \qquad
	\hat{Q} = \int v_g \vec{s} \vec{s} e d\Omega.
\end{equation}

It is follows from the above equations 
\begin{equation}
\frac{\partial^2 E}{\partial t^2} = \nabla_{\vec{x}} \cdot \nabla_{\vec{x}} \cdot	\hat{Q}
\end{equation}
that can be transformed into a wave equation by assuming $e = e_N^{eq}$ \cite{zhang21}.

Zhang \& Guo established that the transient temperature could be lower than the lowest value of initial environment temperature in the hydrodynamic regime within a certain range of time and space. This phenomenon disappears in the ballistic or diffusive regime and in quasi-1D simulations and thus could be used to distinguish hydrodynamic phonon transport from  the ballistic regime.

In the hydrodynamic regime, the phonon MFP is much smaller than the local characteristic length due to the frequent normal scattering \cite{lee15,gz14}. At steady state, the phonon distribution function can be approximated as \cite{guy1,g30} 
\begin{equation}
e \approx e_N^{eq} - \tau_N \vec{v} \cdot \nabla e_N^{eq}.	
\end{equation}

Thus in contrast to the ballistic regime the phonon distribution function varies with the spatial position.

Zhang et al. \cite{guo21a} used the implicit discrete ordinate method \cite{g18a,guo19} is used to solve the stationary phonon BTE. The authors studied phonon dynamics in several situations: 
heat vortices in a 2D ribbon and in 2D porous materials, including porous graphene, using both gray and nongray models.

 Guo et al. \cite{guo21} investigated the phonon vortex dynamics in  a rectangular and a T-type graphene ribbons that is influenced by the temperature, ribbon size and carbon isotope. The wide MFP distribution of resistive phonon scattering processes is the crucial factor that destroys the vortex hydrodynamic effect. The phonon vortex configuration in the T-type graphene ribbon is found to depend on the height-width ratio with primary, secondary and ternary vortexes.

 Shang \& L\"u  \cite{sha18}  studied the hydrodynamic phonon transport in 2D materials. 
  The authors started from
 the BTE with Callaway model (\ref{g_bte}), 
  multiplied each term by $\hbar \omega \vec{v}$ and summed over  wave vectors and phonon  indices  to get
\begin{equation}
	\frac{\partial\vec{q}}{\partial t} + \frac{\hat{\kappa}}{\tau_R} \cdot \nabla T = - \frac{\vec{q}}{\tau_R} - \frac{\vec{q_1}}{\tau_N}, \quad
	\vec{q} = \sum_s \int \frac{d\vec{k}}{(2 \pi)^2} \hbar \omega_{s \vec{k}} \vec{v}_{s \vec{k}} f_{s \vec{k}}
\end{equation} 
$\vec{q}$ is the heat current density,
$\vec{q}_1$ is the second term in the expansion in a small parameter
$\epsilon = \tau_N / \tau_R$ (i.e. the scattering rates of N-process is assumed to be much larger than that of R-process) $\vec{q} = \vec{q}_0 + \epsilon \vec{q}_1 + \epsilon^2 \vec{q}_2 + \cdots$,
and
\begin{equation}
\hat{\kappa} = \tau_r 	\vec{q} = \sum_s \int \frac{d\vec{k}}{(2 \pi)^2} \hbar \omega_{s \vec{k}} \vec{v}_{s \vec{k}} \vec{v}_{s \vec{k}} \frac{\partial f_{s \vec{k}}}{\partial T}	
\end{equation}
is the thermal conductivity tensor.

The authors considered the case of small $\vec{u}$ when $\vec{q}_1 = 0$ and finally derived a 2D Guyer-Krumhansl-like  equation  
 \begin{equation}
 	\label{sha}
 \frac{\partial\vec{q}}{\partial t} + \frac{\kappa_0}{\tau_R} \nabla T +  \frac{\vec{q}}{\tau_R} = \eta \left[ \nabla^2 \vec{q} + 2 \nabla (\nabla \cdot \nabla \vec{q}) \right]	- \zeta \nabla (\nabla \cdot \vec{q}),
 \end{equation}
 where the $\kappa_0$ is the zeroth order thermal conductivity,
 $\eta$ and $\zeta$ are the first and second viscosity coefficients \cite{sha18}.
 
 Shang \& L\"u considered the heat flow through a nano-ribbon with length $L (x \in [0, L])$ and width $w$ 
 with the temperature difference applied along the ribbon. At steady state neglecting $q / \tau_R$ the equation (\ref{sha})
 reduces to
 \begin{equation}
 	\frac{d^2 q}{dy^2} = \frac{\partial T}{\partial x} \frac{\kappa_0}{\eta \tau_R}
 \end{equation}	
 that gives a parabolic heat for the non-slip condition $q(0) = q(w) = 0$ and the total flux that scales as $w^3$ --- a signature of the Poiseuille flow (for diffusive phonon transport, the heat current scales linearly with the ribbon width) \cite{sha18}.
 
 To compute the pattern of vortices in the stationary 2D flow the authors introduced a streamfunction and used a simplified version of the  equation (\ref{sha})
 \begin{equation}
 	\eta \nabla^2 \vec{q} - \frac{\vec{q}}{\tau_R} = C_L v_g^2 \nabla T.
 \end{equation}
 
The presence of the viscous terms in the  equation of the phonon hydrodynamics allowed Huberman \cite{hub18} to introduce a "thermal Reynolds number" $Re_{therm} = {v_{ss} L}/{|\vec{v}|^2 \tau_N}$  where $v_{ss}$ is the speed of second sound.
  However, as noted by Zhang et al. \cite{guo21a}, this definition is  hardly fruitful since the "thermal Reynolds number" has only formal resemblance to the Reynolds number (and its importance) in fluid dynamics: there is no convective term in the Guyer-Krumhansl-like  equation and this equation actually is similar to the Stokes equations --- the low-Reynolds-number form of the Navier-Stokes equations.

 Alternative approach is based on the similarity of phonons and photons \cite{lem07}. They are both bosons --- their equilibrium distribution function is the Bose-Einstein distribution. Both phonons and photons are considered as non-local quasi-particles with zero mass carrying energy $\hbar \omega$ when the length scale exceeds the coherence length for phonons and the wavelength for photons.
 
 It is possible to define the phonon radiative intensity by analogy with the thermally emitted photons
 \begin{equation}
 	I_{\omega} (x, t, \vec{s}) = \frac{1}{4 \pi} D (\omega) n_{\omega} (x, t, \vec{s}) \hbar \omega v_{\omega},
 \end{equation}
\noindent where $D (\omega)$ is the number of modes per unit volume, $v_{\omega} = d\omega / d\vec{k}$ is the group velocity and $n_{\omega}$ is the average number of phonons with angular velocity $\omega$ moving in direction of unit vector $\vec{s}$ averaged over all branches of the dispersion curves.

The internal energy $e$ and the heat flux $\vec{q}$ could be presented as the spectral and the total quantities
\begin{equation}
	e_{\omega} (x, t) = \frac{1}{v_{\omega}} \int\limits_{4 \pi} f (x, t, \omega, \vec{s}) d \Omega, \qquad \vec{q}_{\omega} (x, t) =  \int\limits_{4 \pi} f (x, t, \omega, \vec{s}) \vec{s} d \Omega,
\end{equation}
\noindent and
\begin{equation}
e (x, t) =	\int\limits_0^{\omega_{max}} e_{\omega} (x, t) d \omega,\qquad \vec{q} (x, t) =	\int\limits_0^{\omega_{max}} \vec{q}_{\omega} (x, t) d \omega,
\end{equation}
\noindent where $\omega_{max}$ is the cutoff frequency determined by the crystal lattice, $d \Omega$ is an elementary solid angle around the direction $\vec{s}$.

The propagation of phonons is governed by equation
\begin{equation}
	\frac{\partial n_{\omega}}{\partial t} + v_{\omega} \vec{s} \cdot \nabla n_{\omega} = \frac{d n_{\omega}}{d t} |_{coll} \approx \frac{n_{\omega}^0 - n_{\omega}}{\tau_{\omega}},  
\end{equation}
 \noindent or by the phonon radiative transfer equation \cite{maj93,lem07}
 \begin{equation}
 \label{rad}
 	\frac{1}{v_{\omega}} \frac{\partial I_{\omega}}{\partial t} + \vec{s} \cdot \nabla I_{\omega} = \frac{1}{\Lambda_{\omega}}(I_{\omega}^0 - I_{\omega}),
 \end{equation}
\noindent where $1/\Lambda_{\omega} = \kappa_{\omega}$ is the analogue of the absorption coefficient in the photon transport and $I_{\omega}^0 (T) = D (\omega)  n_{\omega}^0 (T) v_{\omega} /4 \pi$ is the equilibrium radiation intensity.  Integration of (\ref{rad}) over the whole spectrum and all directions leads to 
\begin{equation}
	\frac{\partial e}{\partial t} + \nabla \cdot \vec{q} = \int\limits_0^{\omega_{max}} \left(4 \pi I_{\omega}^0 - \int\limits_{4 \pi} I_{\omega} d \Omega  \right) d \omega.
\end{equation}
 
 When MFP  is small compared to the  length of the sample $L/\Lambda = \kappa_{\omega} L \gg 1$ (the diffusive regime) the radiation intensity is expanded in  a Taylor series
 \begin{equation}
 	I_{\omega} = I_0 + \frac{1}{\kappa_{\omega} L} I_1 + \frac{1}{(\kappa_{\omega} L)^2} I_2 + \dots
 \end{equation}  
 
Substitution of this expansion into the equation ({\ref{rad}) gives to the first order 
\begin{equation}
	I_{\omega} = I_{\omega} ^0 - \frac{1}{\kappa_{\omega}} \vec{s} \cdot \nabla I_{\omega} ^0. 
\end{equation}	
	
It is possible to define the so called Rosseland thermal conductivity \cite{lem07}
\begin{equation}
	\lambda_R = \frac{4 \pi}{3} \int\limits_0^{\omega_{max}} \frac{1}{\kappa_{\omega}} \frac{\partial I_{\omega}}{\partial T} 
d \omega 
\end{equation}	
\noindent and thus get $\vec{q} \approx  - \lambda_R \nabla T$.	
	
The Rosseland approximation ignores the directional nature of the energy propagation. This feature is accounted for in the $P_N$ methods based on the expansion in terms of the orthogonal spherical harmonics truncated at order $N$.

The equation for radiation intensity in the stationary case is written as
\begin{equation}
 \vec{s} \cdot \nabla I_{\omega} + \frac{1}{\Lambda_{\omega}} I_{\omega} = \frac{1}{\Lambda_{\omega}} I_{\omega}^0.	
\end{equation}

The decomposition of the intensity is given as 
\begin{equation}
	I_{\omega} (x, \vec{s}) = \sum_{l = 0}^{\infty} \sum_{m =  - l}^l I_l^m (x) Y_l^m (\vec{s})
\end{equation}
\noindent where $Y_l^m$ are the spherical harmonics defined as
\begin{equation}
	Y_l^m (\Omega) = (- 1)^{\displaystyle \frac{m + |m|}{2}} \left[\frac{(l - |m|)!}{(1 + |m|)!} \right]^{\displaystyle \frac{1}{2}} e^{i m \phi} P_l^{|m|} (\cos \psi)
\end{equation}
\noindent where $\phi$ and $\psi$ are the polar angles (zenith and azimuth), $P_l^{|M|}$ are the associated Legendre polynomials.
Experience shows that  truncation at $N = 1$ (the $P_1$ method)  provides the sufficient accuracy in most cases 
\begin{equation}
I_{\omega} (x, \vec{s}) =   I_0^0 (x) Y_0^0 (\vec{s})	 +  I_1^{- 1} (x) Y_1^{- 1} (\vec{s})	 + I_1^0 (x) Y_1^0 (\vec{s}) + I_1^1 (x) Y_1^1 (\vec{s}),
\end{equation}
\noindent  the next approximation is $P_3$ resulting in the increase of the number of unknown functions from 4 to 16 \cite{lem07}.

The presence of phonons with a broad spectrum means that there is no the single value of the phonon MFP that determines the heat transfer regime. The range of values of the effective MFP of phonons is rather wide. E.g., for silicon at room temperature values from 40 nm to 260 nm are mentioned \cite{don14}.

Computations using the first principal methods predict different MFP distributions. E.g., in Si 80\% of heat is carried by the phonons with MFPs between 0.05 and 8$\mu m$  while in diamond 80\% of heat is carried by phonons with the MFPs between 0.3 and 2$\mu m$ \cite{win15}; more than 95 \% of heat in sapphire is carried by phonons with the MFPs shorter than 1 $\mu m$ \cite{hoog}.

The formation of the nanometre-scale phonon hotspots  is important for the analysis of the conduction  device cooling. This process involves the following  phenomena \cite{sin02}: 
{the boundary scattering that reduces the thermal conductivity;}
{the electron-phonon coupling in the transistor channel;}
{the weak anharmonic coupling between the slow longitudinal optical (LO) phonons generated by the scattering with  high energy electrons  and faster acoustic phonons.}

Dong et al. \cite{don14} in the studies of the heat transfer in the nanostructures distinguishing MFP of phonons in normal ($l_N = v \tau_N$) and resistive ($l_R = v \tau_R$) scatterings and suggested to define an effective length scale as $l = \sqrt{{l_N l_R}/{5}}$
and the Knudsen number as $Kn = l/L$.

The authors also considered different boundary conditions:
{the Maxwell boundary;}
{the backscattering boundary;}
{the MFP-proportional slip boundary.}

\subsection{Guyer-Krumhansl (GK) Equation}
\label{gk}
Guyer \& Krumhansl \cite{guy1,guy2} solved the BTE assuming that the normal scattering rates are  much larger than the resistive scattering rates that is valid at low temperatures.  They developed a phenomenological coupling between the phonons and elastic dilatational fields caused by the  lattice anharmonicity.

When the bulk phonon MFP is large compared to the sample size, the description is equivalent to the "Knudsen flow" or "ballistic transport".  Such heat conduction can occur without
energy dissipation \cite{pop10} because phonons can ballistically travel in straight lines for hundreds of nanometres. 
When boundary scattering dominates over intrinsic
scattering it is called the “Casimir limit” \cite{zim,marc} owing to
the Casimir's investigations  using  cylindrical rods \cite{cas}.

The directions of individual phonons are chaotic.
Anufriev et al. \cite{anu17} demonstrated a method to control the directionality of ballistic phonon transport using the membranes with arrays of holes that form the oriented  fluxes of phonons (silicon films with controlled arrays of holes can be fabricated by patterning the silicon-on-insulator (SOI) layer \cite{ols09,marc}). To show the potential for practical
applications, the authors introduced the thermal lens nanostructures and
demonstrated  the evidence of the nanoscale heat focusing.

Guyer \& Krumhansl established conditions when the Poiseuille flow can considerably contribute to the thermal conductivity.

The Poiseuille regime emerges when energy exchange
between phonons is frequent enough to keep the local
temperature well defined, and Umklapp collisions are
very rare. 
Recently Martelli et al. \cite{mar18} proved the existence of the Poiseuille flow of phonons in the single crystal of  perovskyte $Sr_{1 - x}Nb_xTiO_3$ at low temperatures. The authors  found that the thermal conductivity varies faster than cubic temperature dependence (that corresponds to the ballistic heat conduction regime) in a narrow ($6 K < T < 13 K$) temperature window. 

In the suspended graphene \cite{lee15}  and
other 2D systems \cite{gz14} the Poiseuille phonon flow could be observed at higher temperatures than in the 3D systems, because in the 2D systems
the normal momentum conserving phonon-phonon collisions are two orders of magnitude higher than those in the 3D systems \cite{sel15}.

The Poiseuille flow is realized for the steady heat conduction in a cylinder described by the equation \cite{jou90} 
\begin{equation}
\Lambda^2 \nabla^2 \vec{q} = \lambda \nabla T,
\end{equation}
\noindent with the solution given by  $q (r) = A (R^2 - r^2), \quad A = - \displaystyle  \frac{\lambda  \nabla T}{ \Lambda^2}$.

Usually the Guyer-Krumhansl equation is written in the form
\begin{equation}
	\label{krum0}
\tau \frac{\partial \vec{q}}{\partial t} + \vec{q} = - \lambda \nabla T + \beta^{\prime} \Delta \vec{q} + \beta^{\prime \prime} \nabla \cdot \nabla \vec{q},
\end{equation}
\noindent where $\beta^{\prime}$ and $\beta^{\prime \prime}$ are the Guyer-Krumhansl coefficients (in gas these coefficients are related to the relaxation times of the Callaway collision integral \cite{van17a}) or
\begin{equation}
\label{krum}
\tau \frac{\partial \vec{q}}{\partial t} + \vec{q} = - \lambda \nabla T + \Lambda^2  (\nabla^2 \vec{q} + 2 \nabla \cdot \nabla \vec{q}),
\end{equation}
\noindent where $\Lambda$ is MFP of phonons, $\lambda$ is the Ziman limit for the bulk thermal conductivity $\lambda = \rho c_v \tau \bar{c}^2 / 3$, $\rho$ is the mass density, $\bar{c}$ is the average speed of phonons \cite{alv12}. When $\Lambda = 0$ 
the equation (\ref{krum}) reduces to the Cattaneo relation (\ref{catt}).

Sellitto \& Alvarez \cite{sel12} used the Guyer-Krumhansl equation to study the heat removal from hot nanostructures characterized by a radius $r_0$ and a thickness $h_0$ through the graphene layer with a radius $r_g$ and a thickness $h_g$.

Increase of the integrated circuits (IC) density and of the clock speed makes the heat removal one of the major problem of the chip design \cite{bal09}.
The heat removed through the graphene layer is higher than through the usual materials \cite{sel12} since MFP of phonons in graphene  is very long --- of the order of hundreds nanometres at room temperature \cite{bal08,bal11,gho08,nik09,nik09a}.

The authors considered the simplified case when the heat is carried away  without being transversally transferred to the environment. The radial profile of the heat flux is given (in the first-order approximation in $dr$) by an equation
\begin{equation}
	\label{q}
	\frac{dq}{dr} + q (r) = 0
\end{equation} 
that results in
\begin{equation}
	\label{qr}
	q (r) = \frac{\Gamma}{r}
\end{equation}
where $\Gamma = Q/(2 \pi h_g)$ is a constant, $Q$ is the heat removed from per unit time.

Using this relation in the steady version of equation (\ref{krum}) yields for $r > r_0$ 
\begin{equation}
	\lambda_0 \frac{dT}{dr} = \Gamma \left(\frac{l^2}{r^3} - \frac{1}{r} \right)
\end{equation}
and after integration ($T (r) \equiv T_0$ for $r \le r_0$)
\begin{equation}
	\label{nl}
	T (r) = T_0 \frac{\Gamma}{\lambda_0} \left[\frac{l^2}{2 r_0^2} \left(   1 - \frac{r_0^2}{r^2}\right) + \ln \left(\frac{r_0}{r} \right)\right].
\end{equation}

In the absence of the nonlocal effects the equation (\ref{nl}) is simplified to
\begin{equation}
	\label{fo}
	T (r) = T_0 \frac{\Gamma}{\lambda_0} \ln \left(\frac{r_0}{r} \right).
\end{equation}

To account for the lateral heat transfer from the graphene layer, Sellitto \& Alvarez  \cite{sel12} introduced the heat-exchange coefficient between graphene and environment $\sigma$ and modified the equation (\ref{q}) to
\begin{equation}
	\frac{dq}{dr} + \frac{q}{r} = - \frac{2 \sigma}{h_g} [ T (r) - T_g ].
\end{equation}

The combination of the Guyer-Krumhansl equation with the energy balance produce the {\em parabolic} partial differential equation for the temperature and thus the paradox of the infinite velocity of the perturbations propagation \cite{leb97,leb14}.

Jou \& Cimmelli formulated the simplest heat conduction equation that accounts for the nonlocal effects written as \cite{jou16}
\begin{equation}
\label{krum1}	
\tau \frac{\partial \vec{q}}{\partial t} + \vec{q} = - \lambda \nabla T + \Lambda^2  \nabla^2 \vec{q}.	
\end{equation}

In some cases in the nanosystems it is possible  that $\Lambda^2  \nabla^2 \vec{q} \approx Kn^2 \gg \vec{q}$ since the Knudsen number could be much greater than unity \cite{jou16}. Then the nonlocal term in the equation (\ref{krum1}) is more important than the heat flux $\vec{q}$ itself and the equation reduces to
\begin{equation}
\label{krum2}	
 \nabla^2 \vec{q} = \frac{\lambda}{\Lambda^2} \nabla T.		
\end{equation}

Equation (\ref{krum2}) is similar to the Stokes equation of the hydrodynamics
\begin{equation}
	\nabla^2 = \frac{1}{\eta} \nabla   	
\end{equation}
\noindent that justifies the term "hydrodynamic regime" of the heat conduction and allows to define the "viscosity" of phonons. 
The hydrodynamics of semiconductor heat transport in the nanoscale allows to explain many experimental situations in terms of hydrodynamic concepts, such as friction and vorticity
\cite{tor18}.

The Guyer-Krumhansl equation was later obtained in the framework of nine-moment phonon hydrodynamics by Struchtrup et al. \cite{dre93,str14}. Fryer \& Struchtrup used the phonon BTE and the Callaway model for phonon–phonon interaction; a  for phonon interaction with crystal boundaries similar to the Maxwell boundary conditions in classical kinetic theory was exploited. Macroscopic transport equation for an arbitrary set of moments were developed and closed by means of Grad’s moment method. Sets with 4, 9, 16, and 25 moments were considered and solved analytically for 1D heat transfer and Poiseuille flow of phonons. The results showed the influence of Knudsen number on phonon drag at solid boundaries that the  Knudsen layers near the boundaries reduce the net heat conductivity of solids in rarefied phonon regimes.

Later Mohammadzadeh \& Struchtrup \cite{str17}  derived 
the macroscopic equations for phonon transport from the BTE. The Callaway model with frequency-dependent relaxation time was considered to describe the both resistive and normal processes in the phonon interactions. The Brillouin zone was considered to be a sphere with the diameter that depends on the temperature of the system. A model to describe phonon interaction with crystal boundary was employed to obtain macroscopic boundary conditions, where the reflection kernel is the superposition of diffusive reflection, specular reflection and isotropic scattering. Macroscopic moments were defined using a polynomial of the frequency and wave vector of phonons. As an example, a system of moment equations, consisting of three directional and seven frequency moments, i.e., 63 moments in total, was used to study the 1D heat transfer and the Poiseuille flow of phonons.

The Guyer-Krumhansl equation can be derived from the EIT 
if the non-locality is introduced into the entropy flux as ($\gamma$ is a positive coefficient) \cite{leb11}
\begin{equation}
\vec{J}^s = \frac{\vec{q}}{T} + \gamma (\vec{q} \cdot \nabla \vec{q} + 2 \vec{q}\nabla \cdot \vec{q})
\end{equation}

The non-local character of the heat transfer could lead to unusual results such as the heat flowing from cold regions to hotter ones. Cimmelli et al. \cite{sim15} studied an  axisymmetrical problem of the heat transfer in the  circular thin layer of thickness $h$ surrounding a source of heat --- the hot nanodevice. The authors assumed that the total heat flux $\vec{q}$ arises from two contributions due to both phonons and electrons $\vec{q} = \vec{q_p} + \vec{q_e}$. The evolution of the phonon and electron fluxes is governed by the  Guyer-Krumhansl equation (\ref{krum})
\begin{equation}
\label{kr_p}
\tau_p \frac{\partial \vec{q_p}}{\partial t} + \vec{q_p} = - \lambda \nabla T + \Lambda^2  (\nabla^2 \vec{q_p} + 2 \nabla \cdot \nabla \vec{q_p})
\end{equation}
and
\begin{equation}
\label{kr_e}
\tau_e \frac{\partial \vec{q_e}}{\partial t} + \vec{q_e} = - \lambda \nabla T + \Lambda^2  (\nabla^2 \vec{q_e} + 2 \nabla \cdot \nabla \vec{q_e}).	
\end{equation}
The temperatures of phonons and electrons are assumed to be equal (this restriction has been removed in the later authors' papers \cite{jou13,jou14}).

The radial dependence of the heat fluxes was obtained by considering two concentric circular areas at a radial  from the source equal to $r$ and $r + d r$ to get in the first order approximation in $d r$ 
\begin{equation}
	2 \pi ( r + d r) q ( r + d r) \approx 2 \pi ( r + d r) \left(q(r) + \frac{dq}{dr} dr \right)
\end{equation}
and thus
\begin{equation}
	r \frac{dq}{dr} + q(r) = 0.
\end{equation}
From the last equation it follows
\begin{equation}
	q(r) \equiv q_p(r) + q_e(r) = \frac{\Gamma}{r}
\end{equation}
where $\Gamma = (Q_0/2 \pi h)$ is a constant value, $Q_0$ is the heat produced by the hot source per unit time.

Both fluxes $\vec{q}_p$ and $\vec{q}_e$ are divergence free and 
thus (\ref{kr_p}, \ref{kr_e}) reduce to 
\begin{equation}
	\tau_p \frac{\partial \vec{q_p}}{\partial t} + \vec{q_p} = - \lambda \nabla T + \Lambda^2  \nabla^2 \vec{q_p}, \quad
	\tau_e \frac{\partial \vec{q_e}}{\partial t} + \vec{q_e} = - \lambda \nabla T + \Lambda^2  \nabla^2 \vec{q_e}	
\end{equation}
and  allow to determined the profiles of the phonon and electron fluxes. Finally,  Cimmelli et al. obtained the first-order differential equation for the temperature
\begin{equation}
	\frac{dT}{dr} = \left(\frac{l_p^2}{\lambda_p}\right) \frac{d^2 q_p}{dr^2} + \frac{1}{r} \left(\frac{l_p^2}{\lambda_p}\right) \frac{dq_p}{dr} - \frac{q_p}{\lambda_p}.
\end{equation}

Integration  in the neighbourhood of the source shows the temperature increase with the radial distance. This anomalous temperature hump --- the heat flux against the temperature gradient --- is related to the nonlocal effects. 

To prove the thermodynamic compatibility of this behaviour the authors considered the local balance of the entropy that in the steady case is
\begin{equation}
	\sigma^{(s)} = \nabla \cdot \vec{J^{(s)}}
\end{equation}
where $\sigma^{(s)}$ is the entropy production per unit volume and the entropy flux is
\begin{equation}
\vec{J^{(s)}} = \frac{\vec{q}}{T} + \left(\frac{l_p^2}{\lambda_p T^2} \right) \nabla \vec{q}_p \cdot \vec{q}_p + 	\left(\frac{l_e^2}{\lambda_e T^2} \right) \nabla \vec{q}_e \cdot \vec{q}_e.
\end{equation}

The entropy production was everywhere positive, thus the hump in the temperature distribution is physically possible since it agrees with the second law.

Calvo-Schwartzwalder \cite{cal19a} investigated the 1D by considering a Fourier law with an
 effective thermal conductivity proposed by Alvarez and Jou \cite{alv07} 
\begin{equation}
	\kappa_{eff} (L) = 2 \kappa \left(\frac{L}{l} \right)^2 \left( \sqrt{1 + \left(\frac{l}{L} \right)^2} - 1 \right)
\end{equation}
and a Newton cooling condition at the interface
between the solid and the cold environment. The author found that non-local effects become less important as the Biot number $Bi = h l/\kappa$ ($h$ is the the heat transfer coefficient in the Newton cooling condition $q (0, t) = h (T_e - T (0, t))$) decreases.

The simplified version of GK equation (\ref{krum2})
was used by Jou et al. to model heat transfer in NWs considered as the phonon flow \cite{x25,x27,x28}. The authors used the analogy with the flow of the viscous fluid along the cylindrical duct of radius $R$ under the pressure gradient $\Delta p / L$ with the velocity profile
\begin{equation}
	V (r) = \frac{\Delta p}{4 L \eta} (R^2 - r^2) 
\end{equation}
and the volume flow
\begin{equation}
	Q = \frac{\pi R^4}{8 \eta} \frac{\Delta p}{\eta}
\end{equation}
to obtain for the total heat flow along a  cylindrical conductor
\begin{equation}
	\label{qh}
		Q^h = \frac{\pi R^4 \kappa_0}{8 l^2} \frac{\Delta T}{L}.
\end{equation}
Thus an effective conductivity of NW depends on the ratio $l/R$  as
\begin{equation}
	\label{cnw}
	\kappa_{eff} = \frac{Q^h L}{\pi R^2 \Delta T}  = \frac{\kappa_0}{8 l^2} \frac{R^2}{l^2}. 
\end{equation}

If instead of equation (\ref{krum2}) the full Guyer-Krumhansl equation is used, the effective thermal conductivity is \cite{x27}
\begin{equation}
	\kappa_{eff} = \kappa_0 \frac{\tau_R c_0}{R} \left(1 - \frac{2 J_1 (i z)}{i z J_0 (i z)} \right),
\end{equation}
where $J_0$ and $J_1$ are the cylindrical Bessel functions \cite{leb}, $c_0$ is the Debay velocity, $l = c_0 \sqrt{\tau_r \tau_N}$ is the phonon MFP, $\tau_N$ and $\tau_R$ are the relaxation times. 

The equation (\ref{cnw}) gives a quadratic dependence of the effective thermal conductivity on the nanowire radius while experiments show linear variation \cite{x27}.
Alvarez et al. assumed that heat flux, similar to the velocity in the rarefied gas dynamics, does not obey the no-slip boundary condition on the wall, i.e. takes no-zero value on the wall and as in microfluidics \cite{bru} is proportional to the phonon MFP and to the flux gradient at the wall 
\begin{equation}
	\label{qw}
	q_w = C l \left( \frac{\partial q}{\partial r} \right)_{r = R}.
\end{equation}

This boundary condition is called sometimes "first-order slip condition" \cite{alv12} or "Maxwell slip model" \cite{bur59}.
In the case of the general geometry the boundary condition (\ref{qw})  written as \cite{sel15}
\begin{equation}
		q_w = C l \left( \frac{\partial q}{\partial \xi} \right)_{\gamma}
\end{equation}
where $\xi$ means the normal direction to the wall cross
section (pointing towards the flow), and $\gamma$ is the curve accounting for the outer surface of the transversal section of the system.
Lebon et al. \cite{leb12} investigated the heat slip flow along solid walls in the frame of EIT elevating the heat flux at the boundary to the status of independent variable and formulate boundary conditions obtained from the constraint imposed by the second law of thermodynamics expressing that the rate of entropy production is non-negative.

Xu \cite{xu14} derived the slip boundary condition for the heat flux from the BTE for phonon. Sellitto et al. \cite{a101,a11}  suggested an extended version of this condition ("second-order slip condition" \cite{alv12})
\begin{equation}
	\label{qw2}
	q_w = C l \left( \frac{\partial q}{\partial r} \right)_{r = R} - \alpha l^2 \left( \frac{\partial^2 q}{\partial r^2} \right)_{r = R}.
\end{equation}

The coefficients $C$ and $\alpha$ describe the effect of the interactions of phonons with walls: $C$ describes the specular or diffusive reflections of the phonons while $\alpha$ accounts for the backscattering collisions \cite{a11}. Both coefficients  depend on the temperature and are related to the properties of the wall \cite{a101,a102}.

Sellitto et al. \cite{a102} formulated the relations of the  coefficients $C$ and $\alpha$ to the characteristics of the wall and to the temperature
\begin{equation}
	C = C^{\prime} (T) \left( 1 - \frac{\Delta}{L} \right),\quad 
	\alpha = \alpha^{\prime} (T) \frac{\Delta}{L},
\end{equation}
the functions $C^{\prime} (T)$ and $\alpha^{\prime} (T)$  were determined by the authors for both smooth-walled and rough-walled silicon nanotubes.
When the condition (\ref{qw2}) is used with $\alpha = 2 /9$ the term "1.5-Order Slip-Flow Model" is also used \cite{mit93}.

Zhu et al. \cite{zhu17a} studied the heat transport in a 2-D nanolayer  in which the size along x-axis direction is much larger than MFP of heat carriers using the Guyer-Krumhansl equation
(\ref{krum}) with the second-order slip condition (\ref{qw2}) where
the coefficient $C$ is written  \cite{car15} similar to the relation for gases and for electron
collisions \cite{zim} $C =	2 (1 + P)/(1 - P)$ 
where $P$ is the fraction of carriers reflected back  specular from a solid surface.
Since $P \in [0,1]$, the coefficient $C  \in [2, \infty]$. Small values of $C$ mean that diffusive phonon-wall collisions are predominant over specular ones.  The boundary scattering is the main cause of the non-uniform heat-flux profile in the hydrodynamic regime \cite{sel15}.

The slip-flow contribution to the heat flux is essential in the so-called Knudsen layer with the  thickness of the order of the phonon   MFP. In nano systems with dimension comparable or smaller than the phonon  MFP the influence of the slip boundary condition extends to the whole system.

Alvarez et al. \cite{alv12} stated that the slip boundary condition shold be modified in the case of the high-frequency perturbations: the relaxation of the heat flux should be introduced similar to the Cattaneo law
\begin{equation}
	\tau_w \frac{\partial q_w}{\partial t} + q_w = C l \left( \frac{\partial q}{\partial r} \right)_{r = R} - \alpha l^2 \left( \frac{\partial^2 q}{\partial r^2} \right)_{r = R}
\end{equation}

The relaxation time $\tau_w$ should account for the specular, diffusive and backward reflection of the phonons from the wall and can be determined according to the Matthiessen's rule
\begin{equation}
	\frac{1}{\tau_w} = \frac{1}{\tau_{spec}} + \frac{1}{\tau_{diff}} + \frac{1}{\tau_{back}}.
\end{equation}

Alvarez et al. suggested an estimate of the relaxation time through the total frequency of the phonon-wall collisions assuming that the wall has smooth and rough regions  of width $D$ and $\Delta$,  the relaxation time is expressed as
\begin{equation}
		\frac{1}{\tau_w} = \frac{\bar{c}}{d  - D} \frac{D}{D + \Delta}  + \frac{\bar{c}}{d  - \Delta} \frac{\Delta}{D + \Delta} 
\end{equation}
where the ratios $D/(D + \Delta)$  and $\Delta/(D + \Delta)$  are the probabilities of finding the smooth and rough regions, respectively,   $\bar{c}$ is the mean phonon speed.

Use of the slip boundary condition for the heat flux increases the total heat flow along a  cylindrical conductor: (\ref{qh})  is modified as
\begin{equation}
Q^h = \frac{\pi R^4 \kappa_0}{8 l^2} \frac{\Delta T}{L} \left(1 + 4 C \frac{l}{R} \right)	
\end{equation}
and the effective thermal conductivity (\ref{cnw}) as
\begin{equation}
	\label{eff}
	\kappa_{eff}  = \frac{\kappa_0}{8} \frac{R^2}{l^2}	\left(1 + 4 C \frac{l}{R} \right)
\end{equation}
where $C$ is a constant related to the wall properties.

In  terms of Knudsen number equation (\ref{eff}) is written as
\begin{equation}
	\kappa_{eff}  = \frac{\kappa_0}{8 Kn} \left(1 + 4 C Kn \right).
\end{equation}

For nanowires of small radius $R \ll l$ the equation (\ref{eff}) simplifies to
\begin{equation}
		\kappa_{eff} =  \frac{\kappa_0 C}{2} \frac{R}{l},	
\end{equation}
thus linear dependence  on the nanowire radius is obtained.
It is known \cite{x27} that quantum effects could lead to the crossover from the linear dependence of the effective conductivity $\kappa_{eff}$ on the radius to the quadratic behaviour \cite{a16,a17}.

For the thin film of thickness $h$ Alvarez et al. \cite{x27} derived \begin{equation}
	\kappa_{eff} = \frac{\kappa_0}{12} \frac{h^2}{l^2} \left(1 + 6C \frac{l}{h} \right)
\end{equation}
when the simplified  Guyer-Krumhansl equation is used and
\begin{equation}
	\kappa_{eff} = \kappa_0 \left[1 - \frac{2 l}{h} \tanh \left(\frac{h}{2 l} \right) \right] 
\end{equation}
in the case of the full equation.

\subsection{Ballistic-Diffusive Model}
Ballistic-diffusive (BD) introduced by Chen \cite{che1,che2} is an approximation to the {\em grey} unsteady BTE \cite{mur05} and is close to the  differential approximation  used in the analysis of the radiative heat transfer \cite{mod} that involves replacing the integral equation for the heat transfer by the differential equation for the heat flux  and its modification (MDA) suggested by Olfe \cite{olfe}.

BD  model is based on the splitting of the distribution function,  the internal energy and the heat flux into the two parts
\begin{equation}
f = f_b + f_d,\quad  e = e_b + e_d,\quad  \vec{q} = \vec{q}_b + \vec{q}_d 
\end{equation}
\noindent reflecting the coexistence of two kinds of the heat carriers:
{ballistic phonons that experience mainly the collisions with the boundaries and}
{diffusive phonons that undergo multiple collisions within the core of the system.}
The evolution of the parts of the internal energy  is governed by the balance equations
\begin{equation}
	\frac{\partial e_b}{\partial t} = - \nabla \cdot \vec{q}_b + r_b, \qquad
	\frac{\partial e_d}{\partial t} = - \nabla \cdot \vec{q}_d + r_d .
\end{equation}

The source terms $r_b$ and $r_d$ describe the energy exchange between the ballistic and the diffusive phonon populations and in the absence of sources $r_b = - r_d$.

The evolution equations for the heat flux are
\begin{align}
\label{bd_1}	
	\tau_b \frac{\partial \vec{q}_b}{\partial t}  + \vec{q}_b &= - \lambda_b \nabla T + \Lambda_b^2 (\nabla^2 \vec{q}_b + 2 \nabla \nabla \cdot \vec{q}_b,)\\
\label{bd_2}
		\tau_d \frac{\partial \vec{q}_d}{\partial t}  + \vec{q}_d &= - \lambda_d \nabla T.
\end{align}

The coupled equations (\ref{bd_1}, \ref{bd_2}) lead to the Guyer-Krumhansl equation under the assumption that $\tau_d = \tau, \Lambda_b = \Lambda, \lambda = \lambda_b + \lambda_d$.

The relative contributions of the components of the phonon distribution  depend on the value of the Knudsen number $Kn$ and on the geometry of the considered system \cite{Li18a}. As noted Li \& Cao in the just cited paper, the ballistic transport is responsible for the non-linearity of the temperature distribution and this non-linearity increases with increasing the Knudsen number.

The advantage of the BD model over the BTE is the simplicity --- only time and spatial coordinates are involved.

Similar model called the "two-channel model" is based on the assumption that there are two individual heat conduction channels and division of the phonon population into two parts: the long MFP and the short MFP phonons.

Chen \cite{che1} used the BTE under the RTA assuming that the relaxation time depends on the phonon frequency $\omega$  and does not depend on the on the wave vector that is linearly related to the phonon quasi-momentum $\vec{p} = \hbar \vec{k}$ , i.e. the isotropic scattering is considered  

\begin{equation}
	\frac{\partial f}{\partial t} + \vec{v} \cdot \nabla f = - \frac{f - f_0}{\tau (\omega)}.
\end{equation}
\noindent where $f_0$ is the equilibrium phonon  distribution function, $\vec{v}$ is the phonon group velocity $\vec{v} = \nabla_{\vec{k}} \omega$.
The SMRTA leads to the following expression for the phonon thermal conductivity \cite{nik09}
$\lambda = {1}/{3} v^2 c \tau$,
 where $c$ is the specific heat capacity.

The equations for the ballistic and diffusive parts are 
\begin{equation}
	\frac{1}{|\vec{v}|} \frac{\partial f_b}{\partial t} + \vec{\Omega} \cdot \nabla f_b = - \frac{f_b}{|\vec{v}|}, \quad
	\frac{\partial f_d}{\partial t} + \vec{v} \cdot \nabla f_d = - \frac{f_d - f_0}{\tau},	
\end{equation}
\noindent where $\vec{\Omega}$ is the unit vector in the direction of the phonon propagation.

Rezgui et al. \cite{rez} assumed that $\tau = \tau_R$ obeys to the Matthiessen's rule
\begin{equation}
	\frac{1}{\tau_R} = \frac{1}{\tau_U} + \frac{1}{\tau_i} + \frac{1}{\tau_b}
\end{equation}
where $\tau_U$  is the relaxation time of Umklapp phonon-phonon collisions, $\tau_i$ is the relaxation time of phonon-impurity collisions and $\tau_b$ is the relaxation time of phonon-boundary collision. Using the Taylor expansion it is possible to get
\begin{equation}
\label{rez9}
\frac{f_d (r, \epsilon, t  + \tau_R) - f_d (r, \epsilon, t)}{\tau_R}	 + v \nabla f_d (r, \epsilon, t) = \frac{f_d (r, \epsilon, t) - f_0 (r, \epsilon)}{\tau_R}
\end{equation}
where $\epsilon$ is the kinetic energy; the diffusive and the ballistic  fluxes are
\begin{equation}
q_d (r, t) = \int_{\epsilon} v (r,t)  f_d (r, \epsilon, t) \epsilon D(\epsilon) d\epsilon, \quad
q_b (r, t) = \int_{\epsilon} v (r,t)f_b (r, \epsilon, t) \epsilon D(\epsilon) d\epsilon,
\end{equation}
$D(\epsilon)$ is the density of states, the total heat flux is clearly $q = q_d + q_b$.

Rearranging the terms of equation (\ref{rez9}) Rezgui et al. get
\begin{equation}
	\label{rez10}
	f_0 (r, \epsilon) = \tau_R v \nabla f_d (r, \epsilon, t) + f_d (r, \epsilon, t + \tau_R).
\end{equation}

Multiplying of the equation (\ref{rez10}) by $\epsilon D (\epsilon) v$ and accounting for the relation 
\begin{equation}
\int_{\epsilon} v f_0 \epsilon D(\epsilon) d\epsilon
\end{equation}
yields
\begin{equation}
	q_d (r, t + \tau_r) + \int_{\epsilon} \tau_R v (r, T) \nabla f_d (r, \epsilon, t) \epsilon D(\epsilon) v d\epsilon = 0
\end{equation}
or, assuming that
\begin{equation}
	\nabla f = \frac{df}{dT} \nabla T, \quad \lambda = \int \tau_R v^2 \frac{df_d}{dt} \epsilon D(\epsilon) d\epsilon
\end{equation}
the equation
\begin{equation}
	\tau_R \frac{dq_d}{dt} + q_d = - \lambda \nabla T_d.
\end{equation}

Using the conservation of the total internal energy is defined as $u = u_d + U_b$ in the form ($\dot{q_h}$ is the volumetric heat generation)
\begin{equation}
	\frac{\partial u (r, t)}{\partial t} = C \frac{\partial T (r, t)}{\partial t} = - \nabla q (r, t) + \dot{q_h}.
\end{equation}
The authors wrote out the equation they called {\em extended} ballistic-diffusive model
\begin{equation}
	\tau_R \frac{\partial^2 T_d (r, t)}{\partial t^2} + \frac{\partial T_d (r, t)}{\partial t} = \frac{1}{C} \nabla (\lambda \nabla T_d (r, t)) - \frac{1}{C} \nabla q_b (r, t)
	+ \frac{\dot{q_h}}{C} + \frac{\tau_R}{C}  \frac{\partial \dot{q_h}}{\partial t}.
\end{equation}

Chen \cite{che1} solved the BTE  using the spherical harmonic expansion and  keeping the first two terms
\begin{equation}
	f_b = g_0 + \vec{g}_1 \cdot \vec{\Omega},
\end{equation}
\noindent where $g_0$ is the average of $f_b$ over all directions, $\vec{g}_1$ is related to the heat flux.

Using this solution, multiplication of the equation for diffusive part of the distribution function  by $\vec{\Omega}$ and integration over the solid angle gives 

\begin{equation}
	\frac{1}{|\vec{v}|} \frac{\partial \vec{g}_1}{\partial t} + \nabla g_0 = - \frac{\vec{g}_1}{\Lambda},
\end{equation}
\noindent $\Lambda = |\vec{v}| \tau$ is the phonon MFP.
Thus the heat flux and the internal energy are
\begin{equation}
	\vec{q} = \frac{1}{4 \pi} \int \vec{v} \hbar \omega (f_b + f_d) d^3 \vec{v} = \vec{q}_b + \vec{q}_d, \quad
	e = \frac{1}{4 \pi} \int \hbar \omega (f_b + f_d) d^3 \vec{v} = e_b + e_d.
\end{equation}

Yang et al. \cite{yan05} used the BTE in terms of the phonon intensity 
$
I_{\omega} = \vec{v}_{\omega} \hbar \omega f D(\omega)/4 \pi 
$
 where $\vec{v}_{\omega}$ is the carrier group velocity, $\omega$ is the phonon circular frequency, $D(\omega)$ is the phonon density of states per unit volume, $S_{\omega}$ is the source term that could be determined, e.g., by the electron-phonon scattering
\begin{equation}
\frac{\partial I_{\omega}}{\partial t} + \vec{v}_{\omega} \nabla I_{\omega} = - \frac{I_{\omega} - I_{0 \omega}}{\tau_{\omega}} + S_{\omega}.
\end{equation}

The equations for the ballistic and diffusive parts of the phonon distribution function  were written, respectively, as
\begin{equation}
\frac{\partial I_{b \omega}}{\partial t} + \vec{v}_{\omega} \nabla I_{b \omega} = - \frac{I_{b \omega}}{\tau_{\omega}} + S_{\omega}, \quad 
\frac{\partial I_{d \omega}}{\partial t} + \vec{v}_{\omega} \nabla I_{d \omega} = - \frac{I_{b \omega} - I_{0 \omega}}{\tau_{\omega}}.
\end{equation}

Allen \cite{all18} performed an analysis of the crossover from ballistic to diffusive regime of the heat conduction using the computer simulations and a Fourier-transformed version of the phonon BTE.  

The evolution equation  of the average occupation in the reciprocal space of the phonon mode $Q$ includes several terms (drift, scattering, external)
\begin{equation}
\frac{\partial N_Q}{\partial t} = \left( \frac{d N_Q}{d t} \right)_{drift} + \left( \frac{d N_Q}{d t} \right)_{scatt} + \left( \frac{d N_Q}{d t} \right)_{ext},
\end{equation}
\begin{equation}
\left( \frac{d N_Q}{d t} \right)_{drift} = - \vec{v}_Q \cdot \nabla N_Q =  - \vec{v}_Q \cdot \left[ \frac{d n_Q}{d T} \nabla T + \nabla \Phi_Q \right],
\end{equation}
\begin{equation}
\left( \frac{d N_Q}{d t} \right)_{scatt} = - \sum_{Q^{\prime}} S_{Q, Q^{\prime}} \Phi_{Q^{\prime}},
\end{equation}
\noindent where $n_Q$ is the local equilibrium Bose-Einstein distribution, $\Phi_Q = N_Q - n_Q$ and $S_{Q, Q^{\prime}}$ is the linearised scattering operator. 

Vazquez et al. \cite{vaz20} and Lebon et al. \cite{leb11} have considered the two-temperature variant of BD model. Vazquez et al. used the Guyer-Krumhansl equation to describe both the ballistic and diffusive heat fluxes. Lebon et al. assumed that the distributive function of the diffusive phonons is governed by the Cattaneo equation (\ref{catt}) while the ballistic phonons obeys  the Guyer-Krumhansl equation.

Direct imaging of ballistic and diffusive thermal transport in graphene structures was reported by Pumarol et al. \cite{pum12}. 

Siemens et al. \cite{sim10} used the ultrafast coherent soft X-ray beams  to study heat transfer from the nanoscale hotspot (the highly doped Si resistor near a thin Si membrane). The authors found three times decrease of the energy dissipation away from the heat source compared to the predictions by  the Fourier law. 

\subsection{Unified Nondiffusive-Diffusive Model}
Rama \& Ma \cite{ram14} suggested a two-fluid (two-channel) model similar to the BD model and called it the Unified Nondiffusive Diffusive (UND) model. The authors based this model on the BTE, considered 1D case and  used different accuracy of the spherical harmonic expansions of the phonon distribution function for the low frequency (LF) phonons  with MFP of the same order of magnitude as the length scale of interest and high frequency (HF) phonons.

The phonon spectrum is divided into two parts:
{high-heat-capacity high frequency phonons that are in quasi-equilibrium with local temperature} and 
{low-heat-capacity  low frequency phonons that are farther out of equilibrium.}

The LF modes do not interact with each other due to the small phase-space for for such scattering \cite{maz11} but can exchange energy with the HF modes \cite{ram14}. 

The distribution function for LF phonons $g (x, \vec{k})$ is expanded in the spherical harmonic functions that form a complete orthogonal set 
\begin{equation}
\label{ram1}
	g (x, \vec{k}) = \sum_{i = 0}^{\infty} g_i (x, \vec{k}) P_i (\cos \theta),
\end{equation}
\noindent where $\vec{k}$ is the phonon wave-vector making an angle $\theta$ with the x-axis.

The authors assumed that all LF modes have the same lifetime $\tau$,  the same group velocity $v$ and that the phonon dispersion is isotropic (thus $\vec{v} = v \vec{k}/k$). 

The linearised steady-state BTE for LF part of spectrum is 
\begin{equation}
\label{ram2}
	v \cos \theta \frac{\partial g}{\partial t} = - \frac{g - f_0}{\tau}.
\end{equation}

Substitution of (\ref{ram1}) into the BTE (\ref{ram2}), multiplying by $P_i (\cos \theta) \sin \theta$ ($i = 0, 1, 2,...$) and integration over $\theta$  produces a hierarchy of coupled equations for $g_i$. The first three of these equations are written as
\begin{equation}
	\frac{1}{3} v \frac{\partial gne_1}{\partial x} + \frac{g_0 - f_0}{\tau} = 0,\quad
	\frac{2}{5} v \frac{\partial g_2}{\partial x} + v \frac{\partial g_0}{\partial x}  + \frac{g_1}{\tau} = 0,\quad
	\frac{3}{7} v \frac{\partial g_3}{\partial x} + \frac{2}{3} v \frac{\partial g_1}{\partial x}  + \frac{g_2}{\tau} = 0.
\end{equation} 

Ramu \& Ma truncate the expansion at the second order by requiring $g_3 = 0$ that is the next approximation after Fourier law which consists in setting $g_2 = 0$.
 
After elimination of $g_2$ and $g_0$ Ramu \& Ma get the equation in terms of $g_1$
\begin{equation}
\label{ram5}
	- \frac{3}{5} (v \tau)^2 \frac{\partial^2 g_1}{\partial x^2} + v \tau \frac{\partial f_0}{\partial x} + g_1 =0.
\end{equation}

Since $f (T)$ depends on $x$ only through $T$ it is possible to use 
\begin{equation}
	\frac{\partial f_0}{\partial x} = \frac{\partial f_0}{\partial T} \frac{d T}{d x}.
\end{equation}

Multiplying equation (\ref{ram5}) by $(4 \pi/3) \hbar \omega v$ and summing over all $\vec{k}$ gives 
\begin{equation}
	- \frac{3}{5} (v \tau)^2 \frac{\partial^2 q^{LF}}{\partial x^2} + \frac{1}{3} C^{LF} v^2 \tau \frac{\partial T}{\partial x} + q^{LF} = 0
\end{equation}
\noindent where
\begin{equation}
	C^{LF} = 4 \pi \frac{\partial}{\partial T} \left( \frac{1}{(2 \pi)^3} \int\limits_k \hbar \omega f_0 k^2 d k \right)
\end{equation}
\noindent is the volumetric heat capacity of LF modes.

The expression for the LF heat flux could be written defining the thermal conductivity of LF modes as $\lambda^{LF} = {1}/{3} C^{LF} v^2 \tau$ and MFP of low frequency phonons as $\Lambda^{LF} = v \tau$
\begin{equation}
	q^{LF} = \frac{3}{5} (\Lambda^{LF})^2 \frac{\partial^2 q^{LF}}{\partial x^2} - \lambda^{LF} \frac{\partial T}{\partial x}.
\end{equation}

Similar analysis is performed for the HF phonons
\begin{equation}
	h (x, \vec{k}) = \sum_{i = 0}^{\infty} h_i P_i (\cos \theta)
\end{equation}
\noindent with two distinctions: (1) {the spherical harmonic expansion is truncated at the first order, i.e. $h (x,\vec{k}) = h_0 + h_1 \cos \theta$}; (2) {the Bose statistics is used for the symmetric part of the distribution, i.e. $h_0 (x,\vec{k}) = f_0$.}

Thus the HF heat flux is written as 
\begin{equation}
	q^{HF} = - \lambda^{LF} \frac{\partial T}{\partial x}.
\end{equation}

Using the definition $\lambda = \lambda^{LF} + \lambda^{HF}$ the total heat flux is expressed as
\begin{equation}
		q = \frac{3}{5} (\Lambda^{LF})^2 \frac{\partial^2 q^{LF}}{\partial x^2} - \lambda  \frac{\partial T}{\partial x}.
\end{equation}

\subsection{Two-fluid model}
Two-fluid model \cite{bax81,lai96,ju99} similar to the BD model divides the phonon population into two groups:  
{reservoir group (longitudinal optical (LO), transverse optical (TO), transverse acoustic (TA) phonons)} and 
{propagating group (longitudinal acoustic (LA) phonons that  have a single group velocity).}
All phonons are assumed to have a single overall scattering time \cite{sin02}.

Monte-Carlo computations show that the high-energy electrons scatter preferentially with LO phonons \cite{jac83}  thus the BTE for LO phonons includes the term $S_{e-LO}$ (that is absent in equations for other phonon modes) and could be written in the steady state as 
\begin{multline*}
	\vec{v} \cdot \nabla \phi = - \frac{\phi - \bar{\phi}}{\tau_{imp} (\omega)} - \frac{\phi}{\tau_{an} (\omega)} \int\limits_{\omega^{\prime} < \omega} P (\omega \rightarrow \omega^{\prime}) d \omega^{\prime} -  \int\limits_{\omega^{\prime} > \omega} \frac{\phi}{\tau_{an} (\omega^{\prime})} P (\omega \rightarrow   \omega^{\prime}) d \omega^{\prime} \\ + \int\limits_{\omega^{\prime} < \omega} \frac{\phi^{\prime}}{\tau_{an} (\omega)} P (\omega^{\prime} \rightarrow   \omega) d \omega^{\prime} + \int\limits_{\omega^{\prime} > \omega} \frac{\phi^{\prime}}{\tau_{an} (\omega^{\prime})} P (\omega^{\prime} \rightarrow   \omega) d \omega^{\prime} + S_{e-LO}
\end{multline*}
\noindent where $\vec{v}$ is the group velocity, $\omega$ is the angular frequency, $\phi$ is the number of phonons per unit volume and angular frequency, $\bar{\phi}$ is an average over all directions at a given point, indexes "imp" and "an" refer to the elastic impurity scattering and the anharmonic phonon scattering, respectively.

Multiplication of equations by $\omega$, integration over  frequency and summing branches gives energy balances for the reservoir and propagating groups \cite{sin02}
\begin{equation}
	0 = \frac{\Delta u_{LR}}{\tau} + q^{\star} = - \frac{C_R (T_R - T_L)}{\tau} +  q^{\star}
\end{equation}
\noindent and
\begin{equation}
\nabla \cdot j_P = \frac{\Delta u_P}{\tau} = \frac{C_p (T_P - T_L)}{\tau}
\end{equation}
\noindent where $\Delta u_{LR}$ is the energy transferred out of the reservoir as it relaxes towards equilibrium with the lattice, $q^{\star}$ is the power generated by hot electrons, $C_R$ is the heat capacity of the reservoir phonons, $\Delta u_{PL}$ is the energy gained by the propagating phonons from the phonon reservoir. 

\subsection{Generalized Fourier Law by Hua et al.}
Hua et al. \cite{hua19} developed a generalized Fourier model valid  from ballistic to diffusive regimes. The authors start with the mode-dependent phonon  BTE under the RTA (Bhatnagar-Gross-Krook) model \cite{xu11b,xu14})
\begin{equation}
	\label{hua1}
	\frac{\partial g_{\mu} (\vec{x},t)}{\partial t} + \vec{v}_{\mu} \cdot g_{\mu} (\vec{x},t)  = \frac{g_{\mu} - g_0 (T, \vec{x},t)}{\tau_{\mu}}  + \dot{Q}_{\mu},
\end{equation}
\noindent where $g_{\mu} (\vec{x},t) = \hbar \omega_{\mu} (f_{\mu} (\vec{x},t) - f_0 (T))$ is the deviational energy distribution function for the phonon  state $\mu = (\vec{q}, s)$; here $\vec{q}$ is the phonon wavevector, $s$ is the phonon branch index, 
$f_0$ is the Bose-Einstein distribution, $g_0 (T) = \hbar \omega_{\mu} (f_0 (T) - f_0 (T_0)) \approx C_{\mu} \Delta T$, $C_{\mu}$ is the mode dependent specific heat, $\dot{Q}_{\mu}$ is the heat input rate per mode.  

To close the problem, the energy conservation law is used
\begin{equation}
	\frac{\partial E (\vec{x},t)}{\partial t} + \nabla  \cdot  \vec{q} (\vec{x},t) = Q (\vec{x},t),
\end{equation}
\noindent where the total volumetric energy and the heat flux are
\begin{equation}
E (\vec{x},t) = \frac{1}{V}	\sum_{\mu} \vec{g}_{\mu}, \quad
	\vec{q} = \frac{1}{V}	\sum_{\mu} \vec{g}_{\mu} \vec{v}_{\mu}
\end{equation}
\noindent The sum over $\mu$ denotes a sum over all phonon modes in the Brillouin zone.

To get a generalized constitutive relation between the heat flux and the temperature gradient, the authors rearrange the equation (\ref{hua1}) and performed the Fourier transformation in time to get
\begin{equation}
\label{hua3}
	\Lambda_{\mu x} \frac{\partial \tilde{g}_{\mu}}{\partial x} + 
	\Lambda_{\mu y} \frac{\partial \tilde{g}_{\mu}}{\partial t} + 
	\Lambda_{\mu z} \frac{\partial \tilde{g}_{\mu}}{\partial z} + 
	(1 + i \eta \tau_{\mu}) \tilde{g}_{\mu} = C_{\mu} \Delta \tilde{T} + \tilde{Q}_{\mu} \tau_{\mu},
\end{equation}
\noindent $\eta$ is the Fourier frequency, $\Lambda_{\mu x}, \Lambda_{\mu y}, \Lambda_{\mu z}$ are the directional MFPs along $x, y, z$.

The authors introduced the new independent variables $\xi, \rho, \zeta$
\begin{equation}
	\xi = x, \quad
	\rho = \frac{\Lambda_{\mu y}}{\Lambda_{\mu}} x - \frac{\Lambda_{\mu x}}{\Lambda_{\mu}} y,\quad
    \zeta =  \frac{\Lambda_{\mu z}}{\Lambda_{\mu}} x - \frac{\Lambda_{\mu x}}{\Lambda_{\mu}} z .		
\end{equation}
\noindent where $\Lambda_{\mu} = \sqrt{\Lambda_{\mu x} ^2 + \Lambda_{\mu y} ^2 + \Lambda_{\mu z} ^2}$. 

The equation (\ref{hua3}) after this transformation becomes 
\begin{equation}
\label{hua5}
\Lambda_{\mu \xi} \frac{\partial \tilde{g}_{\mu}}{\partial \xi} + \alpha_{\mu} \tilde{g}_{\mu} = C_{\mu} \Delta \tilde{T} + \tilde{Q}_{\mu} \tau_{\mu},
\end{equation}
	\noindent where $\alpha_{\mu} = 1 + i \eta \tau_{\mu}.$ 
The authors, solving it on the interval $[L1, L2]$, related the temperature gradient to the mode-specific heat fluxes
\begin{equation}
	\label{hua12}
\tilde{q}_{\mu \xi} = - \int\limits_{\Gamma} \lambda_{\mu \xi} (\xi - \xi^{\prime}) \frac{\partial T}{\partial \xi^{\prime}} d \xi^{\prime} + B_{\mu} (\xi, \rho, \zeta, \eta),
\end{equation}
\noindent where $B_{\mu} (\xi, \rho, \zeta, \eta)$ is determined by the boundary conditions and the volumetric heat input rate,
\begin{equation}
	\Gamma \in 
	\begin{cases}
	 [L1, \xi)\qquad if\quad v_{\mu \xi} > 0,\\
	 
	 	(\xi, L_2]\qquad if\quad v_{\mu \xi} < 0,
	 	\end{cases}
\end{equation}
\noindent and the model thermal conductivity along the direction $\xi$ 
\begin{equation}
\lambda_{\mu \xi} = C_{\mu} v_{\mu \xi} \Lambda_{\mu \xi} \frac{\exp(- \alpha_{mu} 
\displaystyle	\left|\frac{\xi}{\Lambda_{\mu \xi}} \right|)} {\alpha_{\mu} |\Lambda_{\mu \xi}|}.	
\end{equation}

The first term in equation (\ref{hua12}) is a convolution a space- and time-dependent thermal conductivity and the temperature gradient along the the direction $\xi$ and reflect the nonlocality of heat conduction.

\subsection{Phonon hydrodynamics}
Guo \& Wang \cite{guo15,guo18} derived the macroscopic equations for the phonon gas motion from the phonon BTE
\begin{equation}
\frac{\partial f}{\partial t} + \vec{v}_g = C (f)
\end{equation}
\noindent where $f = f (\vec{x}, t, \vec{k})$ is the phonon distribution function,
$\vec{v}_g = \nabla_{\vec{k}} \omega$ is the group velocity. 
The scattering term $C (f)$ includes the contribution of 
{the normal scattering (N-process) and}
{the resistive scattering (R-process).}
The energy is conserved in both kinds of collision of pnonons with the wavevectors $\vec{k}_1$ and $\vec{k}_2$
\begin{equation}
	\hbar \omega (\vec{k}_1) + \hbar \omega (\vec{k}_2) = \hbar \omega (\vec{k}_3)
 \end{equation}
while the quasimomentum of phonons 
\begin{equation}
	\label{qm}
	\vec{k}_1 + \vec{k}_2 = \vec{k}_3 + \vec{b}
\end{equation}
is conserved only in the N-processes for which $\vec{b}$ is the reciprocal lattice vector or $\vec{b} = 0$ \cite{zim,sn03}. 

The simplification of the BTE is based on the Callaway's dual relaxation model that assumes that N process and R process proceed separately. At low temperatures the dominant process is  N process while at ordinary temperatures N process is negligible and BTE becomes
\begin{equation}
\frac{\partial f}{\partial t} + \vec{v}_g = - \frac{f - f_R^{eq}}{\tau_R}
\end{equation} 

\noindent where the equilibrium distribution for R processes is the Planck distribution
\begin{equation}
	\label{pla}
f_R^{eq} = \frac{1}{exp \left(\displaystyle \frac{\hbar \omega}{k_B T} \right) - 1}.
\end{equation}

The phonon hydrodynamic model is based on the macroscopic field variables ---
the phonon energy density, the heat flux and the flux of the heat flux:
\begin{equation}
e = \int \hbar \omega f d \vec{k}, \quad
\vec{q} = \int \vec{v}_g \hbar \omega f d \vec{k}, \quad
\hat{Q} = \int \hbar \omega \vec{v}_g \vec{v}_g f d \vec{k}.
\end{equation}

Integration of the BTE over the wave vector space gives the balance equations for the energy density and the heat flux
\begin{equation}
\frac{\partial e}{\partial t} +\nabla \cdot \vec{q} = 0, \quad
\frac{\partial \vec{q}}{\partial t} +\nabla \cdot \hat{Q} = - \frac{\vec{q}}{\tau_R}
\end{equation}

These balance equations are the four-moment field 
equations of the phonon BTE\footnote{There is a misprint in \cite{guo18} --- $\hat{Q}$ is printed as a vector in the equation for the heat flux.}. To close the system of the phonon transport equations  the flux of the heat flux $\hat{Q}$ has to be specified in terms of four basic field variables (the energy density and three components of the heat flux).

The authors develop a perturbation equation to the BTE around the four-moment nonequilibrium phonon distribution function obtained by the maximum entropy principle.
Thus the problem reduces to maximization of the functional
\begin{multline}
	\label{fun}
\Phi = -k_B \int [f \ln f - (1 + f) \ln (1 + f)] d \vec{k}  + \beta \left( e - \int \hbar \omega f d \vec{k} \right) + \\
\gamma_i \left( q_i - \int v_{gi} \hbar \omega f d \vec{k} \right),
\end{multline} 
\noindent where $\beta$ and $\gamma_i$ are the Lagrange multiplies.

The extremum conditions of the functional (\ref{fun}) are  as follows \cite{guo18}:
\begin{equation}
	\label{df}
	\frac{\partial \Phi}{\partial f} = \int \left[ k_B \ln \left(1 + \frac{1}{f}\right) - \beta \hbar \omega - \gamma_ig v_{gi} \hbar \omega \right] d\vec{k} = 0
\end{equation}
\begin{equation}
	\frac{\partial \Phi}{\partial \beta} = e - \int \hbar \omega f d \vec{k}, \quad
	\frac{\partial \Phi}{\partial \gamma_i} = q_i - \int v_{gi} \hbar \omega f d \vec{k}.
\end{equation}

Finally, from equation (\ref{df}) follows the four-moment nonequilibrium phonon distribution function as 
\begin{equation}
	\label{f4}
f_4 = \frac{1}{\exp \left( \displaystyle  \beta \frac{\hbar \omega}{k_B} + \gamma_i \frac{v_{gi} \hbar \omega}{k_B} \right) - 1}
\end{equation} 
where the subscript "4" in the phonon distribution function
represents its dependence on the four basic field variables.
At equilibrium state, the heat flux and its corresponding
Lagrange multiplier vanish and equation (\ref{f4})  reduces
to the Planck distribution equation (\ref{pla}).

The higher-order approximation to the flux of heat flux $\hat{Q}$ is derived from the balance equation
\begin{equation}
\frac{\partial Q_{ij}}{\partial t} + \frac{\partial M_{ijk}}{\partial x_k } = \frac{1}{\tau_R} \left( \frac{1}{3} v_g^2 e \delta_{ij} - Q_{ij} \right),
\end{equation}
\noindent the third-order tensor $\hat{M}$ is defined as
\begin{equation}
M_{ijk} = \int v_{gi} v_{gj} v_{gk} \hbar \omega f d \vec{k}.
\end{equation}
Using the  expansion in the Knudsen number as a small parameter 
$\epsilon = Kn$
\begin{equation}
Q_{ij} = Q_{ij}^{(0)} + \epsilon Q_{ij}^{(1)} + \dots
\end{equation}
\noindent and retaining the zeroth- and first-order terms 
\begin{equation}
Q_{ij}^{(0)} = \frac{1}{3} v_g^2 e \delta_{ij}, \quad
Q_{ij}^{(1)} = - \tau_R \left[\frac{\partial}{\partial t} (Q_{ij}^{(0)}|_{f_4}) +  \frac{\partial}{\partial x_k} (M_{ijk}^{(0)}|_{f_4})   \right]
\end{equation}
\noindent the flux of the heat flux is written as
\begin{equation}
Q_{ij} = \frac{1}{3} v_g^2 \epsilon R{ij} + \frac{2}{15} \tau_R v_g^2 \frac{\partial q_k}{\partial x_k} \delta_{ij} - \frac{1}{5} \tau_R v_g^2 \left( \frac{\partial q_i}{\partial x_j} + \frac{\partial q_j}{\partial x_i} \right).
\end{equation}

Higher-order are
thus derived in the form of the gradient of heat flux, which is
crucial for modelling nanoscale heat transport.

Finally the balance equation is written as 
\begin{equation}
\label{ph}
\tau_R \frac{\partial \vec{q}}{\partial t} + \vec{q} = - \lambda \nabla T + \frac{1}{5} \Lambda^2 \left[ \nabla^2 \vec{q} + \frac{1}{3} \nabla (\nabla \cdot \vec{q}) \right]
\end{equation}
\noindent where $\Lambda = v_g \tau_R$ is MFP.  
Since equation (\ref{ph}) uses the same field variables as the
traditional Fourier’s description, this model avoids the complexity of classical moment methods that 
involve the governing equation of higher-order moments.
In the diffusive limit where both
relaxation and nonlocal effects are negligible, equation (\ref{ph}) reduces to the Fourier’s law.

The equation (\ref{ph}) differs only by the numerical coefficient at the nonlocal term from the Guyer-Krumhansl equation (\ref{krum}). The authors stressed that the structures of these equations  are the same, but the underlying mechanisms of heat transport are different. The nonlocal terms in the Guyer-Krumhansl equation are originate from the  normal scattering and this equation is suitable for the study of the heat transfer at low temperatures. The  nonlocal terms in the phonon hydrodynamic by represent the spatial nonequilibrium effects from the phonon-boundary scattering or from the large spatial thermal variation.

The nonequilibrium phonon distribution function corresponding to the hydrodynamics equation (\ref{ph}) is written as \cite{guo18}
\begin{equation} 
f = f_R^{eq} + \frac{3}{C_V v_g^2} \frac{\partial f_R^{eq}}{\partial T} q_i v_{g i} + \frac{\tau_R}{C_V} \frac{\partial q_i}{\partial x_i} \frac{\partial f_R^{eq}}{\partial T} - \frac{3 \tau_R}{C_V v_g^2} v_{gi} v_{gj} \frac{\partial q_i}{\partial x_j} \frac{{eq}}{\partial T}.
\end{equation}

Guo \& Wang used the phonon hydrodynamics equations to solve a number of examples: {the in-plane phonon transport through a thin film\footnote{This problem was studied in numerous work both experimentally and theoretically --- see, e.g., Hua \& Gao \cite{hua16}. Monte-Carlo method is frequently used to solve the BTE, intrinsic scattering processes such as phonon-phonon and phonon-impurity scatterings being accounted for in the RTA. The lateral boundaries usually are assumed to be diffusive while other boundaries are treated as phonon blackbody, i.e. as absorbing phonons. One of conclusions drawn is that boundary scattering significantly  the in-plane heat transport \cite{fli92,hua16}.};}
{the cross-plane phonon transport through a thin film;}
{the phonon transport through a NW;}
{the 1D transient transport across a thin film;}
{the periodic heating of a semi-infinite surface;}
{the heat conduction in a transient thermal grating.}

\subsection{Relaxon Model}
Recently (2020) Simoncelli et al. \cite{sim20} used the evolution of relaxons to derive two coupled equations for the temperature and for the drift velocity that describe the heat transfer in dielectrics.
The concept of relaxons was introduced  by Cepellotti \& Marzari \cite{cep16} as the collective excitation of the lattice vibrations consisting of a linear combination of the phonon populations.

Each relaxon is characterized by a well-defined relaxation time, drift velocity and MFP. The thermal conductivity is interpreted as the relaxon gas motion.

Simoncelli et al. restricted their analysis to the "simple" {\ } crystals, i.e. such crystals where the phonon interbranch spacings are much larger than their linewidths.
The authors start with the linearised phonon BTE
\begin{equation}
\label{lbte}
\frac{\partial n_{\mu}}{\partial t} + \vec{v_{\mu}} \cdot \nabla n_{\mu} = - \frac{1}{V} \sum_{\mu{\prime}} \Omega_{\mu \mu^{\prime}}  n_{\mu^{\prime}}
\end{equation}
\noindent where the sum is over all possible phonon states $\mu$ ($\mu = (\vec{q},s)$ where $\vec{q}$ varies over the Brillouin zone and $s$ over phonon branches),
$\vec{v_{\mu}}$ is the phonon group velocity, $V$ is the normalization volume, $\Omega_{\mu \mu^{\prime}}$ is the linear phonon scattering operator (the phonon scattering matrix).

The phonon BTE (\ref{lbte}) governs the evolution of the deviation of the population $n_{\mu}$ from the 
 equilibrium, i.e. the Bose-Einstein distribution
\begin{equation}
\frac{1}{\exp \left(\displaystyle \frac{\hbar \omega_{\mu}}{k_B T} \right) - 1},
\end{equation} 
\noindent where $\omega_{\mu}$ is the phonon frequency,
$n_{\mu} = N_{\mu} - \bar{N}_{\mu}$.

One can also consider the phonon drifting distribution ($\vec{u}$ is the drift velocity)
\begin{equation}
	\label{dr}
	N_{\mu}^D = \frac{1}{\exp \left( {\displaystyle  \frac{\hbar (\omega_{\mu} - \vec{q} \cdot \vec{u})}{k_B T}} \right) - 1}.
\end{equation}

Equation (\ref{dr}) depends on time and space implicitly through $T$ and $\vec{u}$.

Simoncelli et al. in order to study the small perturbation of the temperature and drift velocity split the deviation of the distribution function as 
\begin{equation}
	\label{dec}
	n_{\mu} = 	n_{\mu}^T +	n_{\mu}^D 	 + n_{\mu}^{\delta} = \left( \frac{\partial N_{\mu}^D}{\partial T}\right)_{eq}(T - \bar{T}) + \left( \frac{\partial N_{\mu}^D}{\partial u}\right)_{eq}\vec{u} 	 + n_{\mu}^{\delta}.
\end{equation}

Linearisation of the BTE around the constant temperature and drift velocity gradients at steady state gives 
\begin{equation}
	\label{lbte1}
	\frac{\partial \bar{N_{\mu}}}{\partial T} \vec{v}_{\mu} \cdot \nabla T + \vec{v}_{\mu} \cdot \left(\frac{\partial N_{\mu}^D}{\partial \vec{u}}\cdot \nabla \vec{u}   \right)= - \frac{1}{V} \sum_{\mu^{\prime}} \Omega_{\mu \mu^{\prime}} (n_{\mu^{\prime}}^T  +	n_{\mu^{\prime}}^D 	 + n_{\mu^{\prime}}^{\delta}). 
\end{equation}

Equation (\ref{lbte1})  can be recasted \cite{sim20} into symmetric form in terms of
\begin{equation}
	\bar{\Omega}_{\mu \mu^{\prime}} = \Omega_{\mu \mu^{\prime}} \sqrt{\frac{\bar{N}_{\mu^{\prime}} (\bar{N}_{\mu^{\prime}}  + 1)}{\bar{N}_{\mu} (\bar{N}_{\mu}  + 1)}}, \quad
	\tilde{n}_{\mu} = \frac{n_{\mu}}{\sqrt{\bar{N}_{\mu} (\bar{N}_{\mu}  + 1)}}.
\end{equation}

The  scattering operator $	\bar{\Omega}_{\mu \mu^{\prime}}$ is real symmetric and thus can be transformed into the diagonal form
\begin{equation}
\frac{1}{V} \sum_{\mu{\prime}} \Omega_{\mu \mu^{\prime}} \theta_{\mu^{\prime}}^{\alpha}  = \frac{1}{\tau_{\alpha}} \theta_{\mu}^{\alpha},
\end{equation}
\noindent where $\theta_{\mu}^{\alpha}$ is the relaxon (an eigenvector), $\alpha$ is the relaxon index, $\tau_{\alpha}$ is the relaxon lifetime - the inverse eigenvalue. 
Any response $\Delta \bar{n}_{\mu}$ can be represented as a linear combination of the eigenvectors $\theta_{\mu}^{\alpha}$ that are called relaxons. 
\begin{equation}
	\Delta \bar{n}_{\mu} = \sum_{\alpha} f_{\alpha} \theta_{\mu}^{\alpha}.
\end{equation}
The BTE could be formulated in  relaxon (collective excitation) basis $\theta^{\alpha}$  that consists of a few phonons interacting through the scattering between themselves but uncoupled from the phonons belonging to other relaxons
\begin{equation}
	\sqrt{\frac{C}{k_B T^2}} \left(\frac{\partial T}{\partial t} <0|\alpha> + \nabla T + \vec{V}_{\alpha}  \ \right) + \frac{\partial f_{\alpha}}{\partial t} + \sum_{\alpha^{\prime}} \vec{V}_{\alpha \alpha^{\prime}} \cdot \nabla f_{\alpha} = - \frac{f_{\alpha}}{\tau_{\alpha}} .
\end{equation}

Since the eigenvectors of the scattering matrix $\bar{\Omega}_{\mu \mu^{\prime}} $ have the well-defined parity, $\tilde{n}_{\mu}^{\delta}$ is splitted into even $\tilde{n}_{\mu}^{\delta E}$ and odd $\tilde{n}_{\mu}^{\delta O}$ components; equation (\ref{lbte1}) is decoupled into two parts, one for each parity \cite{sim20}:
\begin{itemize}
	\item {the odd part that describes the responce to the temperature gradient
		\begin{equation}
\frac{\vec{v}_{\mu}}{\sqrt{\bar{N}_{\mu} (\bar{N}_{\mu}  + 1)}} \cdot \left(\frac{\partial \bar{N_{\mu}}}{\partial T} \nabla T \right) = - \frac{1}{V} \sum_{\mu{\prime}} \Omega_{\mu \mu^{\prime}}  \tilde{n}_{\mu^{\prime}}^{\delta O}			
		\end{equation}
}
	\item {and the even part that describes the responce to the drift velocity gradient
	\begin{equation}
	\frac{\vec{v}_{\mu}}{\sqrt{\bar{N}_{\mu} (\bar{N}_{\mu}  + 1)}} \cdot \left(\frac{\partial \bar{N_{\mu}^D}}{\partial \vec{u}} \nabla \vec{u} \right) = - \frac{1}{V} \sum_{\mu{\prime}} \Omega_{\mu \mu^{\prime}}  \tilde{n}_{\mu^{\prime}}^{\delta E}	.
	\end{equation}
}
\end{itemize}

The total crystal-momentum flux tensor is written as 
$
	\Pi_{tot}^{ij} = \frac{1}{V} \sum_{\mu} \hbar q^i v_{\mu}^J N_{\mu}
$ or, accounting for the decomposition (\ref{dec})
\begin{equation}
\Pi_{tot}^{ij} = \frac{1}{V} \sum_{\mu} \hbar q^i v_{\mu}^J (\bar{N}_{\mu} + n_{\mu}^T + n_{\mu}^{\delta E})	
\end{equation}
\noindent (only the even part of the phonon distribution function contributes to the crystal-momentum flux tensor). 

The  asymmetric thermal viscosity tensor is formulated as
\begin{equation}
	\eta^{ijkl} = \sqrt{A^i A^k} \sum_{\alpha > 0} w_{i \alpha}^J w_{k \alpha}^l
\end{equation}
\noindent where
\begin{equation}
	A^i = \frac{1}{k_B \bar{T} V} \sum_{\mu} \bar{N}_{\mu} (\bar{N}_{\mu} + 1) (\hbar q^i)^2, \quad
w_{i \alpha}^j = \frac{1}{V} \sum_{\mu} \phi_{\mu}^i v_{\mu}^j \theta_{\mu}^{\alpha}	
\end{equation}
\noindent for relaxon $\theta_{\mu}^{\alpha}$ and the eigenvector $\phi_{\mu}^i$.

The symmetrized viscosity tensor is expressed as
\begin{equation}
	\mu^{ijkl} = \frac{\eta^{ijkl} + \eta^{ilkj}}{2}.
\end{equation}

The thermal conductivity and viscosity  are quantities describing energy and crystal momentum transport due to the odd and even parts of the spectrum.

The thermal conductivity in the harmonic approximation for the heat flux
\begin{equation}
q = \frac{1}{V} \sum_{\mu} \hbar \omega_{\mu} v_{\mu} \Delta n_{\mu} 
\end{equation}
\noindent is written as
\begin{equation}
(\lambda^{ij})^{SMA} = \frac{1}{V} \sum_{\mu} C_{\mu} v_{\mu}^i (\Lambda_{\mu}^j)^{SMA},
\end{equation}
\noindent where $(\Lambda_{\mu}^j)^{SMA}$ is the component of the phonon 
MFP in direction $j$.

Thus, thermal conductivity is provided by photons carrying a specific heat 
\begin{equation}
C_{\mu} \frac{1}{k_B T^2} \bar{n}_{\mu} (\bar{n}_{\mu} + 1)(\hbar \omega_{\mu})^2,
\end{equation} 
\noindent travelling at velocity $v_{\mu}$ and MFP  $(\Lambda_{\mu}^j)^{SMA}$ until thermalised by scattering. 

Capellotti \& Marzari stress that the definition of the phonon lifetime or MFP cannot be extended beyond the SMA since the off-diagonal terms of the scattering operator introduce the coupling between the phonons and the phonon thermalisation cannot be described by the exponential relaxation.

Finally, Cepelotti \& Marzari derived the relation for thermal conductivity
\begin{equation}
\lambda^{ij} = \frac{- 1}{V \nabla_i T} \sum_{\mu} \hbar \omega_{\mu} v_{\mu}^j \Delta n_{\mu}  = \sum_{\alpha} C V_{\alpha}^i \Lambda_{\alpha}^j .
\end{equation}

The authors also showed that Matthiessen rule \cite{zim} for the relaxation time  leads to the overestimation of the thermal conductivity.

Simoncelli et al. have derived two coupled equations for the temperature and drift-velocity  fields
\begin{equation}
	\label{tu1}
C \frac{\partial T}{\partial t} = \sum_{i,j}^3 W_{0j}^i \sqrt{\bar{T} A^j C} \frac{\partial u^j}{\partial r^i} - \sum_{i,j}^3 \lambda^{ij} \frac{\partial^2 T}{\partial r^i \partial r^j} = 0
\end{equation}
\noindent and
\begin{equation}
\label{tu2}
 A^i \frac{\partial u^i}{\partial t} +  \sqrt{\frac{C A^i}{\bar{T}}} \sum_{j}^3 W_{i0}^j \frac{\partial T}{\partial r^j} - \sum_{j,k,l}^3 \mu^{ijkl} \frac{\partial^2 u^k}{\partial r^j \partial r^l}	= - \sum_{j}^3 - \sqrt{A^i A^j} D_U^{ij} u^j,
\end{equation}
\noindent where 
\begin{equation}
	C = \frac{1}{k_B \bar{T}^2 V} \sum_{\mu} \bar{N}_{\mu} (\bar{N}_{\mu} + 1)(\hbar \omega_{\mu})
\end{equation}
\noindent is the specific heat,
$ W_{0j}^i = \sum_{\mu} \phi_{\mu}^0 v_{\mu}^i \phi_{\mu}^j
$ is the the velocity tensor, $\bar{T}$ is the reference temperature, $D_U^{ij}$ is the momentum dissipation rate.

\section{Thermomass Model}
\label{tm}
The thermomass model is based on the ideas of Tolman \cite{tol30} that the heat carriers own the mass-energy duality: the carriers exhibit both energy-like characteristics (in the conversion processes) and mass-like characteristics (in the transport processes)  
\cite{wan11}.  
The thermomass theory applied to the heat conduction in the rigid solid bodies at rest  formally could be classified as an example of the phonon model and its equations are similar to that of the phonon hydrodynamics, but underlying physical ideas are quite different.
The mass of heat is determined by the mass-energy equivalence of Einstein \cite{guo06,don13,wan14}
\begin{equation}
E = M c^2 = \frac{M_0 c^2}{\sqrt{1  -\frac{v^2}{c^2}}}
\end{equation}
\noindent where $E$ is the thermal energy, $M_0$ is the rest mass, $v$ is the velocity of the heat carrier, $c$ is the speed of light in vacuum, $M$  is the relativistic mass.

When $v \ll c$, this equation is simplified to $E \approx (M_0 + M_k) c^2$,  here $M_k$ is the additive mass induced by the kinetic energy. The thermomass (TM) $M_h$ is the relativistic mass of the internal energy $U$: $M_h = {U}/{c^2}$.

The thermomass is very small ($10^{-16}$ kg for 1 J heat) \cite{wan10a,wan11}.
The density of the thermomass contained in the medium is \cite{don11}
$\rho_h = {\rho C_V T}/{c^2}$ 
 where $\rho C_V T$ represents the thermal energy density. 

The thermon is defined as a unit quasi-particle carrying thermal energy. For fluids, thermons are supposed to be attached to the particles of the medium; for solids, the thermon gas the phonon gas that flows through vibrating lattices or molecules \cite{wan11}.
The macroscopic drift velocity of the thermon gas that is considered as a continuum is given as $v_h = {q}/{\rho C T}$.

The total energy of thermon gas in a medium is the sum of the kinetic energy and potential (pressure) energy \cite{wan10} ($V$ is the total volume of the medium)
\begin{equation}
E_T = \int\int\limits_V (\rho_Tu_T d u_T + d p_T) d V.
\end{equation}

\subsection{Equation of State (EOS) of the Thermon Gas}
A general form of the EOS of the thermomass gas can be the written   as $F(p_t, \rho_T, T, \xi) = 0$, where $p_T$  is the thermomass pressure, $\rho_T$ is the density of the thermon gas, $\xi$ the parameter related to the  interaction between thermons; if the latter could be neglected, the EOS has an explicit form \cite{wan11}.

\paragraph{EOS of thermon gas in ideal gas}

Two assumptions are made for the thermon gas as an ideal gas:
(1) {the termons are attached to the gas molecules and satisfy the Maxwell-Boltzmann distribution function;} (2)  {Newtonian mechanics is applicable to the thermon gas.}
The pressure $P$ for a system of $n$ particles with mass $m$ randomly moving in $x$ direction with velocity $u_x$ 
$P = n m u_x^2$ 
 and, accounting for  symmetry in $x, y, z\quad $ directions  ($u_x^2 = u_y^2 = u_z^2 = \frac{1}{3} \bar{u}^2$) \cite{wan14},
\begin{equation}
P = \frac{1}{3} n m_h \bar{u}^2 = \frac{1}{3} n \bar{u}^2 \left(\frac{1}{2} \frac{m \bar{u}^2}{c^2} \right) = \frac{1}{6} \frac{n m \bar{u}^4}{c^2}.
\end{equation}

Using the classical Maxwell-Boltzman distribution function
\begin{equation}
f_M (u) = 4 \pi u^2 \left(\frac{m}{2 \pi k_B T} \right)^{3/2} \exp \left(- \frac{m u^2}{2 k_B T} \right)
\end{equation}
\noindent one gets
\begin{equation}
\bar{u}^4 = \int\limits_0^{\infty} u^4 f_M (u) d u = 15 \left(\frac{k_B T}{m}  \right)^2
\end{equation}
\noindent and finally
$P = {5}/{3} {\rho C_V R T^2}/{c^2}$  where $R$ is the gas constant.
Thus the thermon gas pressure is proportional to the square of temperature.

\paragraph{EOS of thermon gas in dielectrics}

Phonons are thermons for dielectrics. The total energy of the lattice vibrations can be written as $E_D = E_{D0} + E_h = (M_0 + M_h) C_V T$, where $E_h$ is the energy of TM. 
Thus the pressure is $P = {\gamma_G \rho}/{c^2} (C_v T)^2$ 
\noindent where $\gamma_G$ is the Gr\"uneisen parameter ("the Gr\"uneisen anharmonicity constant" \cite{liu05}) that 
 describes overall effect of volume change of a crystal on vibrational properties, which is expressed as $\gamma_G =  V/ C_V (dP/dT )_V$, where $V$ is the volume of a crystal, $C_V$ is
the heat capacity, and $(dP/dT )_V$ is the 
pressure change due to temperature variation at constant volume \cite{gu17}.
The pressure of phonons is proportional to the square of temperature as in ideal gas. For Si at room temperature the 
thermon gas pressure is about $5 \cdot 10^{-3} Pa$ \cite{wan14}.
\paragraph{EOS of thermon gas in metals}

The thermons in metals are attached to electrons. The gas pressure is given as $P = \frac{1}{3} n m_h u_h^2$ 
where $m_h = {\epsilon}/{c^2}$, $\epsilon$ is the energy that includes contributions of both electrons and lattice, $u_h = \sqrt{{2 \epsilon}/{m}}$ is the velocity of randomly moving particles, thus the pressure $P = {2}/{3} {n \epsilon^2}/{c^2 m}$.

The general equation for the thermon pressure is written as 
\begin{equation}
P = \frac{2}{3 m c^2} \int\limits_0^{\infty} \epsilon^2 f(\epsilon, T) Z (\epsilon) d \epsilon,
\end{equation}
\noindent where
\begin{equation}
f (\epsilon, T) = \frac{1}{\exp \left( \displaystyle  \frac{\epsilon - \epsilon_F}{k_B T}  \right) + 1}
\end{equation}
\noindent is the Fermi-Dirac distribution function and 
\begin{equation}
Z (\epsilon) = \frac{1}{2 \pi^2} \left(\frac{2 m}{\hbar^2}  \right)^{\frac{3}{2}} \epsilon^{\frac{1}{2}}
\end{equation}
\noindent is the Sommerfeld electron state density function.

Wang \cite{wan14} finally obtained the expression for the thermon gas pressure
\begin{equation}
P = \frac{5}{12} \frac{\pi^2 n k_B^2}{c^2 m} T^2.
\end{equation}

Wang et al. \cite{wan11} noted that electrical interaction between electrons in metals may be not be negligible and can influence EOS below the Debye temperature.
\subsection{Equations of Motion of Thermon Gas}

The one-dimensional conservation equations of mass and momentum are
\begin{equation}
\frac{\partial \rho_h}{\partial t} + \frac{\partial \rho_h u_h}{\partial x } = \frac{S}{c^2}
\end{equation}
\noindent where $S$ is the internal heat source and
\begin{equation}
\frac{\partial}{\partial t} (\rho_h u_h) + \frac{\partial}{\partial x} (u_h \cdot \rho_h u_h) + \frac{\partial P}{\partial x} + f_h = 0 .
\end{equation}
\noindent where $f_h$ is the resistance.

The continuity equation for the thermon gas is actually the energy conservation equation.
The thermon flow in the solid can  seen as the flow of the compressible fluid in the porous medium and the D'Arcy law ($K$ is the permeability of the porous medium)
\begin{equation}
u = - K \frac{d p}{d x}
\end{equation}can be used to estimate the TM resistance $f_h = \beta_h u_h$, the proportionality constant \cite{guo07,wu09}
\begin{equation}
\beta_h = \frac{2 \gamma \rho^2 C^3 T^2}{c^2 \lambda}.
\end{equation}

The effective resistance force $f_h$ is introduced instead of the viscous term ($\mu_h \nabla^2 \vec{u}_h$) to avoid \cite{wan11}
{determination of the viscosity $\mu_h$ for complex materials} and
{the interaction effects between the thermomass gas and the lattice/solid molecules.}
The thermon gas flow in the solids (phonon flow) is driven by the pressure gradient, thus by the gradient of the square of the temperature \cite{guo07}.

The thermomass is too small to be observed under common conditions but under the extreme conditions of the ultra-fast heating or the ultra-high heat flux the inertia of TM will  cause detectable influences to the heat conduction. 

The conservation equation of the thermon momentum can be written as a heat conduction equation \cite{wan14}
\begin{equation}
\tau_h \left(\frac{\partial q}{\partial t} + 2 u_h \frac{\partial q}{\partial x} - u_h^2 \rho C_V \frac{\partial T}{\partial x}   \right) + \lambda \frac{\partial T}{\partial x} + q  = 0.
\end{equation}

The general heat conduction in the 3D could be written as \cite{don13} 
\begin{equation}
	\label{d11}
	\tau_h \frac{\partial \vec{q}}{\partial t} + 2 (\vec{l} \cdot \nabla) \vec{q} - b \lambda \nabla T + \lambda \nabla T + \vec{q} = 0,
\end{equation}
\noindent where 
\begin{equation}
	b = \frac{q^2}{2 \gamma_G \rho_h^2 C_V^3 T^3}.
\end{equation}

The coefficient $b$ could be expressed through the thermal Mach number $Ma_h$ as the ratio of the phonon gas velocity $u_h$ to the thermal wave speed in the phonon gas $C_h$: $b = Ma_h^2$.

The equation (\ref{d11}) could be re-written in the form similar to the Cattaneo model  by introduction of the material derivative 
\begin{equation}
	\frac{D}{D t} = \frac{\partial}{\partial t} + 2 (\vec{v}_h \nabla)
\end{equation}
\noindent as \cite{don13}
\begin{equation}
	\tau_h \frac{D \vec{q}}{D t} + \vec{q} = - \lambda (1 - b) \nabla T.
\end{equation}

Wang et al. \cite{wan10a} noted that relaxation times in the thermomass theory and in the Cattaneo model have different meaning: in the thermomass theory the characteristic time means the lagging time from the temperature gradient to the heat flux, in the Cattaneo model it is relaxation time to the equilibrium state.

Wang \cite{wan14} developed a special two-step version of the thermomass theory for metals that are subjected to the ultra-fast laser heating under the following assumptions:
{the electrons absorb the laser energy and then transfer it to the lattice;}
{the scattering at the defects and the grain boundaries is ignored;}
{the electron-phonon collisions are presented by an electron-phonon coupling factor.}

Similar to the porous flow hydrodynamics a Brinkman term $\mu \nabla^2 \vec{q}$ ($\mu$ is the viscosity that is defined as $\mu = 2 \tau / C_v T$ \cite{cim10}) could be introduced into the equation of motion (\ref{d11}) that exhibits the additional drag by the walls  and is necessary only if the characteristic length is comparable to the friction boundary layer of the thermomass, i.e. if the Knudsen number is large enough \cite{don11} 
\begin{equation}
	\label{d11a}
	\tau_h \frac{\partial \vec{q}}{\partial t} + 2 (\vec{l} \cdot \nabla) \vec{q} - b \lambda \nabla T + \lambda \nabla T + \vec{q} - \mu \nabla^2 \vec{q} = 0.
\end{equation}

The equation similar to the equation (\ref{d11a}) was suggested by Cimmelli et al. \cite{cim10} as the nonlinear extension of the  Guyer-Krumhansl equation (\ref{krum0}).

Sometimes the Brinkman number $Br = l_B / L$ is used \cite{don14} that compare the viscous friction with the D'Arcy friction; the length scale $l_B = \sqrt{\mu_h / \beta \rho_h}$:
\begin{itemize}
\item {if $Br \ll 1$, the boundary effect region is small compared to the channel width and the velocity profile is nearly uniform across the channel;} 
\item {if $Br \gg 1$, the velocity profile is close to the Poiseuille flow.}
\end{itemize}

The entropy production  is due to the dissipation rate of the mechanical energy of the thermon gas similar to the viscous dissipation in hydrodynamics
\begin{equation}
\frac{d E_h}{d t} + \nabla \cdot \vec{J}_{E_h} = \vec{f}_h \cdot \vec{u_h},
\end{equation}
\noindent where $E_h$ is the mechanical energy of the thermon gas, $\vec{J}_{E_h}$ is the flux of $E_h$.

Thus the entropy production in  thermomass theory can be written as \cite{don12}
\begin{equation}
	\sigma_{TM}^s = - \frac{1}{T} \vec{F}_h \cdot \vec{u}_h = \frac{1}{\lambda T^2} \vec{q} \cdot \vec{q}.
\end{equation}

Dong et al. \cite{don12} formulated the total derivative of the entropy density as 
\begin{equation}
\frac{d s}{d t} = - \nabla \cdot \vec{J}_s + \sigma_{TM} = \frac{\vec{q}}{\lambda T^2} \cdot (\vec{q} + \lambda \nabla T) -  \frac{\nabla \cdot \vec{q}}{T}.
\end{equation}
\noindent where $\vec{J}_s$ is the entropy flux.

The nonequilibrium temperature is introduced by Dong et al. \cite{don13} as
\begin{equation}
	\theta^{-1} = \left(\frac{\partial s}{\partial e}   \right)_{V, \vec{q}} =  \frac{1}{T_{eq}} - \frac{1}{2} \frac{\partial (\tau /(\lambda T_{eq}^2))}{\partial e} \vec{q} \cdot \vec{q},
\end{equation}
\noindent where the subscript $V, \vec{q}$ means the derivative at constant volume and heat flux.
The temperature $\theta$ is lower than the local-equilibrium temperature $T_{eq}$.

Under the assumption that $\tau$ does not depend on $T_{eq}$
\begin{equation}
	\theta = \frac{1}{\displaystyle  \frac{1}{T_{eq}} + \displaystyle  \frac{\tau}{\lambda C_V T_{eq}^3 \vec{q} \cdot \vec{q}}} \approx  T_{eq} [1 - \frac{\tau}{\lambda C_V T_{eq}^2 \vec{q} \cdot \vec{q}} + o (q^2)].
\end{equation}

Dong et al. \cite{don21} exploited  an analogy analysis between non-Fourier heat conduction and non-Newtonian momentum transport. Similar to the assumptions in the thermomass model, the authors derived an equation for momentum transport, which accounts for  the varying effective viscosity in steady flow. This shear thinning effect will be apparent in nano-channel flow where the velocity gradient and the momentum transport flux are huge.

\subsection{Heat flow Choking Phenomenon}
Wang et al. \cite{cho,cho1} for 1D steady heat transfer without internal heat source use the heat conduction
equation in the form
\begin{equation}
\kappa \left(1 - Ma_h^2\right) \frac{dT}{dt} + q = 0 	
\end{equation}
where the thermal Mach number  $Ma_h = {u_h}/{C_h}$, $u$ is the drift velocity of the phonon gas, the thermal sound speed is, e.g., for dielectrics, $C_h = \sqrt{2 \gamma C_V T}.$

Thermon gas is a compressible fluid thus its flow demonstrates  features similar to those of the flow  of compressible gas including the  
 behaviour in the convergent nozzle when the Mach number equals to unity.

In the flow of compressible air driven by the pressure gradient  in the converging nozzle the velocity increases and the pressure decreases in the flow direction. The flow choking  occurs when the current  Mach number  equals unity and there is the pressure jump. 
The drift velocity of thermon gas driven by the temperature gradient increases as it flows in the opposite direction of temperature gradient. The heat flow choking that occurs at the thermal Mach number equalled to unity results in the temperature jump.

The confirmation of the heat flow choking phenomenon was obtained in the experiments on the heat conduction in a single-walled carbon nanotube (CNT) suspended between two metal electrodes by Wang et al. \cite{cho}.  The CNT was electrically heated by the internal Joule heat, the heat flowed from the middle to the two ends of the CNT. The heat flux was governed by the electrode temperature until the thermal Mach number reaches unity at the CNT ends; further 
decrease of the electrode temperature had no effect on the heat flux.

\subsection{Dispersion of Thermal Waves}
Zhang et al. \cite{zha13a} investigated numerically using an implicit finite difference scheme (with central differences for space discretization and backward differences for time) the dispersion of the thermal waves. 

The authors considered the case of the cosine heat flux pulse boundary condition. As the wave moves forward, peaks appear in the rear of 
the thermal wave. The underlying mechanism for the dispersion is that thermal waves travel faster in the the regions with higher temperature.

Computations were performed for the CV, DPL and TM models. Zhang et al. \cite{zha13a} started with the TM model written in the form
\begin{equation}
\label{z4}
\vec{q} + \tau_{TM} \frac{\partial \vec{q}}{\partial t}	- \tau_{TM} \frac{\vec{q}}{T} \frac{\partial T}{\partial t} + \tau_{TM} \frac{\vec{q}}{\rho C_V T} \nabla \cdot \vec{q} -  \tau_{TM} \frac{\vec{q}}{\rho C_V T} \cdot \frac{\vec{q}}{T} \nabla T = -\lambda \nabla T 
\end{equation}
and in order to analyse the origin of the thermal wave dispersion considered special versions of the equation (\ref{z4})
\begin{equation}
	\label{z20}
	\vec{q} + \tau_{TM} \frac{\partial \vec{q}}{\partial t}	 = -\lambda \nabla T, 
\end{equation}	
\begin{equation}
	\label{z21}
	\vec{q} + \tau_{TM0} \frac{\partial \vec{q}}{\partial t}	- \tau_{TM} \frac{\vec{q}}{T} \frac{\partial T}{\partial t}  = -\lambda \nabla T, 
\end{equation}
\begin{equation}
	\label{z22}
	\vec{q} + \tau_{TM0} \frac{\partial \vec{q}}{\partial t} + \tau_{TM} \frac{\vec{q}}{\rho C_V T} \nabla \cdot \vec{q} = -\lambda \nabla T, 
\end{equation}
\begin{equation}
	\label{z23}
	\vec{q} + \tau_{TM0} \frac{\partial \vec{q}}{\partial t} -  \tau_{TM} \frac{\vec{q}}{\rho C_V T} \cdot \frac{\vec{q}}{T} \nabla T = -\lambda \nabla T.
\end{equation}

The equation (\ref{z20}) is used to study the effects of the inertia term of heat flux to time on the dispersion of the TM-wave; this term is different from that of the CV model since the characteristic time $\tau_{TM}$ decreases with the temperature while the relaxation time  remains unchanged in the propagation of the CV-wave. 
The effects of the inertia term of temperature to time, the inertia term of heat flux to space, the inertia term of temperature to space could be investigated using the equations (\ref{z21}), (\ref{z22}), (\ref{z23}), respectively, \cite{zha13a}.

It should be noted that according to the energy conservation equation
\begin{equation}
	\rho C_v \frac{\partial T}{\partial t}   +  \nabla \cdot \vec{q} = 0
\end{equation} 
with the constant thermal properties and no heat sources the equation (\ref{z21}) could be transformed into the equation (\ref{z22})by replacing $\partial T/\partial t$ by $- \nabla \cdot \vec{q} / \rho C_v$.

 Zhang et al.  also showed that the increase of the amplitude of the heat flux pulse and the decrease of the initial temperature enhance the dispersion of TM-wave; the increase of the amplitude of the heat flux pulse and of the relaxation time $\tau_q$ enhance the dispersion of CV-wave and DPL-wave  while the increase of the relaxation time $\tau_T$ weaken the dispersion of the DPL-wave.

\section{Mesoscopic Moment Equations}
 Bergamasso et al.\cite{ber18} developed in the frame of kinetic theory a number of the mesoscopic moment systems (the two-moment and the three-moment systems) using the concept of {\it ghost} moment.
The authors considered the following  flow regimes: 
hydrodynamic ($Kn \le 0.01$), 
slip-flow regime ($0.01 < Kn \le 0.1$), 
transition regime ($0.1 < Kn \le 10$),
free molecular flow ($Kn > 10$). The ghost moments are those that have a higher order with respect to those required to recover the hydrodynamic level (hydrodynamic moments). The authors start from the Fourier equation re-written as 
\begin{equation}
Kn^2 \frac{\partial T}{\partial t} + Kn \frac{\partial}{\partial x} \left(- Kn \kappa \frac{\partial T}{\partial x} \right) = 0
\end{equation}
\noindent and add a ghost moment $\phi = Kn q$, which has units of thermal flux, to increase the order of the physical description 
\begin{equation}\label{17a}
Kn^2 \frac{\partial T}{\partial t} + \frac{Kn}{\rho c_p} \frac{\partial \phi}{\partial x} = 0,
\end{equation}
\begin{equation}\label{17b}
\frac{Kn^2 \kappa}{\rho c_p c^2} \frac{\partial \phi}{\partial x} + Kn \kappa \frac{\partial T}{\partial t} = - \frac{1}{\rho c_p} \phi,
\end{equation}
\noindent where an additional term  involving the time derivative of the ghost moment responsible for the enriched mesoscopic description is introduced into equation (\ref{17b}), $c$ is the constant  velocity of arbitrary value. Equations (\ref{17a}) and (\ref{17b}) are called the Two-Moment Hyperbolic Equation system from which the mesoscopic equation for the temperature follows
\begin{equation}\label{20}
\frac{\kappa Kn^2}{c^2} \frac{\partial^2 T}{\partial t^2} + \frac{\partial T}{\partial t} = \kappa \frac{\partial^2 T}{\partial t^2}.
\end{equation}
Equation (\ref{20}) has  the same form as the Cattaneo equation
where the heat-flux relaxation time $\tau = \kappa Kn^2 / c^2$.

Bergamasso et al. also suggested another two-moment system where the velocity appears in both equations:
\begin{equation}\label{22a}
\frac{Kn^2}{c^2} \frac{\partial \phi}{\partial t} + Kn \rho c_p \frac{\partial T}{\partial x} = 0,
\end{equation}
\begin{equation}\label{22b}
Kn^2 \frac{\partial T}{\partial t} + \frac{Kn}{\rho c_p} \frac{\partial \phi}{\partial x} = - \frac{c^2}{\kappa} T.
\end{equation}

The system of equations (\ref{22a}) - (\ref{22b}), called by the authors the Switched Two-Moment Hyperbolic system, lead, as the system (\ref{17a}) -  (\ref{17b}), to the same equation for the temperature (\ref{20}).

The authors introduce another ghost moment $e$ that has units of temperature and is defined such that $e - T = 0$ and two additional parameters $\gamma$ and $\theta$  to get Three-Moment Hyperbolic Equations system:
\begin{equation}\label{26a}
Kn^2 \frac{\partial T}{\partial t} + \frac{Kn}{\rho c_p} \frac{\partial \phi}{\partial x} = 0,
\end{equation}
\begin{equation}\label{26b}
\frac{Kn^2 \kappa}{\rho c_p c^2} \frac{\partial \phi}{\partial t} + \kappa Kn \frac{\partial e}{\partial x} = - \frac{1}{\rho c_p} \phi,
\end{equation}
\begin{equation}\label{26c}
Kn^2 \frac{\partial e}{\partial t} + \frac{Kn}{\rho c_p} \frac{\partial \phi}{\partial x} = \frac{1}{\gamma} \left( \frac{\theta}{T} - e \right).
\end{equation}

Three-moment system can be reduced to the equation for the temperature
\begin{equation}\label{29}
Kn^2 \left(\gamma + \frac{\kappa}{c^2} \right) \frac{\partial^2 T}{\partial t^2} + \frac{\partial T}{\partial t} = \frac{\kappa}{\theta} \frac{\partial^2 T}{\partial x} + \gamma \kappa Kn^2 \left( - \frac{\partial^2}{\partial x^2} \left(\frac{\partial T}{\partial t} \right) + \frac{Kn^2}{c^2} \frac{\partial^3 T}{\partial t^3} \right). 
\end{equation}
The equation ({\ref{29}) reduces to the equation (\ref{20}) for $\gamma = 0$ and $\theta = 1$.

To analyse the solution, Bergamasso et al. perform the Fourier transform
\begin{equation}
\hat{T (k,t)} = \int_{- \infty}^{\infty} T (x,t) e^{- i k x} d x
\end{equation}
with the inverse Fourier transform
\begin{equation}
T (x,t) = \frac{1}{2 \pi} \int_{- \infty}^{\infty} T (x,t) e^{i k x} d k.
\end{equation}
to get the Fourier equation in terms of the Fourier image
\begin{equation}
\frac{d \hat{T}}{d t} = - \kappa k^2 \hat{T} 
\end{equation}
that has the general solution
$
\hat{T}(k, t) = \hat{T} (k, 0) e^{- \kappa k^2 t}.
$ 

The authors introduce the complex temperature $\Theta$
\begin{equation}
\Theta (k,x,t) = \hat{T} (k,0) e^{i k x - \kappa k^2 t} = \Theta_0  e^{i (k x + \omega t)}, \quad
\omega = i \kappa k^2.
\end{equation}

and assumed that the ghost moment $\Phi$ has the same form
\begin{equation}
\Phi (k,x,t) = \Phi_0 e^{i (k x + \omega t)}
\end{equation}

Substituting of these solutions into the two-moment system (\ref{17a}) -  (\ref{17b}) yields an eigenvalue problem with the characteristic polynomial
\begin{equation}
\frac{\kappa Kn^2}{c^2} \omega^2 - i \omega - \kappa k^2 = 0
\end{equation} 
that has roots
\begin{equation}
\label{cp}
\omega_{1,2} = \frac{i c^2 \pm \sqrt{- c^4  + 4 \kappa^2 Kn^2 k^2 c^2}}{2 \kappa Kn^2}.
\end{equation}

The solution can be written as 
\begin{equation}
\Theta = \Theta_{01} e^{i(k x + \omega_1 t)} + \Theta_{02} e^{i(k x + \omega_2 t)}, \quad
\Phi = \Phi_{01} e^{i(k x + \omega_1 t)} + \Phi_{02} e^{i(k x + \omega_2 t)}.
\end{equation}

The same procedure applied to the three-moment system (\ref{26a}) - (\ref{26c}) result in more complex characteristic polynomial
\begin{equation}
(\gamma  Kn^2 \omega - i)  \left(\frac{\kappa Kn^2}{c^2} \omega^2 - i \omega - \frac{\kappa Kn^2}{\theta} \right) = \kappa \gamma k^2 \omega \left(1 - \frac{1}{\theta} \right).
\end{equation}
Solution of the last equation requires a rather cumbersome algebra.

Authors analyse the two-moment systems. The equation (\ref{cp}) is re-written  
\begin{equation}
\omega_{1,2} = i \frac{c^2 \pm c \sqrt{c^2  - 4 \kappa^2 Kn^2 k^2}}{2 \kappa Kn^2}.
\end{equation}

Two cases are 
\begin{itemize}
\item {${\kappa Kn k}/{c} < \frac{1}{2}$

The argument of the square root is positive. The authors used Taylor expansions for $\omega_1$ and $\omega_2$ and concluded that $\Theta_1$ and $\Theta_2$ depend on time with multiple scales
\begin{equation}
\Theta_1 = \Theta_1 \left(\frac{t}{Kn^2}, t, Kn^2 t, \dots  \right), \quad
\Theta_2 = \Theta_2 (t, Kn^2 t, \dots ).
\end{equation}

Thus the authors concluded that solution can have two modes:
(1) {a fast (advective) mode that goes to zero very quickly when $Kn$ is small;} (2) {a slow (diffusive) mode that does not depend on $Kn$ and recovers the diffusive behaviour of the macroscopic equation.}

}
\item {$ {\kappa Kn k}/{c} > \frac{1}{2}$

The argument of the square root is negative; thus, the square root yields a complex number and oscillations in the solution are expected.

}	
\end{itemize}

\section{Thermodynamic Models}
Thermodynamics models are deduced from the thermodynamic constraints following from the second law of thermodynamics \cite{van01,cim09,van12,van15,sel12a,leb14,jou16,sel16,scj16,rog18}.

\subsection{Jou \& Cimmelli Model}
Jou \& Cimmelli \cite{jou16} introduced an extra internal  variable represented by a second order tensor $\hat{Q}$ and wrote the balance of the heat flux as follows

\begin{equation}
\label{jc1}
\tau_1 \dot{\vec{q}} + \vec{q} = - \lambda \nabla T + \nabla \cdot \hat{Q},
\end{equation}
\noindent where $\tau_1$ is the relaxation time. 

The tensor $\hat{Q}$ is assumed to be symmetric and may be split $\hat{Q} = Q \hat{I} + \hat{Q}_s$, where the scalar $Q$ is one-third of trace of $\hat{Q}$, $\hat{Q}_s$ is the deviatoric part of $\hat{Q}$. In the RTA the evolution equations for the tensor parts are written as 
\begin{equation}
	\tau_0 \dot{Q} + Q = \gamma_0 \nabla \cdot \vec{q},\qquad
	\tau_2 \dot{Q}_s + \vec{Q}_s = 2 \gamma_2 \left[ (\nabla \vec{q})^0 \right]_s
\end{equation}
\noindent where $\left[ (\nabla \vec{q})^0 \right]_s$ is the symmetric traceless part of $\nabla \vec{q}$. 

Under the assumption that $\tau_0$ and $\tau_2$ are small the equation for the heat flux (\ref{jc1}) could be re-written as \cite{jou16} 
\begin{equation}
\tau_1 \dot{\vec{q}} + \vec{q} = - \lambda \nabla T + \gamma_2 \nabla^2 \vec{q}    = \left(\gamma_0 + \frac{1}{3} \gamma_2 \right) \nabla \nabla \cdot \vec{q}.	
\end{equation}

The constitutive equations  \cite{jou16} for the specific entropy si introduced as
\begin{equation}
s = s_{eq} - \frac{\tau_1}{2 \lambda T^2} \vec{q} \cdot \vec{q} -  \frac{\tau_2}{4 \lambda T^2 \gamma_2} \vec{Q} : \vec{Q} -  \frac{\tau_0}{2 \lambda T^2 \gamma_0} Q^2	
\end{equation}
\noindent and for the entropy flux
\begin{equation}
	\vec{J}_s = \frac{\vec{q}}{T} \left(1 + \frac{Q}{\lambda T} \right) + \frac{\vec{Q}_s \cdot \vec{q}}{\lambda T^2} 
\end{equation}
\noindent where $s_{eq}$ is the local equilibrium entropy, $":"$ means the complete contraction of tensors producing a scalar, $\vec{Q}_s \cdot \vec{q}$ means the contraction over the last index of $\vec{Q}_s$ producing a vector.

The first-order flux as sole independent variable is not sufficient for the correct description of the high-frequency processes when the frequency becomes comparable to the inverse of the relaxation time. 

EIT allows to introduce higher order fluxes $\vec{J}_2, \dots, \vec{J}_n$ where $\vec{J}_k$ is the tensor of order $k$ that serves as the flux of the preceding flux $\vec{J}_{k - 1}$. Thus the specific entropy is written as \cite{jou16}
\begin{equation}
s = s (e, \vec{J}_1, \dots \vec{J}_n) = s_{eq} + \frac{\alpha_1}{2} \vec{J}_1:\vec{J}_1 + \dots +	\frac{\alpha_n}{2} \vec{J}_n:\vec{J}_n
\end{equation}
\noindent and entropy flux as
\begin{equation}
\vec{J}_s = \frac{\vec{J}_1}{T} + \beta_1 \vec{J}_2 \cdot \vec{J}_1 + \dots + \beta_{n - 1} \vec{J}_n \cdot \vec{J}_{n -1}
\end{equation}
\noindent where $\alpha_1, \dots \alpha_n$ and $\beta_1, \dots \beta_n$ are functions that can depend on the internal energy, $\vec{J}_k \cdot \vec{J}_{k -1}$ means the contraction over the last $(k - 1)$ indices of $\vec{J}_k$ producing a vector.

The evolution of fluxes is governed by the following equations that are compatible with a positive entropy production
\begin{align*}
	\tau_1 \dot{\vec{J}}_1 + \vec{J}_1 &= \lambda \nabla T + \frac{\beta_1 \tau_1}{\alpha_1} \nabla \cdot \vec{J}_2\\
	\dots\\
	\tau_n \dot{\vec{J}}_n + \vec{J}_n &= \frac{\beta_n \tau_n}{\alpha_n} \nabla \cdot \vec{J}_{n + 1}  +  \frac{\beta_{n - 1} \tau_n}{\alpha_n} \nabla \cdot \vec{J}_{n - 1}.\\
\end{align*}

The authors finally derived a general constitutive equation (that includes the Cattaneo and Guyer-Krumhansl equations as a special cases) as 
\begin{equation}
\tau_R \dot{\vec{q}} + \vec{q} + (\mu \nabla \nabla \vec{q} + \mu^{\prime} \nabla \vec{q}) = - \lambda (1 + \xi \vec{q} \cdot \vec{q}) \nabla T +  l_p^2 (\nabla^2 \vec{q} + 2 \nabla \nabla \cdot \vec{q}),
\end{equation}
\noindent where $\mu, \mu^{\prime}, \xi$ are material coefficients.

\subsection{Kov\'acs \& V\'an Model}
Kov\'acs \& V\'an \cite{van15}  introduced the heat flux and a second order tensorial   variable as additional internal field variables  and assumed the following form of the entropy flux
\begin{equation}
\vec{J} = \hat{b} \cdot \vec{q} + \hat{B} : \hat{Q}, 
\end{equation} 
\noindent where $\hat{b}$ is second order and $\hat{B}$ is third order tensorial functions called current multipliers (Nyiri multipliers).
The authors also assumed a quadratic dependence of the entropy density on the additional field variables
\begin{equation}
	s = s_{eq} (e) -\frac{m_1}{2} \vec{q} \cdot \vec{q} - \frac{m_2}{2} \hat{Q} : \hat{Q},
\end{equation}
\noindent where $m_1$ and $m_2$ are constant positive material coefficients.

The entropy inequality could be written as \cite{van15}
\begin{multline*}
	\frac{\partial s}{\partial t} + \nabla \cdot \vec{q} = - \frac{\nabla \cdot \vec{q}}{T}   - m_1 \vec{q} \cdot \frac{\partial \vec{q}}{\partial t} 
	- m_2 \hat{Q} : \frac{\partial \hat{Q}}{\partial t} + \hat{b} : \nabla \vec{q} + \vec{q} \cdot (\nabla  \cdot \hat{b}) + \hat{B} : \nabla \hat{Q} + \hat{Q}:(\nabla \cdot \hat{B}) \\
	= \left( \hat{b} - \frac{1}{T} \hat{I} \right) : \nabla \vec{q} + \left( \nabla \cdot \hat{b} - m_1 \frac{\partial \vec{q}}{\partial t} \right) \cdot \vec{q} + 
	\left( \nabla \cdot \hat{B} - m_2 \frac{\partial \hat{Q}}{\partial t} \right) : \hat{Q} + \hat{B} : \nabla \hat{Q} \ge 0, 
\end{multline*} 
\noindent where $\hat{I}$ is the unit tensor.

The authors identified four generalized forces and four fluxes in this equations and assumed the linear relationships between them that in the 1D case are
\begin{equation}
m_1 \frac{\partial q}{\partial t}	- \frac{\partial b}{\partial x} = - l_1 q,
\end{equation}
\begin{equation}
m_2 \frac{\partial Qq}{\partial t}	- \frac{\partial Q}{\partial x} = - k_1 Q + k_{12} \frac{\partial q}{\partial x},
\end{equation}
\begin{equation}	b - \frac{1}{T} = - k_{1} Q + k_2 \frac{\partial q}{\partial x},
\end{equation}
\begin{equation}
B = n \frac{\partial Q}{\partial x}.	
\end{equation}

Material coefficients ($l_1, k_1, k_2, k_{12}, k_{21}$) are subjected to following restrictions from the second law of thermodynamics: 
\begin{equation}
l_1 \ge 0,\qquad k_1 \ge 0,\qquad k_2 \ge 0,\qquad k_1 k_2 - k_{12} k_{21} \ge 0.	
\end{equation}

Using these constraints  and eliminating the internal variables, the  authors get the general constitutive equation for the heat flux as
\begin{equation*}
m_1 m_2 \partial_{tt} q + (m_) l_1 + m_1 k_1) \partial_t q - (m_1 n + m_2 k_2) \partial_{xxt} q + n k_2 \partial_x^4 q - (l_1 + K) \partial_{xx} q
\end{equation*}
\begin{equation}
 + k_1 l_1 q = m_2 \partial_{xt} \frac{1}{T} + k_1 \partial_x \frac{1}{T} - \partial_x^3 \frac{1}{T}.
\end{equation}

Choosing some of the materials coefficients to be equal to zero, one can get a number of known models such as
{Fourier: $n = m_1 = m_2 = k_2 = 0$ and ($k_{12} = 0$ or $k_{21} = 0$)}; {Cattaneo:	
$n = m_2 = k_2 = 0$ and ($k_{12} = 0$ or $k_{21} = 0$)};
{Ballistic-diffusion:
$n = k_2 = 0$}; {Jeffrey's type:
$n = m_1 =  k_2 = 0$ and ($k_{12} = 0$ or $k_{21} = 0$)};
{Guyer-Krumhansl: $n =  m_2 = 0$}; {Cahn-Hillard:	
	$n = m_1 = m_2 = 0$ }.

\subsection{Rogolino et al. Models}
Rogolino et al. \cite{rog18}  based their analysis on the work by  V\'an \&  F\"ul\"op \cite{van12} that used two assumptions:
(1) {the deviation from the equilibrium state is described by the heat flux and the second-order tensorial internal variable;}
(2) {the deviation from the classical form of the entropy current is described  by two tensorial functions, called the current multipliers.}

Rogolino et al. developed two versions of the generalized heat conduction equations. The first one ignores the nonlocal effects, the second equation is able to describe the heat conduction in the presence of the nonlocal effects.

The authors used in the first case  as the basic field variables the specific internal energy $e$, the heat flux $\vec{q}$ and the flux of heat flux $\vec{\Phi_q}$. In the second case they considered the consequences in the first case by the spatial and time derivatives of the equations and obtained the higher-order equation for the heat flux.
Rogolino et al. proved that the entropy flux is nonlocal in both cases while the entropy is local in the first case and nonlocal in the second case.

The authors assumed the following form of the balance laws (in 1D  case)
\begin{equation}
\frac{\partial e}{\partial t} + \frac{\partial \vec{q}}{\partial x} = 0, \quad	
\frac{\partial \vec{q}}{\partial t} + \frac{\partial \vec{\Phi_q}}{\partial x} = r_q, \quad	
\frac{\partial \vec{\Phi_q}}{\partial t} + \frac{\partial \vec{\Psi}}{\partial x} = r_{\Phi}			
\end{equation}
\noindent where $r_q$ and $r_{\Phi}$ are the production rates of the heat flux $\vec{q}$ and flux of heat flux $\vec{\Phi}_q$, $\vec{\Psi}$ is the flux of $\vec{\Phi}_q$. 

For the closure of the system of these equations  it is necessary to use the constitutive equations for the flux $\vec{\Psi}$ and the source terms $r_q$ and $r_{\Phi}$. 
This system is the 1D version of the 13-moments system of extended irreversible thermodynamics \cite{jou,jou1} that is related to the Grad's 13-moments method.

The entropy $s$ and the entropy flux $\vec{J}$ depend on the system state 
\begin{equation}
	Z = \left[ e, \frac{\partial e}{\partial x}, \vec{q},\frac{\partial \vec{q}}{\partial x}, \vec{\Phi}_q, \frac{\partial \vec{\Phi}_q}{\partial x}  \right].
\end{equation}

The authors using the entropy constraints and Lagrange-Farkas multipliers $\lambda$, $\alpha$ and $\beta$ calculate the entropy inequality and found that  these multipliers are determined by  the derivatives of the entropy with respect to the basic variables
\begin{equation}
	\lambda = \frac{\partial s}{\partial e},\quad \alpha = \frac{\partial s}{\partial q},\quad \beta = \frac{\partial s}{\partial \Phi_q}.
\end{equation}

The general expression for the entropy flux is written as 
\begin{equation}
\vec{J} = \vec{J}_0 (e, \vec{q}, \vec{\Phi}_q) + \frac{\partial s}{\partial \vec{\Phi}_q}	
\end{equation}
\noindent and the entropy inequality is given by
\begin{equation}
\label{ineq}
\frac{\partial}{\partial x} \left( \frac{\partial s}{\partial \Phi_q} \right) \Psi - \frac{\partial s}{\partial  e} \frac{\partial q}{\partial x} - \frac{\partial s}{\partial  q} \frac{\partial \Phi_q}{\partial x} + \frac{\partial J_0}{\partial x} + \frac{\partial s}{\partial q} r_q + \frac{\partial s}{\partial \Phi_q} r_{\Phi} \ge 0.
\end{equation}

The solution of the inequality (\ref{ineq}) requires the additional assumptions. Rogolino et al. choose as entropy a quadratic function of  $q$ and $\Phi_q$
\begin{equation}
	\label{ent}
	s = \bar{s} (e) - m \frac{q^2}{2} - M \frac{\Phi_q^2}{2}.
\end{equation}
\noindent where $m$ and $M$ are the constant positive coefficients and assumed the following compatibility condition to be valid
\begin{equation}
	\frac{\partial J_0}{\partial  \Phi_q} = \frac{\partial s}{\partial q}.
\end{equation}

The (\ref{ent}) ensures the principle of maximum entropy at the equilibrium.

The authors required that the entropy flux reduces to the classical value $J =\partial s / \partial e$ at  equilibrium and  get the simplified form of inequality (\ref{ineq}) as
\begin{equation}
	- M \frac{\partial \Phi_q}{\partial x} \Psi + q \left( \frac{\partial}{\partial x} \left(\frac{\partial s}{\partial e} \right) - m  r_q \right) + \Phi_q \left( \frac{\partial}{\partial x} \left(\frac{\partial s}{\partial q} \right) - M  r_{\Phi} \right) \ge 0.
\end{equation}

The authors introduce  coefficients $l_1$, $l_2$, $l_3$ and after elimination of the production rates $r_q$, $r_{\Phi}$ and the highest order flux $\Psi$ get the system of equations
\begin{equation}
\label{s1}	
	\frac{\partial e}{\partial t} + \frac{\partial q}{\partial x} = 0,
\end{equation}
\begin{equation}
\label{s2}			
\tau_q	\frac{\partial q}{\partial t} + q + \tau_q \frac{\partial \Phi_q}{\partial x} = l_2 \frac{\partial}{\partial x} \left( \frac{1}{T} \right),
\end{equation}
\begin{equation}
\label{s3}				
\tau_{\Phi}	\frac{\partial \Phi_q}{\partial t} + \Phi_q - \tau_{\Phi} l_1 M \frac{\partial^2 \Phi_q}{\partial x^2} = - l_3 m \frac{\partial q}{\partial x}.
\end{equation}
\noindent where $\tau_q = m l_2$ and $\tau_{\Phi} = M l_3$. 

System of equations (\ref{s1} - \ref{s3}) includes as special cases a number of known models: (1) {If $l_3 = 0$ then equation (\ref{s3}) yields $\Phi_q = 0$ and equation (\ref{s2}) reduces to the Maxwell-Cattaneo-Vernotte equation (\ref{catt}); (2) 	{If $\tau_{\Phi}$ is negligible than from (\ref{s3}) follows that $\Phi_q = 0$ and  equation (\ref{s2}) leads to 1D Guyer-Krumhansl equation (provided $\tau l_3 = 3 l^2$); (3)
{If in the last equation $q$ is also negligible, one gets the equation of Green-Naghdi type.

Rogolino et al. derived  two versions of the  heat equation (1) {the $2^{nd}$  order in space and $1^{st}$ in time equation neglecting the nonlocal effects;}
(2) {the $4^{th}$ order in space and $2^{nd}$ order in time equation incorporating the nonlocal effects.}

\subsection{EIT Ballistic-Diffusive Model}
EIT Ballistic-Diffusive Model by Lebon et al. \cite{leb11} in contrast to the BD model introduced by Chen \cite{che1,che2}  based on the mixture of the kinetic and macroscopic approaches followed the purely macroscopic approach. EITBD model also relays on the coexistence of two kinds of heat carriers:
{ballistic phonon that originate at the boundaries and collide mainly with the walls; } and
 {diffusive phonons that undergo multiple collisions within the core of the system.}

The internal energy and the heat flux are splitted  into  two parts
\begin{equation}
	e = e_b + e_d,\quad  \vec{q} = \vec{q}_b + \vec{q}_d.
\end{equation}

The state variables of EIT are selected as
\begin{enumerate}
	\item {($e_b, \vec{q}_b$) provide the description of the ballistic motion of phonons;}
	\item {($e_d, \vec{q}_d$) provide the description of the diffusive motion of phonons.}
\end{enumerate}

The authors introduce the ballistic and diffusive quasi-temperatures $T_b = e_b / c_b$ and $T_d = e_d / c_d$ that do not represent temperatures in the usual sense but serve as the measure if the internal energies \cite {leb14}. Assuming that the heat capacities are equal, the total quasi temperature is introduced $T = e /c = T_b + T_d$.

Evolution of the internal energies is governed by the balance equations
\begin{equation}
	\frac{\partial e_b}{\partial t} = - \nabla \cdot \vec{q}_b + r_b, \quad
	\frac{\partial e_d}{\partial t} = - \nabla \cdot \vec{q}_d + r_d ,
\end{equation}
the total energy $e = e_b + e_d$ satisfies the first law of thermodynamics
\begin{equation}
	\frac{\partial e}{\partial t} = - \nabla \cdot \vec{q} + r.
\end{equation}

To describe the ballistic phonons the authors use 
Guyer-Krumhansl equation
\begin{equation}
	\tau_b \frac{\partial \vec{q}_b}{\partial t} + \vec{q}_b = - \lambda_b \nabla T + l_b^2  (\nabla^2 \vec{q}_b + 2 \nabla \cdot \nabla \vec{q}_b),
\end{equation}
while the evolution of diffusive phonons is governed by the Cattaneo equation
\begin{equation}
	\tau_d \frac{\partial \vec{q}_d}{\partial t} + \vec{q}_d = - \lambda_d \nabla T.
\end{equation}
\section{Nonlocal (Fractional) Models}

\subsection{Fractional Fourier Model}
Deng et al. \cite{den19} studied steady heat transfer using the 2D fractional Helmholtz equation in the fractal media
\begin{equation}
\frac{\partial^{2 \alpha} T}{\partial x^{2 \alpha}} + \frac{\partial^{2 \alpha} T}{\partial y^{2 \alpha}} + k^2 = f,
\end{equation}
\noindent where $0 < \alpha \le 1$, $0 < \beta \le 1$.

He \& Liu \cite{he13} used the fractional form of the Fourier law
\begin{equation}
\lambda^{2 \alpha} \frac{d^{\alpha} T}{d x^{\alpha}} = q
\end{equation}
\noindent to study the heat transfer in the silk cocoon hierarchy.

Similar approach was used by
Beybalaev  \cite{bei09} to study the heat conduction in the fractal medium and 
 Beybalaev et al. \cite{bei} to study of the ground freezing. 
 
 He et al. \cite{he11,he16} used the stationary space-fractional equation
 \begin{equation}
 \frac{\partial^{\alpha}}{\partial x^{\alpha}} \left( \lambda \frac{\partial^{\alpha} T}{\partial x^{\alpha}} \right) = 0
 \end{equation}
 \noindent and Wang et al. \cite{wan12} nonstationary equation
 \begin{equation}
 \frac{\partial T}{\partial t} + \frac{\partial^{\alpha}}{\partial x^{\alpha}} \left( \lambda \frac{\partial^{\alpha} T}{\partial x^{\alpha}} \right) = 0
 \end{equation}
 \noindent to study the heat transfer in the fractal medium of the polar bear hair.
  
   Sierociuk et al. \cite{sie13} exploited the time fractional Fourier equation to study the heat transfer in the nonuniform semi-infinite beam.

Cheng \& Pang \cite{che16} suggested a new definition of the fractional  Laplacian using the Riesz fundamental solution, The implicit fractional Laplacian is defined by the integro-differential operator that satisfies the Fourier transform
\begin{equation}
F[(- \Delta)^s u (x), k] = ||k||^s 	F[u (x), k], \qquad x, k \in R^3
\end{equation}
 where the Fourier transform and the inverse transform are defined as
 \begin{equation}
 \hat{u} (k) = 	F(u (x), k) = \int\limits_{R^3} u (x) e^{- i \vec{k} \cdot \vec{x}} d \vec{x}
 \end{equation}
 \noindent and
 \begin{equation}
 u (x) = F^{-1} (\hat{u} (x), x) = \frac{1}{(2 \pi)^3}	\int\limits_{R^3} \hat{u} (k) e^{ i \vec{k} \cdot \vec{x}} d k.
 \end{equation}
 \noindent and  that has a fundamental solution of the form
 \begin{equation}
 	u^{\star} (x) = c_6 (s) \frac{1}{||x||^{3 - s}}, \qquad x \in R^3
 \end{equation}
 \noindent where
 \begin{equation}
 c_6 (s) = \frac{\Gamma(\frac{3 - s}{2})}{2^s \pi^{3/2} \Gamma (\frac{s}{2})}
 \end{equation}
 \noindent is the normalizing function, $\Gamma$ is the Euler gamma function.

\subsection{Nonlinear Diffusivity}
Fa \cite{fa05} used the Fokker-Planck equation in the Stratonovich approach that follows from the Langevin equation with a multiplicative  replacing the probability distribution function by the temperature

\begin{equation}
	\frac{\partial T (x, t)}{\partial t} = \kappa a (t^{\star}) \frac{\partial}{\partial x} \left(D (x) \frac{\partial}{\partial x} (D (x) T (x, t))  \right)
\end{equation}
\noindent where $a (t) = 1 - \exp (- t^{\star})$.

Fa used the dimensionless variables
\begin{equation}
	\xi = \frac{x}{L}, \quad t^{\star} = \frac{t}{\tau}, \quad \theta = \frac{T - T_0}{T_1 - T_0}, \quad \phi = \frac{q}{C_V (T_1 - T_0)} 
\end{equation}
\noindent where $\tau = l / v$, $l$ is MFP, $v$ is the velocity of sound.

Using transformations $d u / d \xi = 1 / D (\xi)$ and $d s / d t^{\star} = a (t^{\star})$,
\quad $u (0) = 0, \quad s (0) = 0)$ Fa \cite{fa05} obtained the solution the problem of the  conduction in a slab with the different sides temperatures
\begin{equation}
	\theta (\xi, t^{\star}) = \frac{D_0}{D (\xi)} \left[1 - \frac{u (\xi)}{u_1} - \frac{2}{\pi}  \sum_{m = 1}^{\infty}  \displaystyle \frac{\sin \left(\frac{m \pi u (\xi)}{u_1}   \right) \exp \left(- \displaystyle \frac{Kn^2 (m \pi)^2 s (t^{\star})}{3 u_1^2}   \right)}{m} \right] 	
\end{equation}
\noindent where $D_0$ is value of $D (\xi)$ at $\xi = 0$ and $u_1$ is value of $u (\xi)$ at $\xi = 1, \quad Kn = l / L$. 

The author stated that solution is close to the results of the simulation of the Radiative Phonon Transport model. When $D = 1$ the model by Fa reduces to the model suggested by Naqvi \& Waldenstream \cite{nw}.

Falcini et al. \cite{fal19} used a limit case Stefan problem --- the moisture infiltration into a porous medium --- to study three effects that can extend the applicability of the Fourier equation to account for the anomalous heat diffusion:
 {time and space nonlocality},
{non-linear spatial dependent thermal diffusivity}
 and action of their combinations.
The authors considered the  equation
\begin{equation}
\label{fal}
	\frac{\partial^{\gamma} T}{\partial t^{\gamma}} = \frac{\partial}{\partial x} \left(\kappa (x) \frac{\partial^{\alpha} T}{\partial x^{\alpha}} \right)
\end{equation}
\noindent where $ \alpha \quad (1 < \alpha \le 1)$ is degree of the space nonlocality, $\gamma \quad  (1 < \gamma \le 1)$ is the measure of the memory effect, nonlinear diffusivity
\begin{equation}
	\kappa (x) = \kappa_0 x^{\displaystyle \frac{2 \beta - 1}{\beta}},
\end{equation}
\noindent thus $\beta \quad  (1 < \beta \le 1)$ could be considered as a measure of nonlinearity.

The relation between the characteristic time and length scales in the normal heat conduction (normal diffusion) $l \propto \sqrt{\tau}$. In the general anomalous case
this relation is written as $l \propto \tau^n$ where $n \ne 1/2$.

Falcini et al. \cite{fal19} showed that three non-Fourier effects combine and compete to determine the time exponent
\begin{equation}
	n =  \frac{\gamma}{1 + \alpha + \left(\displaystyle \frac{1}{\beta} - 2 \right)} .
\end{equation}
\noindent Thus deviations from the Fourier law behaviour could be described by different effects, e.g., the  roles of fractional derivatives and suitable nonlinearity of the thermal diffusivity are interchangeable \cite{fal19}:
\begin{itemize}
	\item {subdiffusion ($n < 1/2)$:
	{memory only}; 
	{nonlinearity only, $\beta \in (0, 0.5]$};
	{memory + nonlocality, $1 + \alpha > 2 \gamma$};
	{memory + nonlinearity, $\beta \gamma < 1 / 2$};
	{nonlocality + nonlinearity, $\alpha + 1/\beta > 3$}
	
}
\item {superdiffusion ($n > 1/2$):
	 {nonlinearity only, $\beta \in (0.5, 1]$};
 {memory + nonlocality, $1 + \alpha < 2 \gamma$};
 {memory + nonlinearity, $\beta \gamma> 1 / 2$};
 {nonlocality + nonlinearity, $3 > \alpha + 1 / \beta > 2$}		
}
\item {superdiffusion+ ($n > 1)$:
{nonlocality + nonlinearity, $\alpha + 1 / \beta < 2$}.		
}
\end{itemize}

\subsection{Fractional Pennes model}
The time-fractional generalization (using the
Caputo fractional derivative  of order $\alpha \in (0,1]$) of the Pennes bioheat equation (\ref{pennes}) 

 \begin{equation}
 \label{pennes_t}
 \varrho c \frac{\partial^{\alpha} T}{\partial t^{\alpha}} = \nabla \cdot \lambda \nabla T +  c_{\mathrm{b}} \omega_b (T_{\mathrm{a}} - T) + \dot{q}_{\mathrm{met}} + Q^{\mathrm{ext}},
 \end{equation}

 \noindent  was used by R.S.~Damor et al. \cite{dam14} to study the hyperthermia \index{hyperthermia} and the anomalous diffusion in the skin tissue with the constant and the sinusoidal heat flux at the boundary \cite{dam13,dam15} and by Ezzat et al. \cite{ezz14} to study the temperature transient in the skin exposed to the instantaneous surface heating.

As noted by L.L. Ferras et al. \cite{ferr15}, the equation (\ref{pennes_t}) is not dimensionally consistent and one has to either redefine the coefficients of this equations or to introduce a factor $\tau^{1 - \alpha}$ to get a "new" thermal conductivity.

Singh et al. \cite{sin11} used the space-time fractional bioheat equation
\begin{equation}
\rho C \frac{\partial^{\beta}}{\partial^{\beta}} = \lambda \frac{\partial^{\alpha}}{\partial^{\alpha}} + Q_p,\qquad 0 < \beta \le 1 < \alpha \le 2
\end{equation}
\noindent to study the heat transfer in tissues during the thermal therapy. 

\subsection{Zingales's Fractional Order Model}
Zingales \cite{zin14} (see also \cite{zin13}) considered two components of the heat transfer in the rigid solid bodies at rest
\begin{enumerate}
	\item {A short-range heat flux governed by the conventional Fourier law. }
\item {A long-ranged heat transfer between the elementary volumes located at points $\vec{x}$ and $\vec{y}$ that is proportional to (1)  {the product of interacting masses;}
(2) {the temperature difference $T (\vec{x}) - T (\vec{y})$;}
(3) {the distance-decaying function $g (|| \vec{x} - \vec{y} ||)$ .}
}
\end{enumerate}

Zingales assumed that the function $g$  decays as a power-law of the distance
\begin{equation}
g (|| \vec{x} - \vec{y} ||) = \frac{1}{d_n (\bar{\alpha})} \frac{1}{|| \vec{x} - \vec{y} ||^{n + \alpha}}
\end{equation}
\noindent where $d_n (\bar{\alpha})$ is the normalizing coefficient related to the decaying exponent $\alpha$ and to the dimension of the topological space of the body $n$.

Finally the energy balance equation is written in the form
\begin{equation}
\rho C \frac{\partial T}{\partial t} = -\nabla \cdot \vec{q} + \rho^2 \lambda_{\alpha} D_x^{\alpha} T
\end{equation}
\noindent where $D_x^{\alpha}$ is the Marchaud fractional derivative of order $\alpha$ defined as
\begin{equation}
D_x^{\alpha} T = \frac{1}{d_n (\bar{\alpha})} \int\limits_{V_y} \frac{T (\vec{x}) - T (\vec{y})}{|| \vec{x} - \vec{y} ||^{n + \alpha}} d V_y . 
\end{equation}

\subsection{Fractional Cattaneo and SPL Models}
The hyperbolic models of non-Fourier heat conduction  suffer from the unrealistic singularity of temperature gradient across thermal wavefront; the fractional calculus  can remove the thermal
wave singularity \cite{akb17}. 
Sometimes the fractional version of Cattaneo equation is called "nonlocal" {\ } Cattaneo-Vernotte equation, reflecting the basic properties of the fractional derivatives \cite{bur10}.

Liu et al. \cite{liu16} used the modification of the Cattaneo model by Christov to develop the space  fractional equation with the Riesz derivative.

A time-fractional SPL model for the bioheat transfer is formulated as \cite{yu16a}
\begin{equation}
\label{spl_t}
\varrho c \left(\frac{\partial T}{\partial t} + \tau \frac{\partial^{1 + \alpha} T}{\partial t^{1 + \alpha}} \right) = \nabla \cdot \lambda \nabla T +  c_{\mathrm{b}} \omega_b (T_{\mathrm{a}} - T) + \dot{q}_{\mathrm{met}} + Q^{\mathrm{ext}}.
\end{equation}

Computations show that the  fractional SPL equation give the same temperature distribution as the DPL model \cite{yu16a}.

Fabrizio \cite{fab14b} formulated the fractional Cattaneo equation as
\begin{equation}
	\frac{\gamma \alpha}{\Gamma (1 - \alpha)} \int_{- \infty}^t \frac{\vec{q} (x,t) - \vec{q} (x,\tau)}{(t - \tau)^{1 + \alpha}} d\tau = \vec{q} (x,t) + \lambda \nabla T (x,t).
\end{equation}

Jiang \& Qi \cite{jia12} derive the fractional thermal wave model of the bioheat transfer changing the Cattaneo relation to
\begin{equation}
 \frac{\tau^{\alpha}}{\alpha !} D_t^{\alpha} \vec{q} + \vec{q} = - \lambda \nabla T,
\end{equation}
\noindent where $D_t^{\alpha}$ is a modified Riemann-Liouville derivative of order $\alpha$.
 
Qi et al. \cite{qi13} studied laser heating generalizing the Cattaneo relation as 
\begin{equation}
\tau^p D_t^p \vec{q} + \vec{q} = - \lambda \nabla T;\qquad 0 < p < 1,
\end{equation}
\noindent where $D_t^p$ is the Caputo derivative of order $p$; the factor $\tau^p$ is introduced to keep  the dimension in order.

Povstenko \cite{pov05,pov09} and Jiang \& Xu \cite{jia10} proposed the time fractional Fourier law as the constitutive relationship
\begin{equation}
	\vec{q} = - \kappa D^{1 - \alpha} \nabla T
\end{equation}
\noindent where $D^{1 - \alpha}$ the Caputo fractional derivative.
Povstenko \cite{pov11} showed that many fractional generalizations of the Cattaneo relation could be obtained from the generalization where the kernels are functions of the Mittag-Leffler type.

Xu et al. \cite{xu13} formulated the fractional Cattaneo equation using two fractional derivatives (Caputo) of different order. 
The authors started with the generalized constitutive equation 
\begin{equation}
\frac{\partial^{\beta  - 1} \vec{q}}{\partial t^{\beta - 1}} + \tau  \frac{\partial^{\alpha  - 1} \vec{q}}{\partial t^{\alpha - 1}}  = - \lambda \nabla T;\qquad 0 < \beta \le \alpha \le 2.
\end{equation}
\noindent and using the Laplace transform obtained a time-nonlocal constitutive relation
\begin{equation}
	\vec{q} (t) = - \frac{\lambda}{\tau} \int\limits_0^t (t - t^{\prime})^{\alpha - 2} E_{\alpha - \beta, \alpha - 1} \left(- \frac{(t - t^{\prime})^{\alpha - \beta}}{\tau} \right) \nabla T (t^{\prime}) d t^{\prime}
\end{equation}
\noindent where the Mittag-Leffler function is defined as
\begin{equation} \label{ml}
	E_{\mu, \nu} (z) = \sum_{k = 0}^{\infty} \frac{z^k}{\Gamma (\mu k + \nu)}.
\end{equation}
Xu et al. \cite{xu13} wrote the generalized Cattaneo heat equation  as
\begin{equation}
	\frac{\partial^{\beta} T}{\partial t^{\beta}} + \frac{\partial^{\alpha} T}{\partial t^{\alpha}} = \mathcal{D} \Delta T,
\end{equation}
\noindent where the generalized thermal diffusivity $\mathcal{D}$ has dimension $[\mathcal{D}] = m^2 / s^{\beta}$. 

The fractional SPL model was used by Mishra \& Rai \cite{mis16} to study the heat transfer in the thin films and by Moroz et al. \cite{mor19}  to simulate the heat conduction in the segnetoelectric material (triglicinsulfate).

Christov \cite{chr16} developed the equation for the transient heat conduction with  the damping term expressed via the Caputo-Fabrizio fractional derivative that is the extension of the Caputo fractional derivative with the singular kernel
\begin{equation}
D_t^{\alpha} f(t) = \frac{M (\alpha)}{1 - \alpha}  \int\limits_0^t exp \left[ - \frac{\alpha (t-s)}{1 - \alpha} \right] \frac{d f(t)}{d t} d s, 
\end{equation}
\noindent where $M (\alpha)$ is a normalization function such as M (0) = M (1) =1 $.$
Numerical solution of this problem was considered by Alkahtani \& Atangana \cite{alk17}.

Yang et al. \cite{yan15} (see also \cite{yan16}) for the study of fractional heat transfer introduced a new fractional derivative without the singular kernel that is the extension of the Riemann-Liuville fractional derivative with the singular kernel
\begin{equation}
D_{\alpha^+}^{(\nu)} T = \frac{\mathfrak{R} (\nu)}{1 - \nu} \frac{d}{dx} \int\limits_a^x \exp  \left(- \frac{\nu}{1 - \nu} (x - x^{\prime}) \right) T d x^{\prime} ,
\end{equation}
\noindent where $\mathfrak{R} (\nu)$ is the normalization function.

 Ghazizadeh \& Maerafat \cite{gha10} used developed by Odibat \& Shawagfeh \cite{odi07} the generalized Taylor's formula
 \begin{equation}
 q (\vec{r}, t + \tau) = q (\vec{r}, t) + \frac{\tau^{\alpha}}{\Gamma (1 + \alpha)}, \quad 0 < \alpha \le 1.
 \end{equation}
 
 Inserting this equation into the energy balance equations and eliminating the heat flux the authors get the "nonlocal" Cattaneo equation
 \begin{equation}
 \rho C \frac{\partial T}{\partial t} + \frac{\rho C \tau^{\alpha}}{\Gamma (1 + \alpha)} \frac{\partial^{\alpha + 1} T}{\partial t^{\alpha + 1}} = \lambda \nabla^2 T.
 \end{equation}

\subsection{Fractional DPL Model}
Ji et al. \cite{ji18,ji19} studied the heat transfer in thin films using the following form of the time fractional DPL (${}_0^CD_t^{\alpha}$ is the Caputo fractional derivative) 
\begin{equation}
\rho C \left(\frac{\partial T}{\partial t} + \frac{\tau_q^{\alpha}}{\Gamma (1 - \alpha)} {}_0^CD_t^{1 + \alpha} T \right) = \lambda \left( T + \frac{\tau_T^{\alpha}}{\Gamma (1 - \alpha)} {}_0^CD_t^{\alpha} T\right).
\end{equation}

Xu et al. \cite{jing15} studied the bioheat transfer using the following equation in the Caputo derivatives of orders $\alpha$ and $\beta$ 
\begin{equation}
\vec{q} + \tau_q^{\alpha} \frac{\partial^{\alpha} \vec{q}}{\partial t^{\alpha}} = - \lambda \left(\nabla T + \tau_T^{\beta} \frac{\partial^{\beta}}{\partial t^{\beta}} \nabla T \right).
\end{equation}
\noindent The authors have changed the relaxation times of DPL model $\tau_q$ and $\tau_T$ to $\tau_q^{\alpha}$ and $\tau_T^{\beta}$ to maintain the dimensions in order.

Xu et al. used the non-linear least-square method to estimate simultaneously two relaxation times and two or up to degrees of fractionality. 

Liu et al. \cite{liu18} used the convected derivative introduced by Christov \cite{chr09} and the Caputo fractional derivative of order $\alpha$ to formulate a model 

\begin{equation}
\tau_q \left[ \frac{\partial^{\alpha} \vec{q}}{\partial t^{\alpha}} + \vec{v} \cdot \nabla \vec{q} + \vec{q} \cdot \nabla \vec{v} + (\nabla \cdot \vec{v}) \vec{q} \right] + \vec{q} = - \lambda \nabla \left(1 + \tau_T \frac{\partial^{\alpha}}{\partial t^{\alpha}} \right) T.
\end{equation}

\subsection{Fractional TPL Model}
\label{ftpl1}

Ezzat et al. \cite{ez12,ez13} obtained the Fractional TPL model (FTPL) taking a Taylor
series expansion on the both sides  the constitutive relation of the TPL model (\ref{tpl}) and retained the terms up to $2 \alpha_F$-order for the relaxation time  $\tau_q$ and up to $\alpha_F$-order for  $\tau_T$ and $\tau_v$ to get
\begin{equation}
\label{ftpl}
\left(1 + \frac{\tau_q^{\alpha_F}}{\alpha_F !} \frac{\partial^{\alpha_F}}{\partial t^{\alpha_F}} + \frac{\tau_q^{2 \alpha_F}}{(2 \alpha_F) !} \frac{\partial^{2 \alpha_F}}{\partial t^{2 \alpha_F}}  \right) \vec{q} = 
- \left[ \left(\tau_v^{\star} + \lambda \frac{ \tau_T^{\alpha_F}}{\alpha_F !} \frac{\partial^{\alpha_F}}{\partial t^{\alpha_F}}  \right) \nabla T  + \lambda^{\star} \nabla v\right]
\end{equation}
where
\begin{equation}
	0 < \alpha \le 1, \quad
\tau_v^{\star} = \lambda + \lambda^{\star}  \frac{\tau_v^{\alpha_v}}{\alpha_F !} \frac{\partial^{\alpha_F - 1}}{\partial t^{\alpha_F - 1}}.	
\end{equation}

The FTPL model as well as ordinary TPL one was used to study the problems of thermoelasticity \cite{ez12,ez13} and 
piezoelectric thermoelasticity problems. 

Ezzat et al. \cite{ez12} used the energy conservation equation for the homogeneois isotropic thermoelastic solid
\begin{equation}
	- \nabla \cdot \vec{q} + \rho Q = \rho C_E \dot{T} + \gamma T_0 \dot{e}_{ii}
\end{equation}
 where $Q$ is the heat source, $C_E$ is the specific heat at constant strain, $e_{ii}$ are the components of the strain tensor, $\gamma = (3 \nu + 2 \mu) \alpha_T$, $\nu, \mu$ are the Lame's constants, $\alpha_T$ is the coefficient of linear thermal expansion.

Taking the divergence and the time derivative of the equation (\ref{ftpl}) Ezzat et al. get
\begin{equation}
	\left(1 + \frac{\tau_q^{\alpha_F}}{\alpha_F !} \frac{\partial^{\alpha_F}}{\partial t^{\alpha_F}} + \frac{\tau_q^{2 \alpha_F}}{(2 \alpha_F) !} \frac{\partial^{2 \alpha_F}}{\partial t^{2 \alpha_F}}  \right) \nabla \cdot  \dot{\vec{q}} = 
	- \left[ \left(\tau_v^{\star} + \lambda \frac{ \tau_T^{\alpha_F}}{\alpha_F !} \frac{\partial^{\alpha_F + 1}}{\partial t^{\alpha_F + 1}}  \right) \nabla^2 \dot{T}  + \lambda^{\star} \nabla^2 \dot{v} \right].
\end{equation}
and finally obtain the modified fractional heat transport equation
\begin{multline}
	\label{e_5}
	\left(1 + \frac{\tau_q^{\alpha_F}}{\alpha_F !} \frac{\partial^{\alpha_F}}{\partial t^{\alpha_F}} + \frac{\tau_q^{2 \alpha_F}}{(2 \alpha_F) !} \frac{\partial^{2 \alpha_F}}{\partial t^{2 \alpha_F}}  \right) (\rho C_E \ddot{T} + \gamma T_0 \ddot{e} - \rho \dot{Q}) = \\
	\tau_v^{\star} \nabla^2 \dot{T} + \lambda \frac{ \tau_T^{\alpha_F}}{\alpha_F !} \frac{\partial^{\alpha_F + 1}}{\partial t^{\alpha_F + 1}} \nabla^2 T + \lambda^{\star} \nabla^2 T.
\end{multline}

The authors consider several limiting cases:
\begin{itemize}
	\item {When $\alpha_F = 1$ and the thermal conductivity $\lambda$ is much smaller than $\lambda^{\star}$, $q = - \lambda^{\star} \nabla v$, $\tau_v^{\star} \lambda^{\star} \tau_v$ and (neglecting $\tau_q^2$) equation (\ref{e_5}) simplifies to
\begin{equation}
\left(1 + \tau_q \frac{\partial}{\partial t} \right) (\rho C_E \ddot{T} + \gamma T_0 \ddot{e} - \rho \dot{Q}) = 
\lambda^{\star} \left(1 + \tau_v \frac{\partial}{\partial t} \right) \nabla^2 T.	
\end{equation}	

This equation is an {\em extension} of the Green-Naghdi theory of type II. The heat conduction equation of the  Green-Naghdi theory of type II \cite{gr2}
\begin{equation}
 \rho C_E \ddot{T} + \gamma T_0 \ddot{e} - \rho \dot{Q} = 
\lambda^{\star} \nabla^2 T	
\end{equation}
is obtained if $\tau_T = \tau_v = \tau_q = 0.$
}
	\item {When $\alpha_F = 1, \tau_T = \tau_v = \tau_q = 0$ (hence $\tau_v^{\star} = \lambda$)} the equation (\ref{e_5}) reduces to the equation of the  Green-Naghdi theory of type III \cite{cho07,ez09}
	\begin{equation}
	 \rho C_E \ddot{T} + \gamma T_0 \ddot{e} - \rho \dot{Q} = 
	\lambda^{\star} \nabla^2 T + \lambda \nabla^2 T.	
	\end{equation}
	\item {When $\alpha_F = 1$ and $\lambda^{\star} = 0$ the equation (\ref{e_5}) reduces to the heat conduction equation of the DPL model \cite{ts96,ez03}.}
	\item {When $\alpha_F = 1$ and $\lambda^{\star} = 0$, $\tau_T = \tau_v =0$, $\tau_q > 0$, the equation (\ref{e_5}) reduces (neglecting $\tau_q^2$) to the Lord-Shulman model \cite{ls67}.}
	\item {When $\alpha_F = 1$  the equation (\ref{e_5}) reduces to the TPL model.}
	\item {When $0 < \alpha_F \le 1$, $\tau_T = \tau_v = 0$ and neglecting $\tau_q^{2 \alpha_F}$ the fractional model due to Sherief et al. \cite{sher} is obtained.}
\end{itemize}

\paragraph{Nonlocal Fractional TPL Model}

Akbarzadeh et al. \cite{akb17} suggested the nonlocal
fractional three-phase-lag (NL FTPL) heat conduction model. The
constitutive equation of this model is written as
\begin{equation}
	\label{fntpl} 
	\vec{q} (\vec{r} + \vec{\lambda}_q, t + \tau_q) = -  [\lambda J^{\alpha_F - 1} \nabla T (\vec{r} + \vec{\lambda}_T, t + \tau_T) + \lambda^{\star} J^{\alpha_F - 1} \nabla v (\vec{r} + \vec{\lambda}_v, t + \tau_v)]
\end{equation}
where $0 \le \alpha_F \le 2$, $J^{\alpha}$ is the Riemann–Liouville fractional integral.

Alternative forms of constitutive equation (\ref{fntpl})  can be derived by the Taylor series expansion of the correlation lengths $\lambda_q,\lambda_T,\lambda_v$ in the space domain
and of the relaxation times  $\tau_q, \tau_T, \tau_v$ in the time domain.

The heat conduction equation is obtained by the elimination of the heat flux from the constitutive equation and the energy conservation equation. In the case of second order Taylor expansion for both the correlation lengths and the relaxation times it is written (in the absence of volumetric heat source) as \cite{akb17}
\begin{multline}
	\label{a11}
\left( 1 + \vec{\lambda}_q \cdot \nabla + \left( \frac{\vec{\lambda}_q \cdot \vec{\lambda}_q}{2} \right) \nabla^2 + \tau_q \frac{\partial}{\partial t} + \frac{1}{2} \tau_q^2 \frac{\partial^2}{\partial t^2} \right)	\left( \frac{1}{\kappa} \tau_q^2 \frac{\partial^2 T}{\partial t^2} \right) = \\
\nabla \cdot \left( F_T + \frac{\lambda^{\star}}{\lambda} F_v
 \right) J^{\alpha - 1} \nabla T,
\end{multline}
where
\begin{equation}
	F_T = 1 + \vec{\lambda}_T \cdot \nabla + \left( \frac{\vec{\lambda}_T \cdot \vec{\lambda}_T}{2} \right) \nabla^2 + \tau_T \frac{\partial}{\partial t} + \frac{1}{2} \tau_T^2 \frac{\partial^2}{\partial t^2}
\end{equation}
and
\begin{equation}
	F_v = 1 + \vec{\lambda}_v \cdot \nabla + \left( \frac{\vec{\lambda}_v \cdot \vec{\lambda}_T}{2} \right) \nabla^2 + \tau_v \frac{\partial}{\partial t} + \frac{1}{2} \tau_v^2 \frac{\partial^2}{\partial t^2}.
\end{equation}

Akbarzadeh et al. \cite{akb17} also derived the wave-like NL FTPL for a homogenous medium by simplifying the NL FTPL heat conduction equation (\ref{a11}) using  the first-order Taylor expansion
of $\lambda_q$  in space and the second-order Taylor  expansion of
$\tau_q$ and the first-order Taylor expansion of $\tau_T$ and $\tau_v$ in time:
\begin{multline*}
	\label{a11}
	\left( 1 + \vec{\lambda}_q \cdot \nabla +  \tau_q \frac{\partial}{\partial t} + \frac{1}{2} \tau_q^2 \frac{\partial^2}{\partial t^2} \right)	\left( \frac{1}{\kappa} \tau_q^2 \frac{\partial^2 T}{\partial t^2} \right) = \\
	\left[ 1 +   \tau_T \frac{\partial}{\partial t}  + \frac{\lambda^{\star}}{\lambda} \left( 1 +   \tau_v \frac{\partial}{\partial t} \right)
	\right] J^{\alpha - 1} \nabla^2 T.
\end{multline*}

The authors noted that $\tau_q, \tau_T, \tau_v$ present delayed thermal responses due to the collisions of electrons and phonons, temporary
momentum loss, normal relaxation in phonon scattering, and
internal energy relaxation \cite{tz10,akb14}  and time-fractional derivative  removes thermal wave singularity across thermal wavefront
and is used to present the subdiffusion $(0 \le \alpha < 1)$,
normal diffusion $(\alpha =  1)$, and superdiffusion phenomena
$(1 < \alpha < 2)$.

\subsection{Fractional Boltzmann and Fokker-Planck  equations}

\subsubsection{Li \& Cao Model}

Li \& Cao \cite{li19} considered the fractional linearised BTE (the Bhatnagar-Gross-Krook equation) suggested by Goychuk \cite{goy17} 
\begin{equation}
\label{goy}
\frac{\partial f}{\partial t} + \vec{v} \cdot \nabla f = \tau^{\alpha} D_t^{\alpha} \left( \frac{f_{MB} - f}{\tau} \right),
\end{equation}
\noindent where $f = f (\vec{x},t,\vec{v})$ is the single-particle distribution function, $\vec{v}$ is the particle velocity, $f_{MB}$ is the Maxwell-Boltzmann distribution, $D_t^{\alpha}$ is the Riemann-Liuville fractional derivative, $0 < \alpha \le 1$. 

The standard Bhatnagar-Gross-Krook model leads to the following expression for the thermal conductivity $\lambda = {5 k_B^2 T \rho \tau}/{2 m^2}$  where $k_B$ is the Boltzmann constant, $\rho$ is the density, $\tau$ is the relaxation time, 
$m$ is mass of the particle.

The equation (\ref{goy}) does not have a stationary solution in the non-equilibrium cases because $\tau^{\alpha -1} D_t^{\alpha} (f_{MB} - f)$ is zero only if $f = f_{MB}$ \cite {li19}, thus 
Li \& Cao consider the quasi-stationary heat conduction. 

The energy density and the heat flux are written as, respectively,
\begin{equation}
	e (\vec{x}, t) = \frac{m}{2} \int (\vec{v} - \vec{c})^2 f d \vec{v}, \quad
	q (\vec{x}, t) = \frac{m}{2} \int (\vec{v} - \vec{c})^2 (\vec{v} - \vec{c}) f d \vec{v},
\end{equation}
\noindent where $\vec{c} = \int \vec{v} d \vec{v}$.
Starting with the relation 
\begin{equation}
	\tau^{\alpha - 1} D_t^{\alpha} \vec{q} + \partial_t \vec{q} = 
	- \frac{m}{2} \int (\vec{v} - \vec{c})^2 (\vec{v} - \vec{c}) \vec{v} \cdot \nabla f d \vec{v} - \partial_t \vec{q}
\end{equation}
the authors use the local equilibrium assumption $\nabla f \approx \nabla f_{MB}$ to derive 
\begin{equation}
\label{lc12a}
		\tau^{\alpha - 1} D_t^{\alpha} \vec{q} + \partial_t \vec{q} = - \frac{m}{2} \int |\vec{v} - \vec{c}|^2 (\vec{v} - \vec{c}) \vec{v} \cdot \nabla f_{MB} d \vec{v}.
\end{equation}
For the pure conduction case ($\vec{c} = 0, \nabla \cdot \vec{v} = 0$) the equation (\ref{lc12a}) simplifies to
\begin{equation}
	\label{lc13a}
	\tau^{\alpha - 1} D_t^{\alpha} \vec{q} + \partial_t \vec{q} = - \left[ \frac{m}{2} \int |\vec{v}|^2 \vec{v} \vec{v} \cdot \partial_T  f_{MB} d \vec{v} \right] \nabla T.
\end{equation}

The equation (\ref{lc13}) could be written as
\begin{equation}
	\tau^{\alpha - 1} D_t^{\alpha} \vec{q} + \partial_t \vec{q} = \lambda \nabla T
\end{equation}
\noindent and in the case of time-independent temperature lead to
\cite{li19}
\begin{equation}
	\label{lc13}
	\vec{q} = \left( (\vec{q} + \tau^{\alpha - 1} D_t^{\alpha} \vec{q}) |_{t = 0} \right) E_{1 - \alpha, 1} \left[ - \left(\frac{t}{\tau} \right)^{1 - \alpha} \right]-
	\end{equation}
	\begin{equation}
		 \lambda \left( \frac{t}{\tau} \right) E_{1 - \alpha, 1} \left[ - \left(\frac{t}{\tau} \right)^{1 - \alpha} \right] \nabla T
\end{equation}
\noindent where $E_{\mu,\nu} (z)$ 
is the Mittag-Leffler function (\ref{ml}).

Thus the heat flux is the sum of two contributions $\vec{q} = \vec{q}_0 + \vec{q}_T$. 

The initial effects are responsible for first part of the heat flux 
\begin{equation}
\vec{q}_0 = 	(\vec{q} + \tau^{\alpha - 1} D_t^{\alpha} \vec{q}) |_{t = 0} E_{1 - \alpha, 1} \left[ - \left(\frac{t}{\tau} \right)^{1 - \alpha} \right],
\end{equation}
and the second part is induced by the temperature gradient
\begin{equation}
\vec{q}_T = - \lambda \left( \frac{t}{\tau} \right) E_{1 - \alpha, 1} \left[ - \left(\frac{t}{\tau} \right)^{1 - \alpha} \right] \nabla T.	
\end{equation}

The long-time behaviour of the heat flux is dominated by the temperature gradient $\vec{q} \propto t^{\alpha} \nabla T$.

Multiplying equation (\ref{goy}) by $|\vec{v} - \vec{c}|$ and integrating it yields the energy conservation equation
\begin{equation}
	\label{lc43}
	\tau^{\alpha}  D_t^{\alpha + 1} + \tau \partial^2_t T = \kappa \nabla^2 T + \frac{T|_{t = 0} \tau^{\alpha}}{t^{\alpha + 1} \Gamma (- \alpha)},
\end{equation}
which reduces to the Cattaneo model for $\alpha = 0$.

When the initial value term in equation (\ref{lc43}) is neglected, this equation is belong to the so-called class GCE II \cite{com97}. However, the constitutive relation for the equation
(\ref{lc43}) does not follows from equation (\ref{lc13}) but is written as 
\begin{equation}
	\vec{q} + \tau^{1 - \alpha} D_t^{1 - \alpha}  \vec{q} = - \tau^{- \alpha} D_t^{- \alpha} \kappa \nabla T.
\end{equation}

Li \& Cao also considered other versions of the  BTE containing in contrast to the equation (\ref{goy})  the standard collision source term without any memory kernel that lead to other constitutive relations introduced in the class CGE II:
\begin{align*}
	\tau^{-\alpha} D_t^{1 - \alpha} f + \vec{v}_g \cdot \nabla  f &= \frac{f_0 - f}{\tau}\\
	\tau^{-\alpha} D_t^{1 - \alpha} f + \tau^{\alpha} D_t^{\alpha}(\vec{v}_g \cdot \nabla  f) &= \frac{f_0 - f}{\tau}\\
	\frac{\partial f}{\partial t}+ \tau^{\alpha} D_t^{\alpha}(\vec{v}_g \cdot \nabla  f) &= \frac{f_0 - f}{\tau}\\
\end{align*}
\begin{align}
	\label{lc45}
	\vec{q} + \tau^{1 - \alpha} D_t^{1 - \alpha} &= - \lambda \nabla T, \\
	\vec{q} + \tau^{1 - \alpha} D_t^{1 - \alpha} &= - \tau^{\alpha} D_t^{\alpha} (\lambda \nabla T), \\
	\vec{q} + \tau \frac{\partial \vec{q}}{\partial t} &= - \tau^{\alpha} D_t^{\alpha} (\lambda \nabla T). 
\end{align}

In the case of the phonon heat transport equation (\ref{goy}) is re-written as 
\begin{equation}
	\label{lc23}
	\frac{\partial f_p}{\partial t} + \vec{v}_g \cdot \nabla f_p = \tau^{\alpha} D_t^{\alpha} \left( \frac{f_0 - f_p}{\tau} \right),
\end{equation}
where $f_p = f_p (\vec{x},t,\vec{k})$ is the phonon distribution function, $\vec{k}$ is the wave vector, $\vec{v}_g$ is the group velocity,
\begin{equation}
	f_0 = \frac{1}{exp \left( \frac{\hbar \omega}{k_B T}\right) - 1}  
\end{equation} 
is the Planck distribution, $\omega$ is the angle velocity.

Multiplying of equations (\ref{lc45} - 7.8) by $\hbar \omega$  and integrating over the wave vector space lead to the fractional order energy conservation equations
\begin{equation}
	\tau^{- \alpha} D_t^{1 -\alpha} e = -\vec{\nabla} \cdot \vec{q},
\end{equation} 
\begin{equation}
	\tau^{- \alpha} D_t^{1 -\alpha} e = - \tau^{\alpha} D_t^{\alpha} \vec{\nabla} \cdot \vec{q},
\end{equation}
\begin{equation}
	\partial_t e = - \tau^{\alpha} D_t^{\alpha} \vec{\nabla} \cdot \vec{q}.
\end{equation}

\section{Elasticity and thermal expansion coupling}

Elasticity is not connected to the Fourier heat conduction without thermal expansion. However, with the nonzero thermal expansion the position dependent temperature influences the strains and displacements. The coupled equations of the Fourier heat conduction, of the elastic mechanics and of the kinematics relationships lead, after elimination of the mechanic and the kinematic quantities, to the equation for the temperature that contains the higher derivative corrections to the Fourier equation. 

F\"ul\"op et al. \cite{fu18} analysed the case of the small strain regime of the homogeneous isotropic solid material at rest with constant thermodynamic properties. The constitutive relations for the Hooke-elastic material  are written as
\begin{equation}
\hat{\sigma}^d = E^d \hat{D}^d, \; 	\hat{\sigma}^s = E^s \hat{D}^s,
\end{equation} 
\noindent where 
$E^d = 2 G, \; E^s = 3 K$,
superscripts $^d$ and $^s$ denote the deviatoric (traceless) and the spherical parts, respectfully, of the stress tensor $\hat{\sigma}$ and strain tensor $\hat{D}$:
\begin{equation}
	\hat{D}^d = \hat{D} - \hat{D}^s, \qquad \hat{D}^s = \frac{1}{3}(tr \hat{D}) \hat{I}, 
\end{equation}
\noindent where $\hat{I}$ is the unit tensor, thus
\begin{equation}
	\hat{\sigma} =  E^d \hat{D}^d + E^s \hat{D}^s = 2G \hat{D} + (3 K - 2 G) \hat{D}^s.
\end{equation}

Stress induces the time derivative in the velocity field $\vec{v}$ of the medium
\begin{equation} 
	\rho \frac{d \vec{v}}{d t} = \vec{\nabla} \cdot \hat{\sigma},
\end{equation}
\noindent the symmetric part $\vec{L}^{sym}$ of the velocity gradient $\vec{L}$ for the case of the constant thermal expansion coefficient $\alpha$
\begin{equation}
	\vec{L}^{sym} = \frac{d \hat{D}}{d t} + \alpha \frac{d T}{d t} \hat{I}. 
\end{equation}

The balance of the specific  internal energy
\begin{equation}
	e = c T + \frac{3 K \alpha}{\rho} T tr \hat{D}^s + e_{el}, \qquad e_{el} = \frac{G}{\rho} tr (\hat{D}^d)^2 + \frac{3 K}{2 \rho} tr (\hat{D}^s)^2 
\end{equation}
\noindent  is written as \cite{fu18}
\begin{equation}
	\rho \frac{d e} {d t} = tr (\hat{\sigma} \vec{L}) - \vec{\nabla} \cdot \vec{q}, 
\end{equation}
\noindent where the heat flux for the constant thermal conductivity $\lambda$
\begin{equation}
	\vec{q} = \nabla T.
\end{equation}

Elimination of the stress tensor and the subsequent elimination of the strain tensor leads to the equation for the temperature that accounts for effect of the coupling of the elasticity and thermal expansion to the heat conduction \cite{fu18}
\begin{equation}
	\frac{d^2}{d t^2} \left( \gamma_1 \frac{d T}{d t} - \lambda \Delta T \right) = c^2_2 \Delta \left( \gamma_1 \frac{d t}{d t} - \lambda \Delta T \right) + \frac{(3 K \alpha)^2 T_0}{\rho} \Delta \frac{d T}{d t},
\end{equation}
\noindent where
\begin{equation}
\gamma_1 = \rho c - 9 K \alpha^2 T_0, \qquad c_2^2 = \frac{4 G + 3 K}{3 \rho}. 	
\end{equation}
 
Coupling of the thermal expansion and elasticity  result in the modification of the thermal diffusivity $\kappa = \lambda /(\rho c)$ to the effective diffusivity $\kappa_{eff} = \lambda / \gamma_2$,  
\begin{equation}
	\gamma_2 = \rho c -\frac{6 E^d E^s \alpha^2 T_0}{2 E^d + E^s}.
\end{equation}

The shift is rather small, for example, it is about 1 \% for steel and copper and about 6 \% for aluminium at room temperature \cite{fu18}. \

 Bargmann \& Favata \cite{bar14} developed the continuum thermomechanical analysis of the laser-pulsed heating in polycrystals that includes the highly nonlinear strongly coupled system of governing equations describing heat conduction, species diffusion and finite elastic effects.

Laser pulse transfers the significant amount of energy that is tightly packed in time and space and causes coupled thermal and mechanical effects. These effects are either desirable (e.g., the attenuation of defects in metals and semiconductors due to the annealing) or undesirable (for example, laser-induced damage).
The authors considered three evolution phenomena:
{thermal conduction (thermal waves),}
{lattice straining (mechanical waves),}
{defect dynamics}
and formulated the continuum model based on the balance of the internal energy, the linear momentum, concentration, order-parameter and the entropy inequality.

The energy equation is written (for each body part $P$) as
$\dot{\mathcal{E}} (P) = \mathcal{Q} (P) + \Pi^s (P) + \mathcal{T} (P)$ 
where
\begin{equation}
	\mathcal{E} (P) = \int_P \rho_0 \epsilon dV,
\end{equation}
$\epsilon$ is the volume density of internal energy, the heat inflow
\begin{equation}
	\mathcal{Q} (P) = \int_{\partial P} \vec{Q} \cdot \vec{n} dA + \int_P \rho_0 r dV,
\end{equation}
$\vec{Q}$ is the heat flux, $r$ is the heat source, the stress power
\begin{equation}
\Pi^s (P) = \int_P \hat{P} : \dot{\vec{F}} dV,	
\end{equation}
$\hat{P}$ is the Piola-Kirchhoff stress tensor, $\vec{F}$ is the deformation gradient, the energy flow due to the species transport
\begin{equation}
\mathcal{T} (P) = \int_{\partial P} \mu \vec{D} \cdot \vec{n} dA + \int_P \mu d dV,	
\end{equation}
$\mu$ is the chemical potential, $\vec{D}$ is the defect flux, $d$ is the defect density. 

The authors formulated the energy balance equation as
\begin{equation}
	\rho \dot{\epsilon} = - div \vec{Q} + \rho r + \hat{P}:\dot{\vec{F}} + \mu \dot{n} - \vec{D} \cdot \nabla \mu.
\end{equation}

There are two contributions into the energy equation due to the coupling of different phenomena: the first containing  Piola-Kirchhoff stress tensor and the deformation gradient called Gough-Joule effect (deformations lead to heating), the second one is responsible for heating due to the species diffusion \cite{bar14}.

\subsection{Non-Fourier Thermoelasticity}
 
 The classical thermoelasticity theory (Biot's theory) is based on the Fourier equation \cite{biot}.
 One of the first studies of the "thermoelasticity with second sound" was performed  by Chandrasekharaiah \cite{cha86}. Later he used a generalized linear thermoelasticity theory to study the phenomena in piezoelectric media \cite{pte3}.
 The reviews of the early studies in the hyperbolic
 thermoelasticity was published by Chandrasekharaiah
 \cite{cha98} and Hetnarski and Ignaczak \cite{het99}. 
 
 Lord \& Shulman \cite{ls67} and Green \& Lindsay \cite{gl72} have extended the  thermoelasticity by introducing the thermal relaxation times in the constitutive equations.
 Lord \& Shulman  have considered isotropic solids and introduced one relaxation time parameter into  the Fourier heat conduction equation, thus the heat equation associated with this theory is hyperbolic. A direct consequence  is that the paradox of infinite speed of propagation inherent in both the uncoupled and coupled
 theories of classical thermoelasticity is eliminated. 
  The non-Fourier model with one relaxation time  was  extended to the anisotropic case by Dhaliwal \& Sherief \cite{dha}. 
  Green \& Lindsay  have developed a temperature rate-dependent thermoelasticity. 
 In this theory, both the equations of motion and the heat conduction equations are hyperbolic. 
 
 Another 
 generalization  is known as the theory of thermoelasticity with two
 relaxation times.  Also, Green \&  Naghdi \cite{gr1,gr2,gr3} have proposed three  formulations of thermoelasticity based on inequality of entropy balance \cite{gre72a}.
 Borjalilou et al. used the dual-phase-lag model to solve the problem of the thermoelasticity --- they analysed the damping of  the micro-beams \cite{bor19}. Luo et al. \cite{luo21} studied numerically the transient
 thermoelastic responses of a slab with temperature-dependent thermal conductivity.
   The triple-phase-lag model has been used for the thermoelasticity problems (e.g., \cite{cho07,kum16}), including magneto-thermoelastic problems in a piezoelastic medium \cite{tiw}.
 
\paragraph{Fractional Thermoelasticity}
Recently Sheordan \& Kundu \cite{te3} reviewed  the  fractional order generalized thermoelasticity theories and considered
\begin{itemize}
	\item {Sherief's model \cite{sher}.
The heat conduction equation is written as 	
\begin{equation}
	q_i + \tau_0 \frac{\partial^{\alpha}}{\partial t^{\alpha}} q_i = -\kappa_{ij}T_{,j}, \quad 0 < \alpha \le 1;
\end{equation}
}
	\item {Youssef's model \cite{you10}. The heat conduction equation is written as
	\begin{equation}
			q_i + \tau_0 \frac{\partial^{\alpha}}{\partial t^{\alpha}} q_i = -\kappa_{ij} I^{\alpha - 1} T_{,j},\quad 0 < \alpha \le 2,
	\end{equation}
where
\begin{equation}
	 I^{\alpha - 1} f (x,t) = \frac{1}{\Gamma (\alpha)} \int_0^t (t - s)^{\alpha - 1} f (x, s)ds;
\end{equation}
}
\item {Ezzat's model \cite{ezz10}
\begin{equation}
		q_i + \frac{\tau_0}{\Gamma (\alpha + 1)} \frac{\partial^{\alpha}}{\partial t^{\alpha}} q_i = -\kappa_{ij}T_{,j}, \quad 0 < \alpha \le 1;
\end{equation}
}
\item {Ezzat's TPL model (see section \ref{ftpl1}).}
\end{itemize}

\section{Conclusions}
The review is attempted to cover all major models designed to study the heat transfer beyond the Fourier conduction model when either temporal or spatial non-locality (or both) is essential. Thermodynamic and fractional models seem to be the most perspective. The relaxon model is very attractive for the study of heat conduction in dielectrics and semiconductors. The heat transfer in biological tissues is studied in majority of cases using the DPL model, in spite of its drawbacks \cite{gha15,sho21}.

\bibliographystyle{acm}
\bibliography{non_f}
\end{document}